 \documentclass[%
  aip,jcp,
 amsmath,amssymb,
  reprint,%
 ]{revtex4-1}

\usepackage{graphicx}
\usepackage{dcolumn}
\usepackage{bm}

\begin{document}

\preprint{APS/123-QED}

\title{Defects in Crystals of Soft Colloidal Particles}

\author{Marjolein de Jager}
\author{Joris de Jong}
\author{Laura Filion}%
\affiliation{Soft Condensed Matter, Debye Institute of Nanomaterials Science, Utrecht University, Utrecht, Netherlands}


\date{\today}

\begin{abstract}
In this paper we use computer simulations to examine point defects in systems of ``soft'' colloidal particles including Hertzian spheres, and star polymers.
We use Monte Carlo simulations to determine the deformation of the different crystals associated with vacancies and interstitials and use thermodynamic integration to predict the equilibrium concentrations of such defects. 
We find that the nature of the lattice distortion is mainly determined by the crystal structure and not by the specifics of the interaction potential. 
We can distinguish one-, two-, and three-dimensional lattice distortions and find that the range of the distortion generally depends on the dimensionality.
We find that in both model systems the deformation of the body-centered cubic (BCC) crystal caused by an interstitial is one-dimensional and we show that its structure is well described as a crowdion. Similarly, we show that the one-dimensional deformation of the hexagonal (H) crystal of Hertzian spheres caused by a vacancy can be characterized as a voidion. Interestingly, with the exception of the FCC crystal in the Hertzian sphere model, in all cases we find that the interstitial concentration is higher than the vacancy concentration. Most noteworthy, the concentration of interstitials in the BCC crystals can reach  up to  1\%.

\end{abstract}

\maketitle


\section{\label{sec:level1}Introduction}

In equilibrium, all crystals contain a finite concentration of point defects, such as vacancies and interstitials. The way these defects manifest themselves can play an important role in the mechanical, optical, and electronic properties of the crystalline material. 

One of the few three-dimensional models in which point defects have been examined is the face-centered cubic (FCC) crystal of single-component hard spheres. Studies have shown that in this system point defects cause relatively local lattice distortions and occur in low concentrations, $10^{-4}$ vacancies and $10^{-8}$ interstitials per lattice site near melting \cite{bennett1971studies,pronk2001point}. 
For a couple of decades, this uneventful defect behavior was expected for other crystals as well, discouraging the examination of point defects in other three-dimensional model systems.

However, relatively recent work on simple cubic (SC) lattices have shown that at least for some colloidal crystals, even simple point defects can extend over large areas and occur in high concentrations \cite{smallenburg2012vacancy,van2017phase,van2018revealing}. In particular, it was found that these SC crystals contain large numbers of vacancies, up to 0.06 per lattice site, that cause one-dimensional lattice distortions along the three main crystalline lattice directions. These one-dimensional deformations are aptly named voidions, as a counterpart to the crowdion -- a largely one-dimensional lattice distortion caused by an interstitial.
Since its proposed existence by Paneth in 1950 \cite{paneth1950mechanism}, the crowdion has been explored in multiple atomic BCC crystals \cite{han2002self,nguyen2006self,derlet2007multiscale}. Moreover, crowdions have recently been predicted to exist in relatively high quantities in the colloidal BCC crystal of repulsive point Yukawa particles \cite{alkemade2021point}. 
This all leads to the question of the importance of the crystal structure in determining the defect behavior and whether other non-FCC crystal structures show interesting defect behaviors as well.

One easy way to access more exotic crystal phases is to add a ``soft'' repulsion between spherical particles -- for instance, charged colloids \cite{kremer1986phase, robbins1988phase,monovoukas1989experimental,yethiraj2003colloidal,smallenburg2011phase}, star polymers \cite{watzlawek1999phase, ren2016star}, colloids covered by a thick polymer layer \cite{riest2015elasticity, salerno2014coating}, or microgel particles \cite{hashmi2009mechanical, mohanty2014effective, bergman2018new, louis2000can, rovigatti2019connecting, pamies2009phase}. 
In this paper, we use computer simulations to explore both the structure and concentration of point defects in the crystal phases of two of these model systems: star polymers and Hertzian spheres. 
We find similar deformations of the FCC and BCC crystals of both models, indicating that the nature of the lattice distortion is mainly determined by the crystal structure and not by the specifics of the model.  The one exception observed was the SC phase in the Hertzian model whose vacancies were extended in two dimensions, in contrast the the one-dimensional distortions seen in the previous studies of SC crystal of repulsive particles \cite{van2018revealing}. 
We did, however, observe a number of other ``one-dimensional'' defects, including  voidions in the hexagonal (H) crystal of Hertzian spheres and crowdions in  BCC crystals of both models, with the concentration of the latter reaching up to  0.01 interstitials per lattice site.


\section{Models}
In this paper we treat two soft colloidal models: Hertzian spheres and star polymers.
The first model describes elastically deformable spheres of diameter $\sigma$ that have a softly repulsive interaction given by 
\begin{equation}
	\label{eq:pothertz}
	\phi(r) = \begin{cases}
		\epsilon \left( 1-\frac{r}{\sigma} \right)^{5/2}  & \quad r\leq\sigma,\\
		0       & \quad r>\sigma,
	\end{cases}
\end{equation}
where $r$ is the center-of-mass distance between the two Hertzian spheres \cite{landaubook}. Here $\epsilon>0$, such that the interaction is repulsive. 
The phase behavior of Hertzian spheres is fully characterized by the dimensionless temperature $k_BT/\epsilon$ and density $\rho\sigma^3$, with $k_B$ the Boltzmann constant and $T$ is the temperature.
In Ref. \onlinecite{pamies2009phase}, P\`amies \textit{et al.} predicted that Hertzian spheres exhibit a stable FCC, BCC, H, SC, and body-centered tetragonal (BCT) phase and computed the phase diagram using free-energy calculations. See Fig. \ref{fig:crystals} for schematic representations of these crystal structures.

The second model describes colloids consisting of a negligible small central particle with polymer chains -- the so-called ``arms'' -- attached to it. 
The purely entropic interaction between two star polymers at distance $r$ is given by
\begin{equation}
\label{eq:potstarpol}
\begin{split}
	&\phi(r) = \\
	& \;\; \frac{5}{18} k_BT f^{3/2} \begin{cases}
		-\ln{\frac{r}{\sigma}} + \frac{1}{1+\sqrt{f}/2}  & r\leq\sigma,\\
		\frac{\sigma/r}{1+\sqrt{f}/2} \exp\left[ -\sqrt{f}(r-\sigma)/2\sigma \right]      & r>\sigma,
	\end{cases}
\end{split}
\end{equation}
where $\sigma$ is the corona diameter and $f$ is the arm number \cite{watzlawek1999phase}. 
The phase behavior of star polymers is fully characterized by the arm number and packing fraction $\eta=\pi\sigma^3 N/6V$, with $N$ the number of particles and $V$ the volume.
In Ref. \onlinecite{watzlawek1999phase}, Watzlawek \textit{et al.} predicted that star polymers are able to from stable FCC, BCC, body-centered orthorhombic (BCO), and cubic diamond crystals (see Fig. \ref{fig:crystals}) and computed the phase diagram using free-energy calculations.

\newcommand{\figwidthSC}{0.18\linewidth}
\newcommand{\figwidthBCC}{0.195\linewidth}
\newcommand{\figwidthHa}{0.18\linewidth}
\newcommand{\figwidthHb}{0.225\linewidth}
\begin{figure*}
\begin{tabular}{lllll}
	\;\; a) SC & \;\; b) BCC/BCT/BCO & \;\; c) FCC & \;\; d) diamond & \; e) H\\[-0.2cm]
	\includegraphics[width=\figwidthSC]{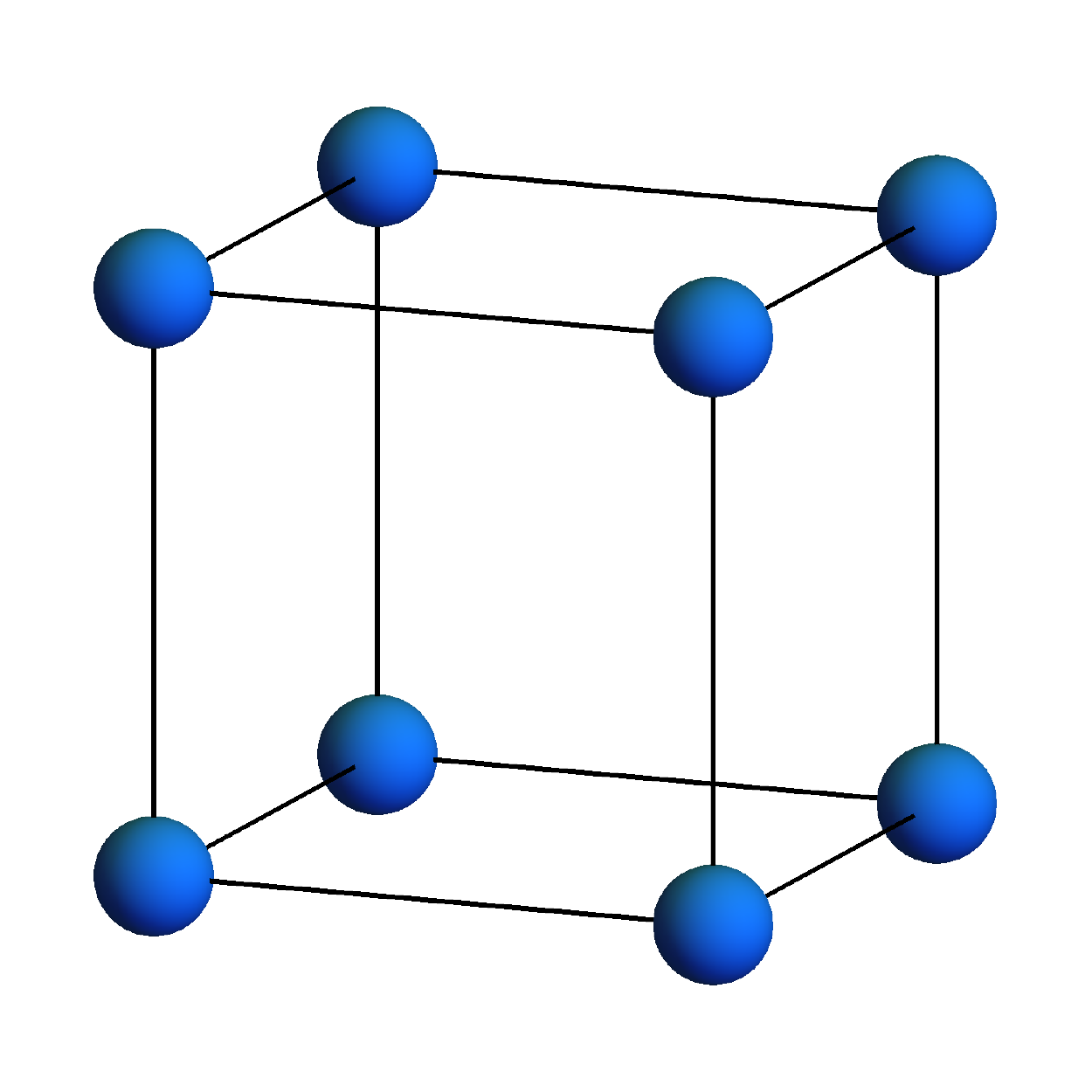} 
    & \includegraphics[width=\figwidthBCC, trim= 0.8cm 1.5cm 0.0cm 0.0cm,clip]{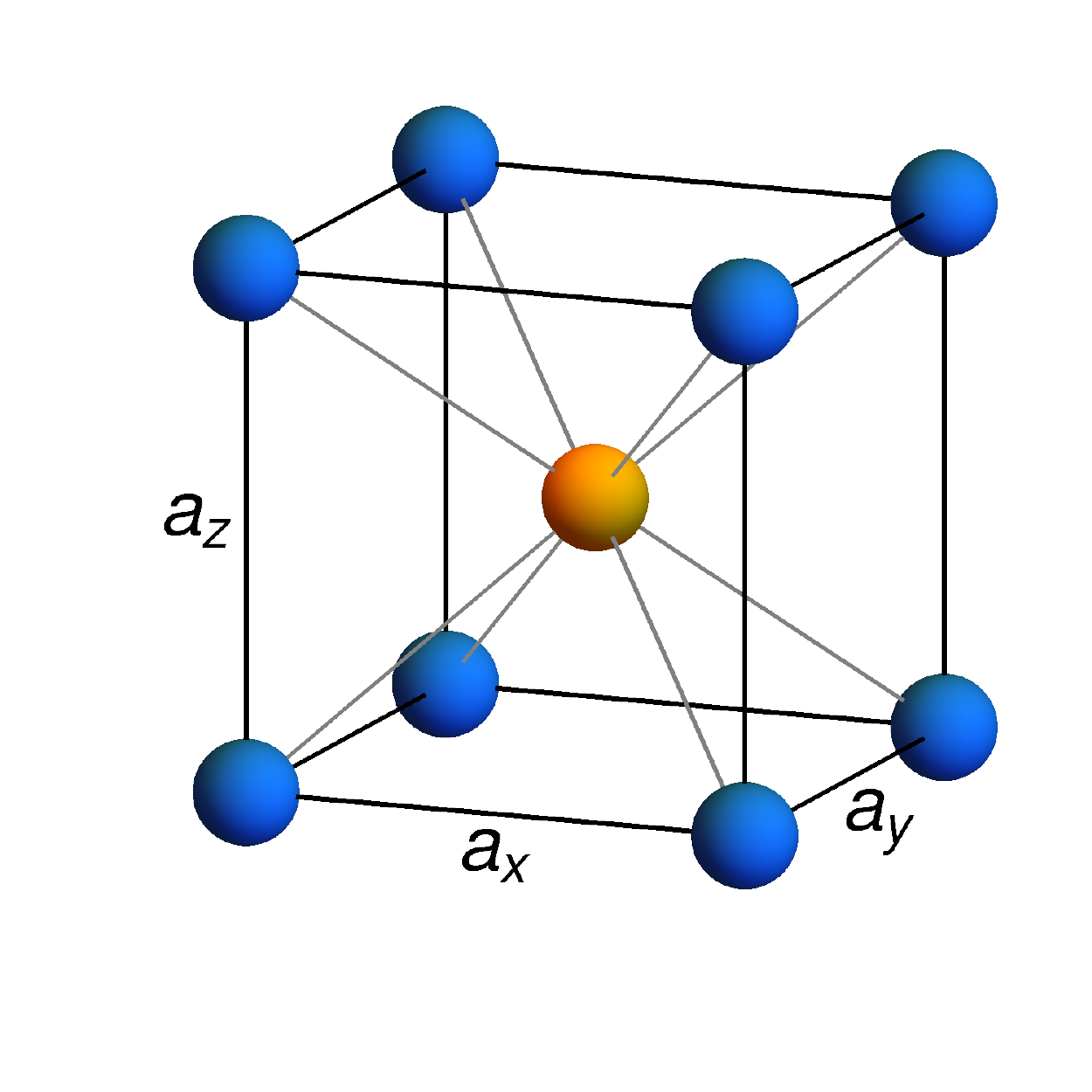} 
    & \includegraphics[width=\figwidthSC]{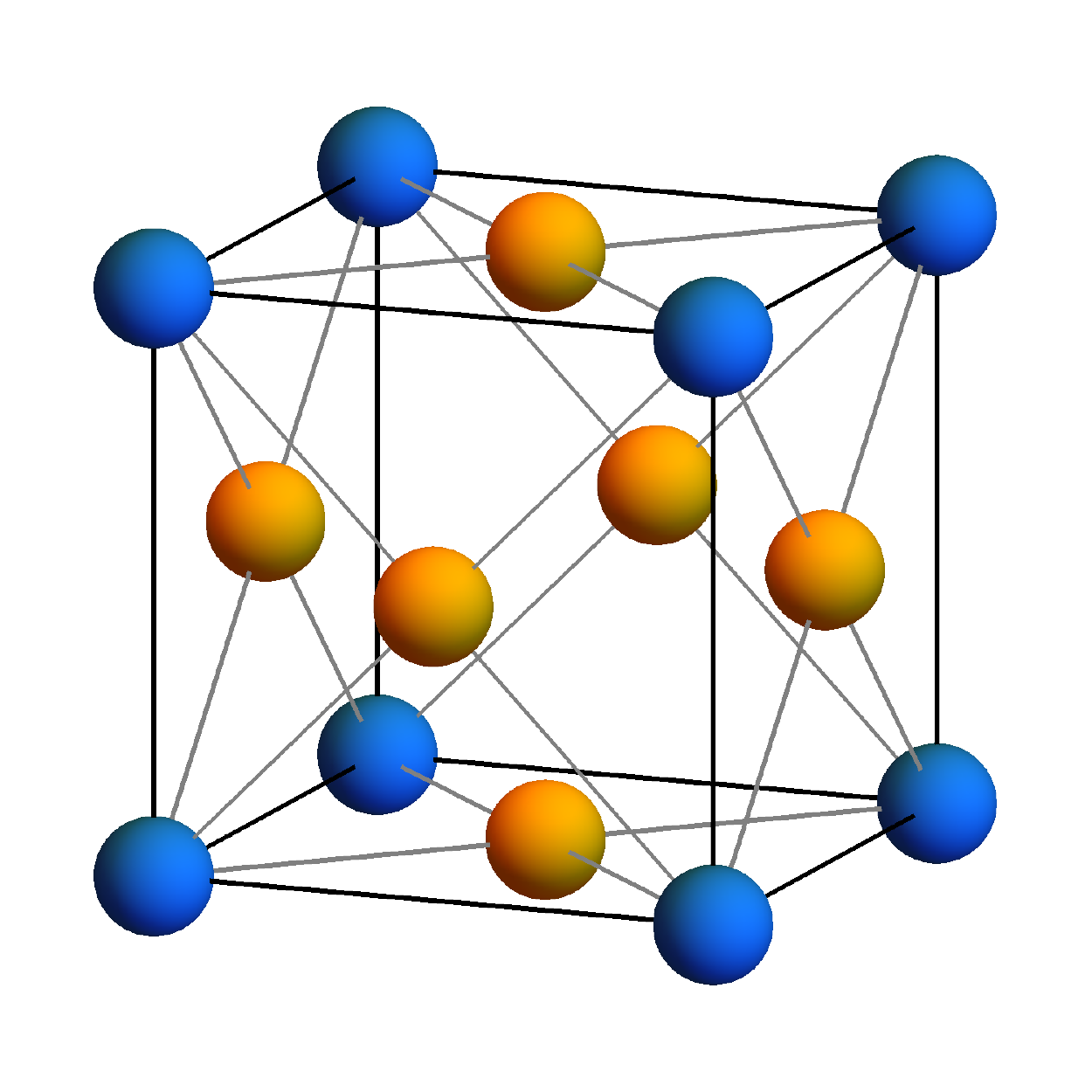}
    & \includegraphics[width=\figwidthSC]{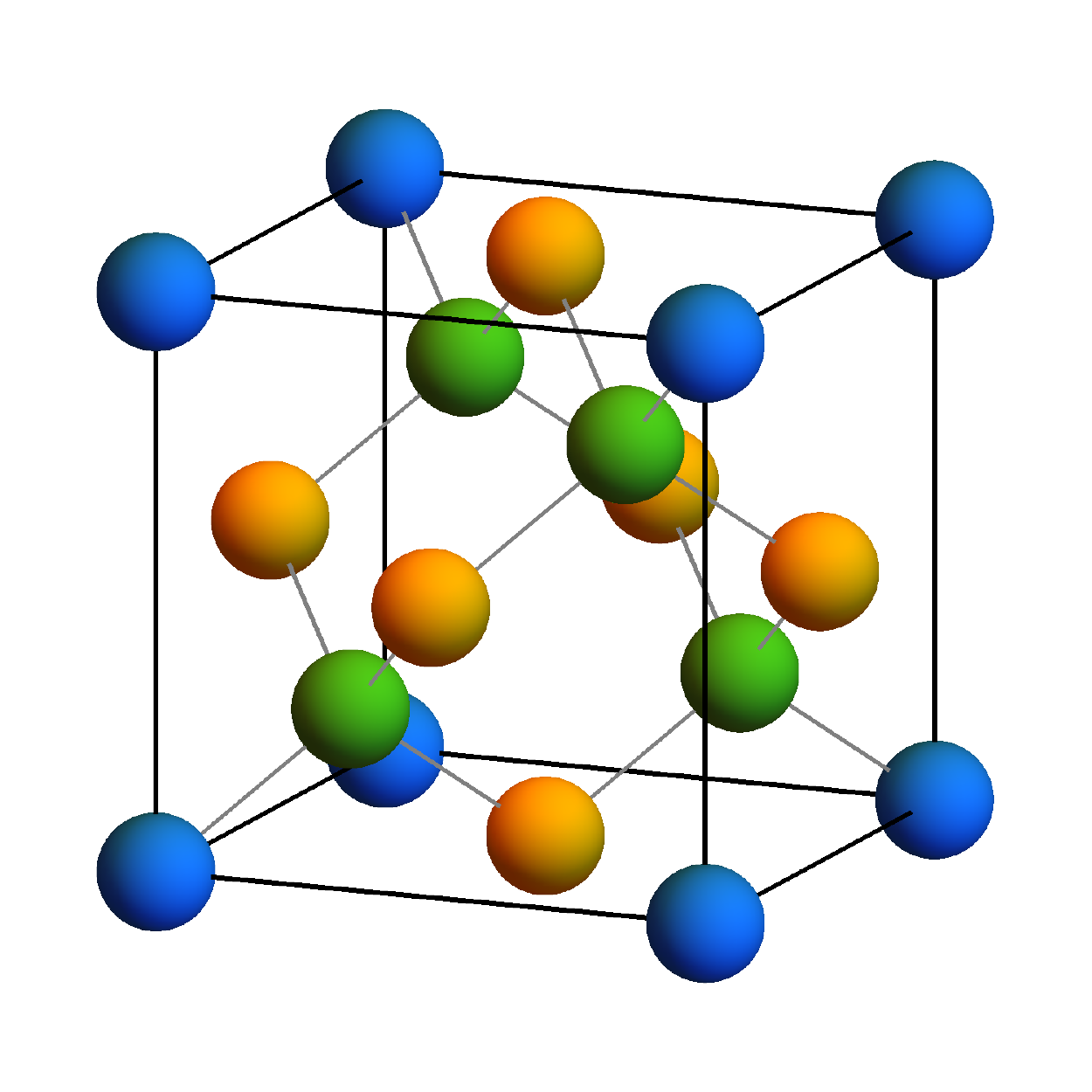}
    & \includegraphics[width=\figwidthHb, trim= 0cm 0.8cm 0.0cm 2.0cm,clip]{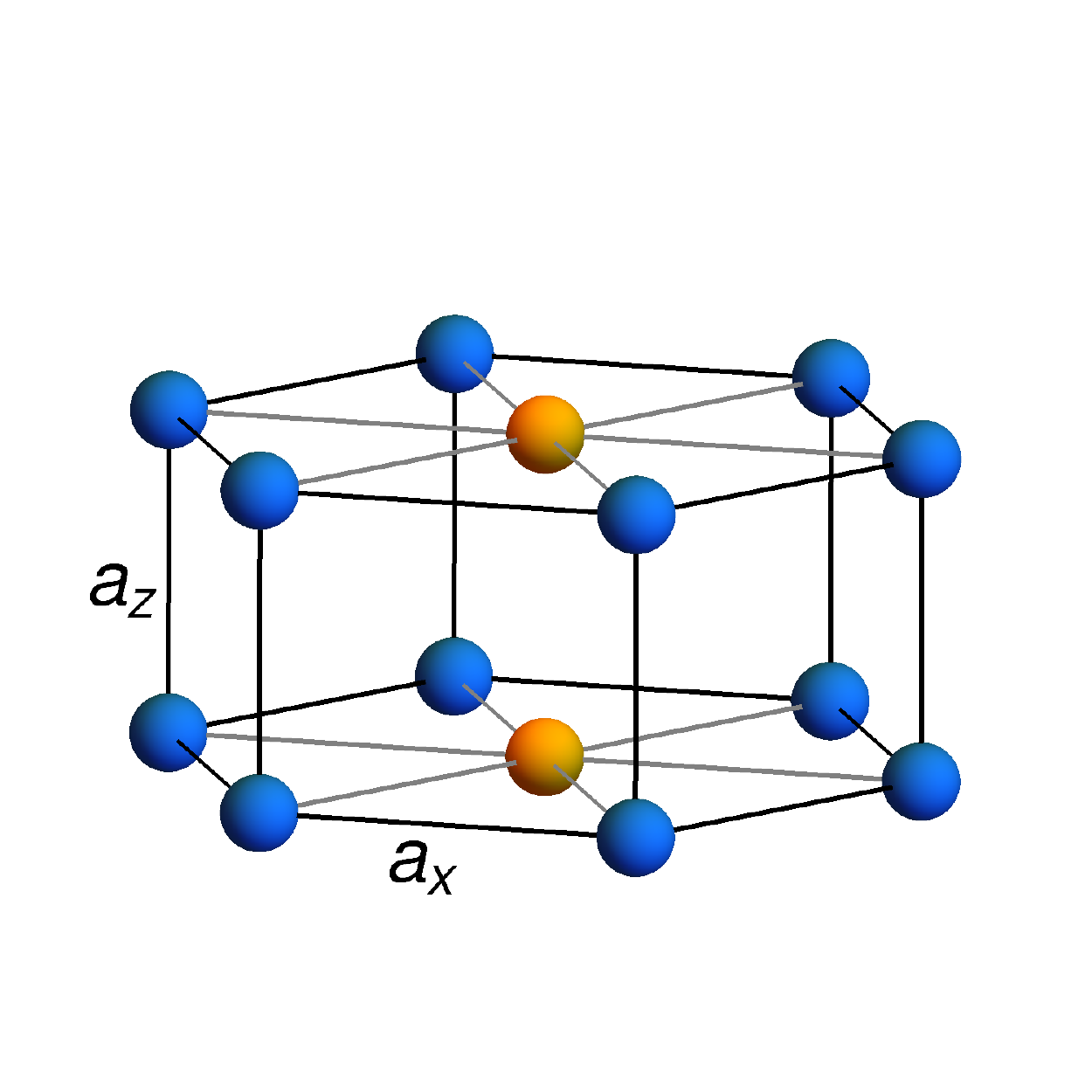}
	\end{tabular}
	\caption[width=1\linewidth]{Schematic representation of the different crystal structures considered in this paper. 
	The colors of the particles have no meaning, but are chosen for clarity purposes.
	Note that for 
	the SC, BCC, FCC, and diamond crystals,
	the corner particles (blue) form a cube ($a_x=a_y=a_z$), whereas for the BCT and BCO crystals they form a rectangular cuboid with $a_x=a_y\neq a_z$ and $a_x\neq a_y\neq a_z$, respectively. 
	For the H crystal, the ratio $a_z/a_x$ is a free parameter.
	}
	\label{fig:crystals}
\end{figure*}


\section{Methods}

\subsection{Lattice deformation}
To examine the manifestation of point defects in the different colloidal crystals, we predicted the average structural deformation associated with a vacancy and with an interstitial, and predicted the equilibrium concentrations of these defects.
For determining the structural deformation, we performed Monte Carlo (MC) simulations in the $NVT$-ensemble \cite{frenkelbook} with a single vacancy or interstitial present. We simulated (approximately) cubic boxes with periodic boundary conditions, and prevented the defect from hopping by confining all particles to their Wigner-Seitz cells \cite{van2017phase, van2020high}. 
The interaction potential of star polymers was truncated and shifted such that the shift was never more than $10^{-5}k_BT$.

First, we looked for any symmetry breaking in the deformations by either slowly quenching the system or going to the zero temperature limit. When, in this limit, the particles belonging to the same neighbor shell are not displaced in a similar, symmetric fashion, we say that the symmetry is broken. In practice, it is sufficient to just look at the nearest and next-nearest neighbor shells.
For the Hertzian spheres, we accessed the zero temperature limit by slowly increasing $\epsilon/k_BT$ during the simulation, while adjusting the maximum displacement of the particles, $\Delta r_\text{max}$, to maintain an acceptance ratio larger than 30\%. Once $\Delta r_\text{max}$ became smaller than $10^{-4}\sigma$, we stopped increasing $\epsilon/k_BT$ and measured the average location of each particle. 
For star polymers, we quenched equilibrated systems by accepting only moves that lowered the potential energy of the system. During the quenching, $\Delta r_\text{max}$ was decreased to compensate for the decrease in accepted moves. As the system can be quenched into one of the many local minima instead of the desired global minimum, we performed multiple quenching simulations for each crystal and took the one with the lowest energy.

Next, when no symmetry breaking was found, we simply calculated the average deformation by measuring the average location of each particle during a simulation. However, when a deformation had a broken symmetry, we averaged over configurations where the deformation had the same orientation. This was done by rotating each snapshot of the crystal after the simulation, such that the deformation had the same orientation in each snapshot. For instance, if the deformation was found to be one dimensional, we rotated all snapshots so that the one-dimensional deformation pointed in the same direction before averaging.

\subsection{Defect concentration}

To predict the defect concentrations, we assumed that the defects do not interact and that their effect on the pressure $P$ of the system is negligible. 
Note that there are methods for calculating true equilibrium concentrations of defects \cite{swope1992thermodynamics}, but we chose to follow the method used by Pronk and Frenkel in Ref. \onlinecite{pronk2001point}.
Within these approximations the vacancy concentration is given by 
\begin{equation}
\label{eq:nvac}
    \langle n_\text{vac}\rangle \equiv \left\langle \frac{M-N}{M}\right\rangle =  \exp{ \left[ -\beta\mu_\text{vac} \right]},
\end{equation}
where $M$ is the number of lattice sites, $\beta = 1/k_B T$, and $\mu_\text{vac}=f_\text{vac}(\rho_M,T) + \mu_\text{df}(P,T)$ with $\rho_M=M/V$ the density of the lattice sites, $f_\text{vac}(\rho_M,T)$ the free energy associated with creating a single vacancy, and $\mu_\text{df}(P,T)$ the chemical potential of the defect-free crystal \cite{pronk2001point}.
Similarly, the interstitial concentration is given by
\begin{equation}
\label{eq:nint}
    \langle n_\text{int}\rangle \equiv \left\langle \frac{N-M}{N}\right\rangle = \exp{\left[ -\beta\mu_\text{int} \right]},
\end{equation}
where $\mu_\text{int} = f_\text{int}(\rho_M,T) - \mu_\text{df}(P,T)$ with $f_\text{int}(\rho_M,T)$ the free energy associated with creating a single interstitial.

We determined the defect-free chemical potential using
\begin{equation}
\label{eq:mu0}
    \mu_\text{df}(P,T) = F_\text{df}(M,V,T)/M + P/\rho_M,
\end{equation}
where we calculated the Helmholtz free energy of the defect-free crystal $F_\text{df}(M,V,T)$ using Einstein integration with finite-size corrections\cite{frenkel1984new,polson2000finite,frenkelbook}, and the pressure $P$ using an $NVT$ MC simulation combined with the virial equation.

Following Ref. \onlinecite{alkemade2021point}, we determine $f_\text{vac (int)}(\rho_M,T)$ using thermodynamic integration in combination with MC simulations. For $f_\text{vac}$, we measured the free-energy difference between a normal particle at a given lattice site and an ideal particle at the same lattice site, and combined this with the free-energy cost of removing the ideal gas particle
\begin{equation}
\label{eq:fremove}
f_\text{remove} = k_BT\ln\left(\frac{V_\text{WS}}{\Lambda^3}\right),
\end{equation}
where $V_\text{WS}=1/\rho_M$ is the volume of the Wigner-Seitz cell and $\Lambda$ is the thermal DeBroglie wavelength.

Similarly, for $f_\text{int}$, we measured the free-energy difference between a system where one of the lattice sites contains two particles, and a system where that lattice site contains one normal particle and one ideal particle \cite{alkemade2021point}. This was then combined with the free energy of inserting the ideal gas particle into the relevant Wigner-Seitz cell $f_\text{insert} = - f_\text{remove}$.


\section{Results}

\subsection{Hertzian Spheres}
We start our investigation by exploring the deformations of the different crystal phases of Hertzian spheres. 
For each crystal phase, we focus on one state point where the crystal is stable, and examine the behavior as a function of temperature and density starting from this state point. 
The temperature at these state points is $k_BT/\epsilon=0.0020$ for all crystal phases and the densities are given in Tab. \ref{tab:hertzstatepoints} together with the number of particles simulated.

\begin{table}
	\centering
	\begin{tabular}{|p{0.18\linewidth}<{\raggedright\arraybackslash}||p{0.13\linewidth}<{\centering} p{0.13\linewidth}<{\centering}|}
		\hline
		Crystal  & $\rho\sigma^3$ & $N$\;\,\\
		\hline \hline
		FCC  & $1.75$  & 1372\;\,\\
		BCC  & $3.00$  & 1024$ ^*$\\ 
		H    & $4.10$  & 1440\;\,\\ 
		SC   & $5.00$  & 1331\;\,\\
		BCT  & $6.46$  & 1240\;\,\\ 
		\hline
	\end{tabular}
	\caption{The number of Hertzian spheres $N$ simulated for each crystal phase together with the density $\rho\sigma^3$. $ ^*$For the interstitial in the BCC crystal we used $N=2000$.}
	\label{tab:hertzstatepoints}
\end{table}

\newcommand{\widthpdB}{0.484\linewidth}
\begin{figure}
\begin{tabular}{ll}
	a) & b)\\[-0.2cm]
	\includegraphics[width=\widthpdB]{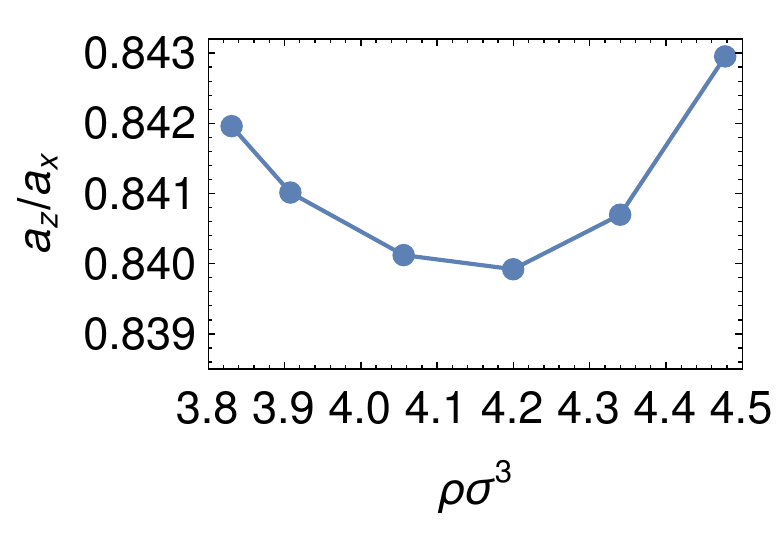} 
	& \includegraphics[width=\widthpdB]{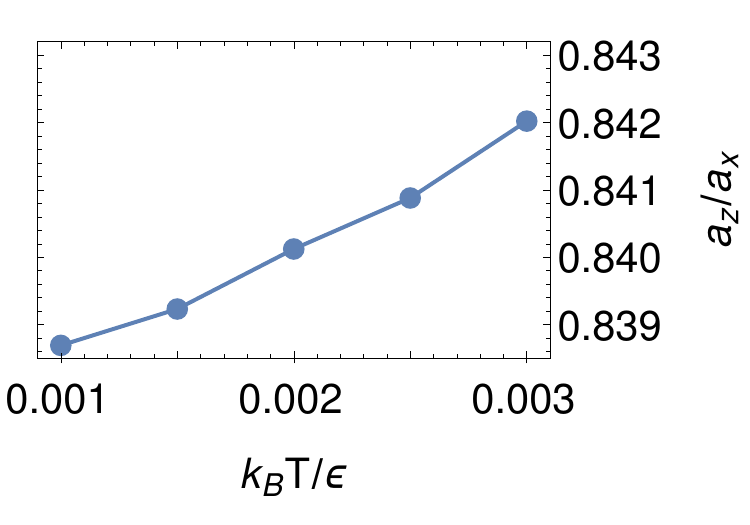} \\[-0.2cm]
	c) & d)\\[-0.2cm]
	\includegraphics[width=\widthpdB]{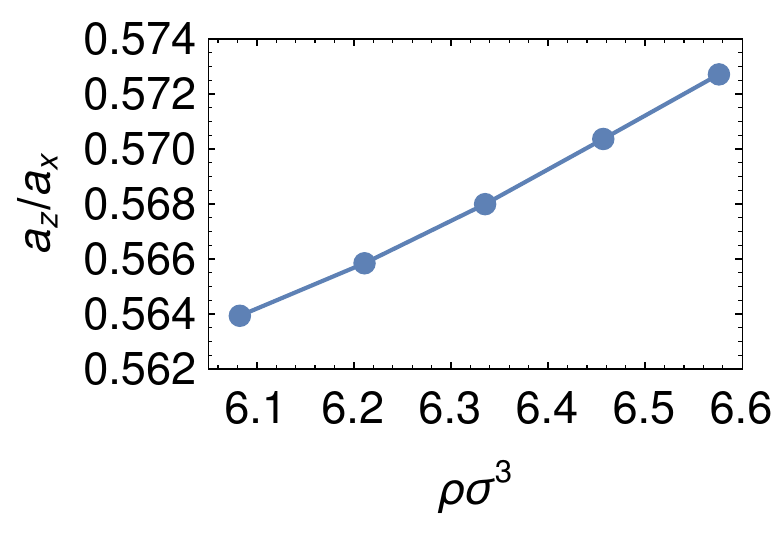} 
	& \includegraphics[width=\widthpdB]{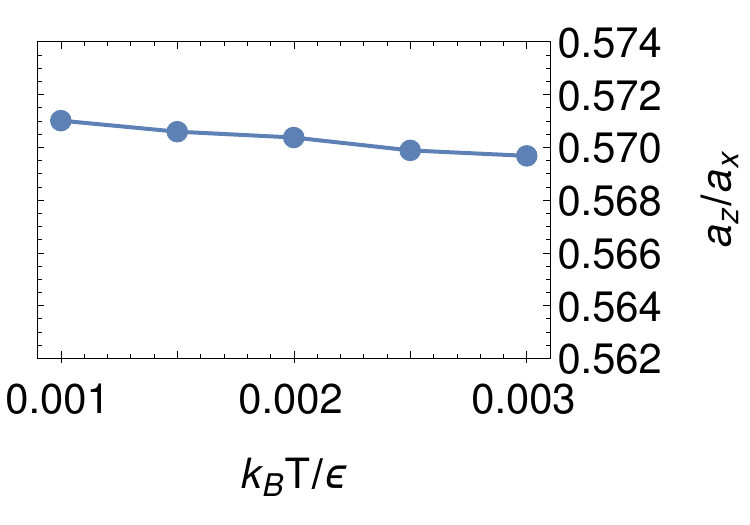} \\
	\end{tabular}
	\vspace{-0.3cm}
	\caption[width=1\linewidth]{
	The ratio $a_z/a_x$ as a function of the density $\rho\sigma^3$ and temperature $k_BT/\epsilon$ for a-b) the H and c-d) the BCT crystals of Hertzian spheres. 
	The results of a,c) are obtained at $k_BT/\epsilon=0.0020$, of b) at $\rho\sigma^3=4.06$ (pressure $P\sigma^3/\epsilon=2.2$), and of d) at $\rho\sigma^3=6.46$ (pressure $P\sigma^3/\epsilon=5.75$).
	}
	\label{fig:hertzunitshape}
\end{figure}

Since the H and BCT crystals have one free parameter, the ratio $a_z/a_x$ with $a_{x(z)}$ as indicated in Fig. \ref{fig:crystals}, we needed to determine the equilibrium value of this ratio first. To this end, we performed MC simulations in the $NPT$-ensemble with disconnected volume changes, i.e. where the volume changes are performed by independently changing the box size in the xy-direction and z-direction. 
We started with an initial guess for $a_z/a_x$ based on the findings of P\`amies \textit{et al.} \cite{pamies2009phase}, and measured the average value after the system has relaxed.
The resulting ratios as a function of the density and temperature are given in Fig. \ref{fig:hertzunitshape}. 
Notice that, for the H crystal, the difference in $a_z/a_x$ is on the order of $0.5\%$. We thus assume that it is negligible, and use $a_z/a_x=0.841$ henceforth. 
For the BCT crystal, the ratio as a function of the temperature differs at most on the order of $0.2\%$. However, for the simulations of various densities the difference is on the order of $1\%$, which we consider to be significant. Thus, for further simulations we take the appropriate $a_z/a_x$ for each density.


\newcommand{\figwidth}{0.28\linewidth}
\newcommand{\legendsizeA}{0.07\linewidth}
\newcommand{\legendsizeD}{0.06\linewidth}
\newcommand{\figwidthB}{0.275\linewidth}
\newcommand{\legendsize}{0.055\linewidth}
\begin{figure*}
\begin{tabular}{lllll}
	& a) & \,\, b) & c) & \\[-0.3cm]
	\includegraphics[width=\legendsizeA,trim= 0cm -0.5cm 0.0cm 0.0cm]{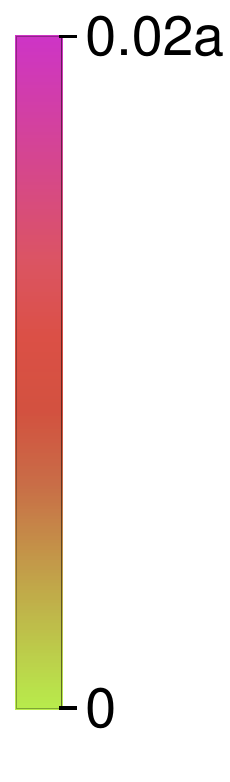} 
    & \includegraphics[width=\figwidth]{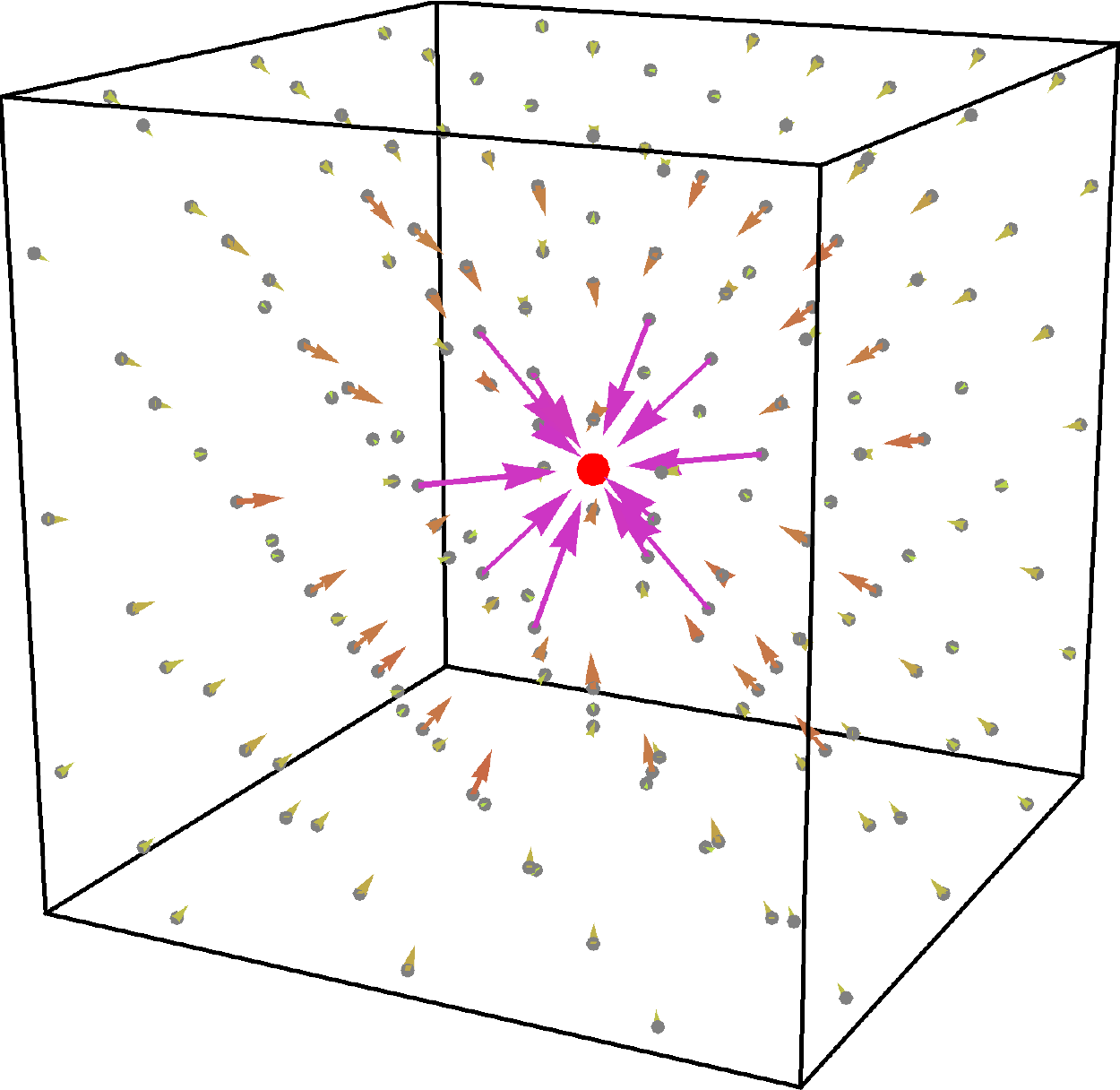} 
    & \includegraphics[width=\figwidthB, trim= 0cm -3.0cm 0.5cm 0.0cm,clip]{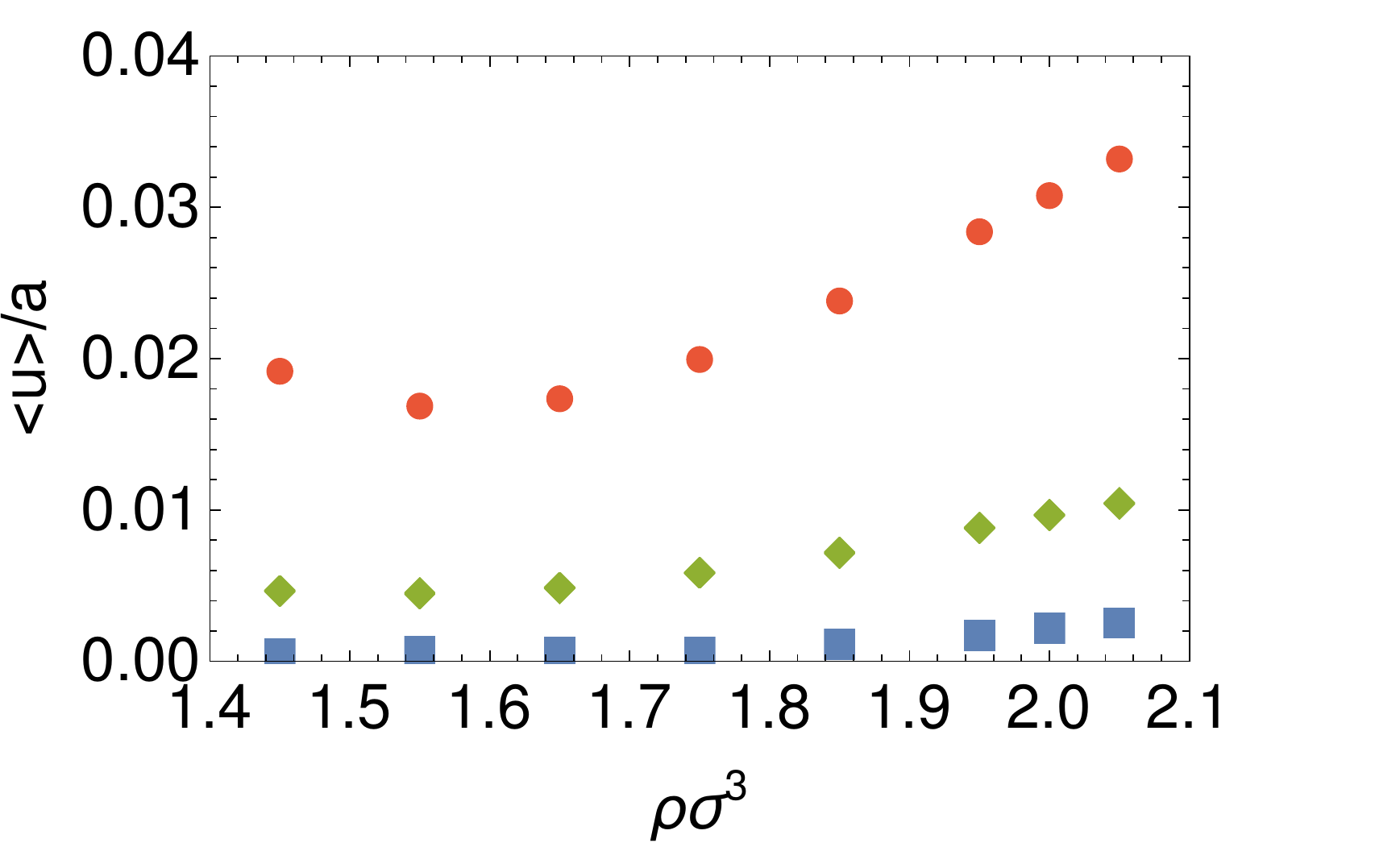}
    & \includegraphics[width=\figwidthB, trim= 0.5cm -2.1cm 0.0cm 0.0cm,clip]{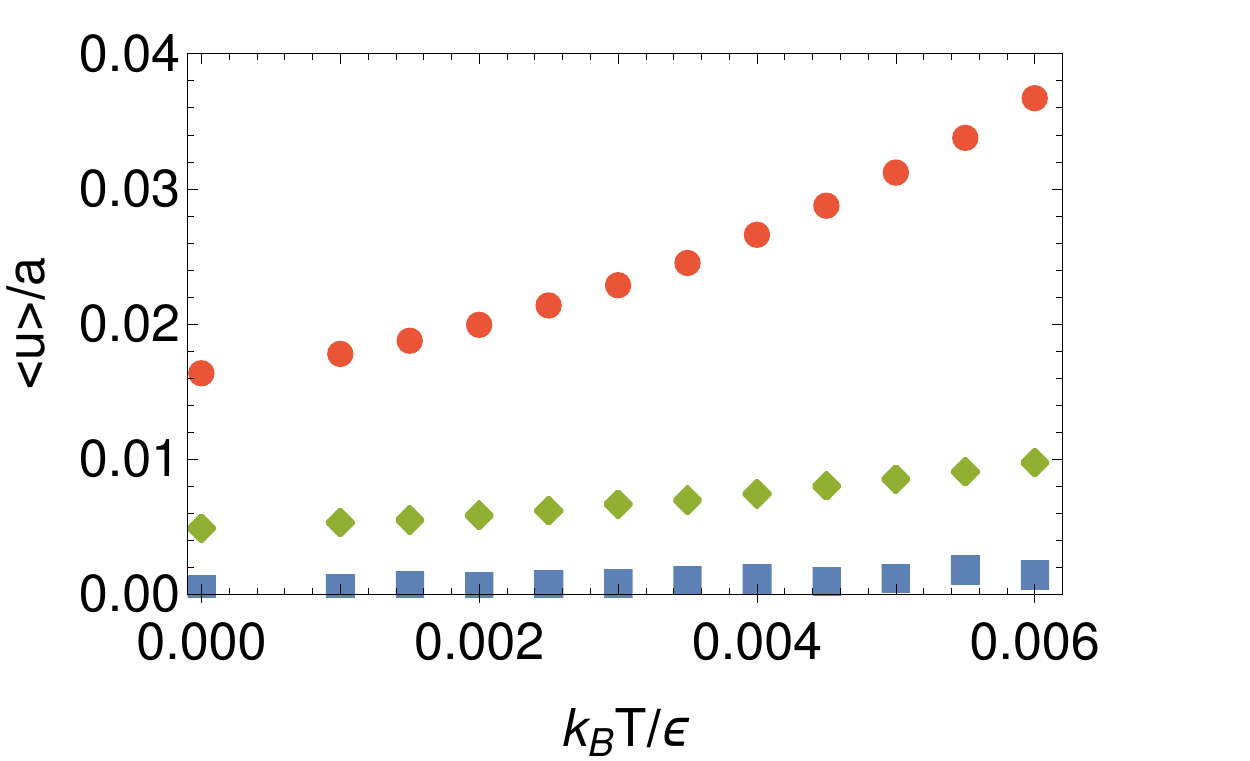}
    & \hspace{-0.2cm}\includegraphics[width=\legendsize,trim= 0.0cm -7.0cm 0.0cm 0.0cm]{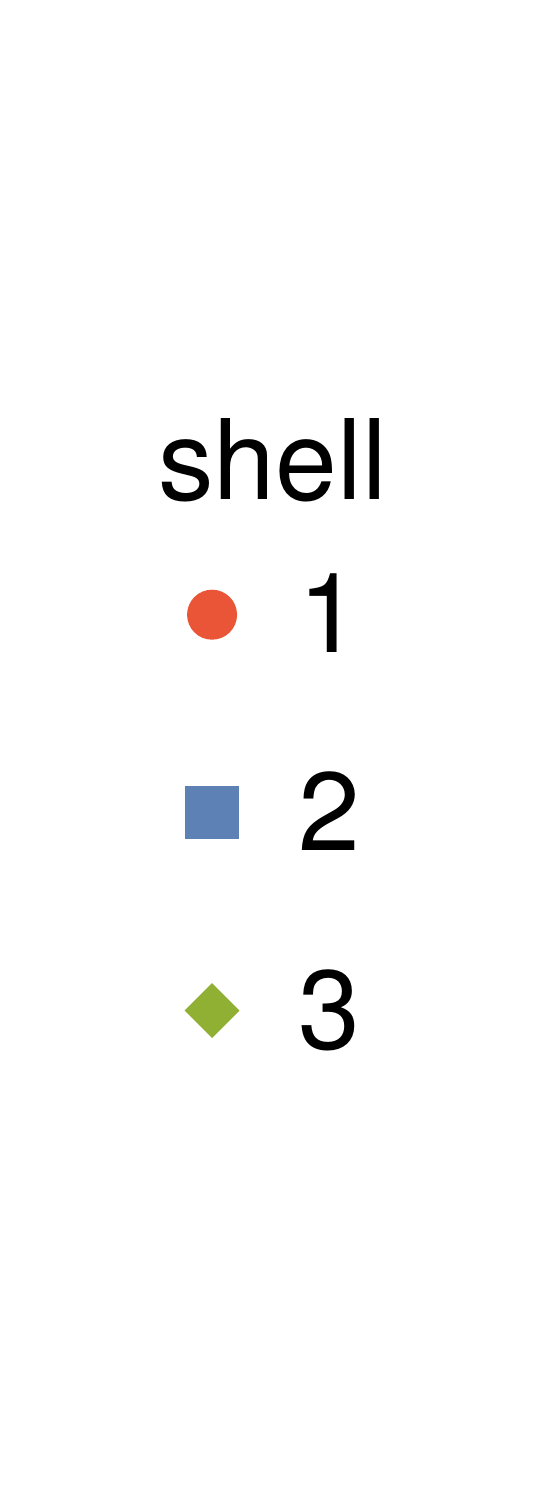} \\
    & d) & \,\, e) & f) &\\[-0.3cm]
    \includegraphics[width=\legendsizeD,trim= 0cm -0.5cm 0.0cm 0.0cm]{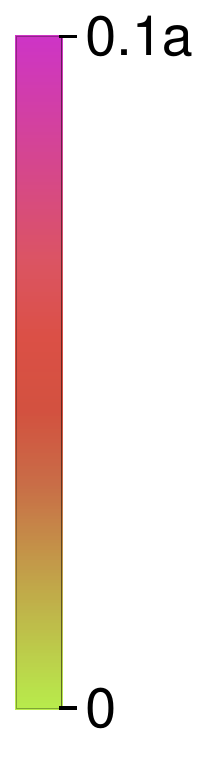} 
    & \includegraphics[width=\figwidth]{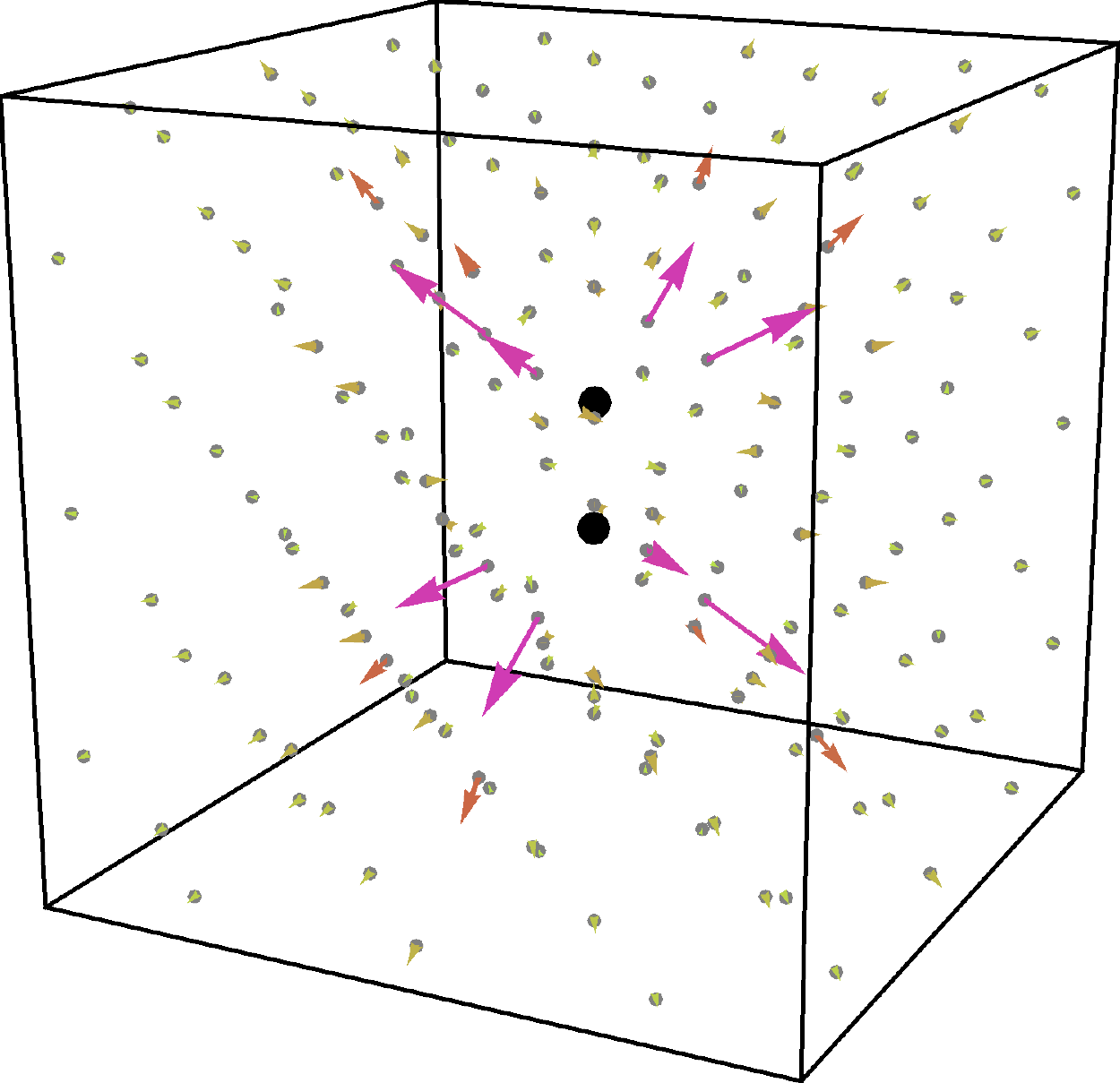} 
    & \includegraphics[width=\figwidthB, trim= 0cm -2.1cm 0.5cm 0.0cm,clip]{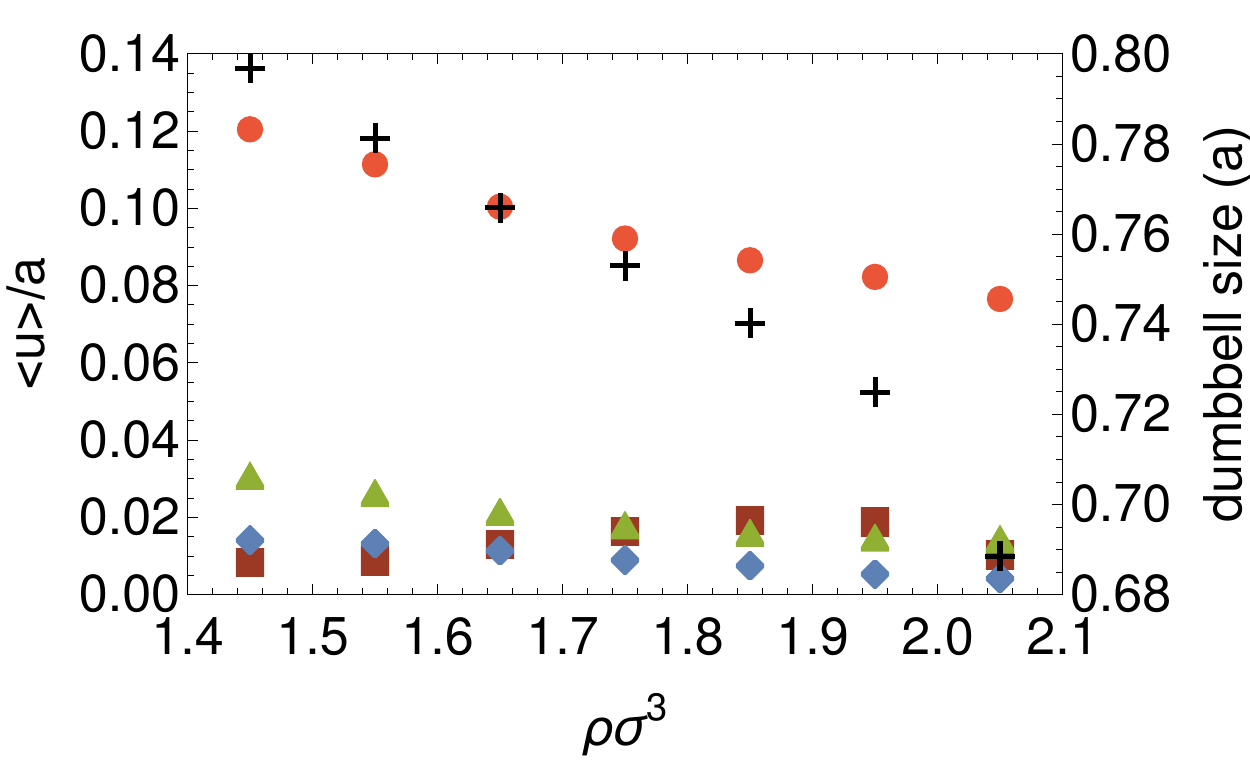}
    & \includegraphics[width=\figwidthB, trim= 0.5cm -2.1cm 0.0cm 0.0cm,clip]{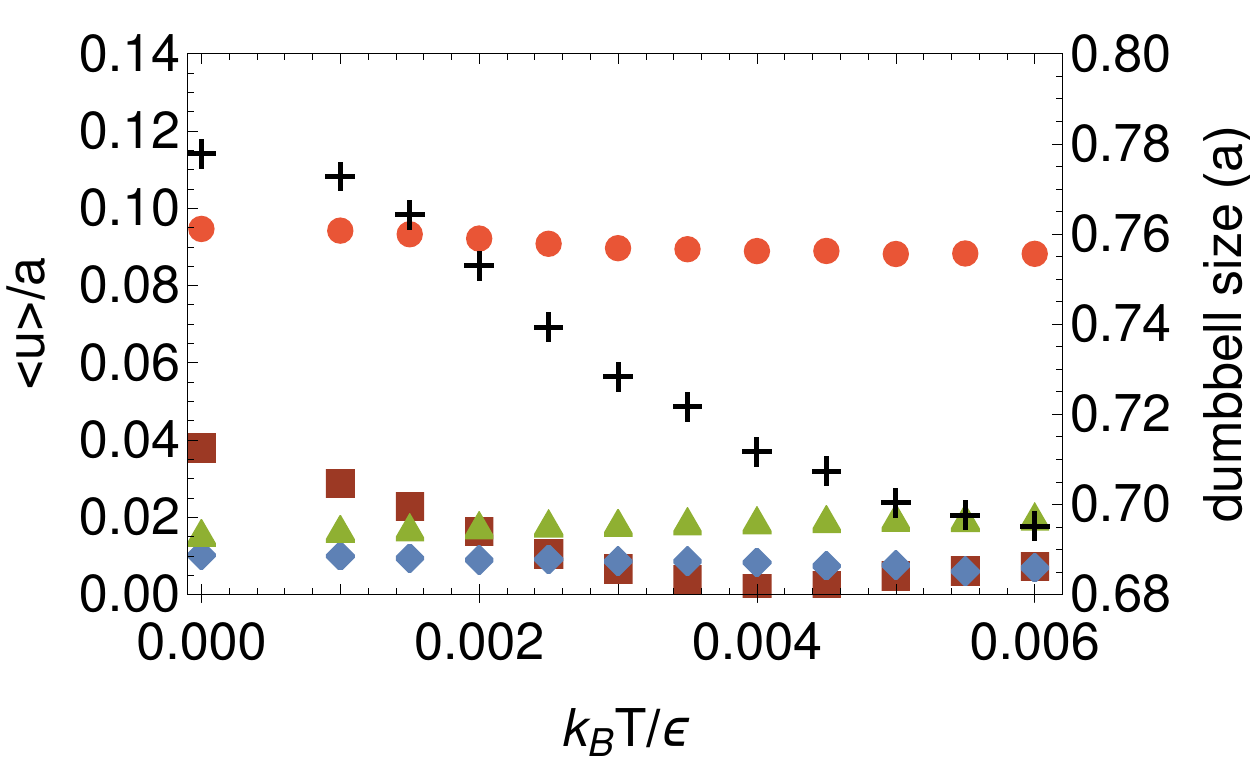}
    & \includegraphics[width=\legendsize,trim= 0.0cm -7.8cm 0.0cm 0.0cm]{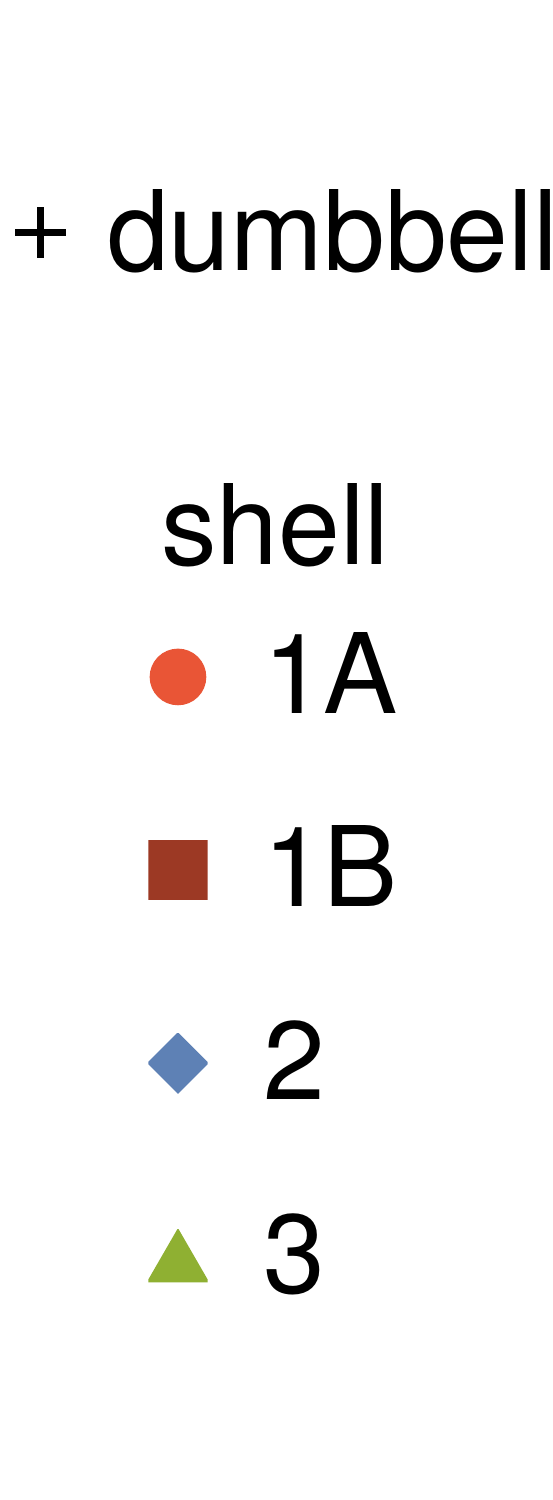}
	\end{tabular}
	\caption[width=1\linewidth]{
	Average deformation of the FCC crystal of Hertzian spheres associated with a-c) a vacancy and d-f) an interstitial. In a,d) the deformation at density $\rho\sigma^3=1.75$ and temperature $k_BT/\epsilon=0.0020$ is shown. The vacancy is indicated by the red dot, and the interstitial and its companion by the two black dots. The gray points represent the lattice sites and the arrows represent the deformation. The size of the arrows is exaggerated, but the color indicates the displacement in terms of the nearest-neighbor distance $a$. 
	b-c,e-f) Average displacement $\langle u\rangle/a$ for the first three neighbor shells (1=nn, 2=nnn, and 3=nnnn) as a function of the density (at $k_BT/\epsilon=0.0020$) and temperature (at $\rho\sigma^3=1.75$). 
	When the displacement of the particles in a neighbor shell has a broken symmetry, the label ``A'' indicates the most displaced particles and ``B'' the others.
	The distance between the interstitial and its companion - the ``dumbbell'' - is given in e-f) as well (right axis). 
	}
	\label{fig:hertzfcc}
\end{figure*}

\begin{figure*}
\begin{tabular}{lllll}
	& a) & \,\, b) & c) & \\[-0.3cm]
	\includegraphics[width=\legendsizeA,trim= 0cm -0.5cm 0.0cm 0.0cm]{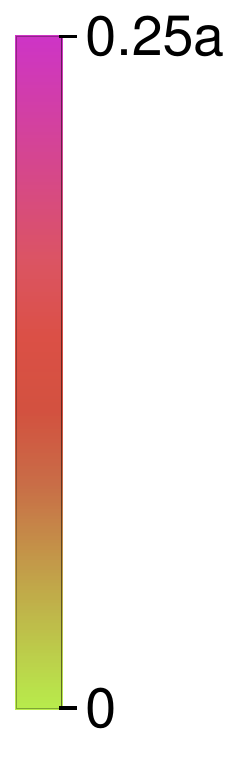} 
    & \includegraphics[width=\figwidth]{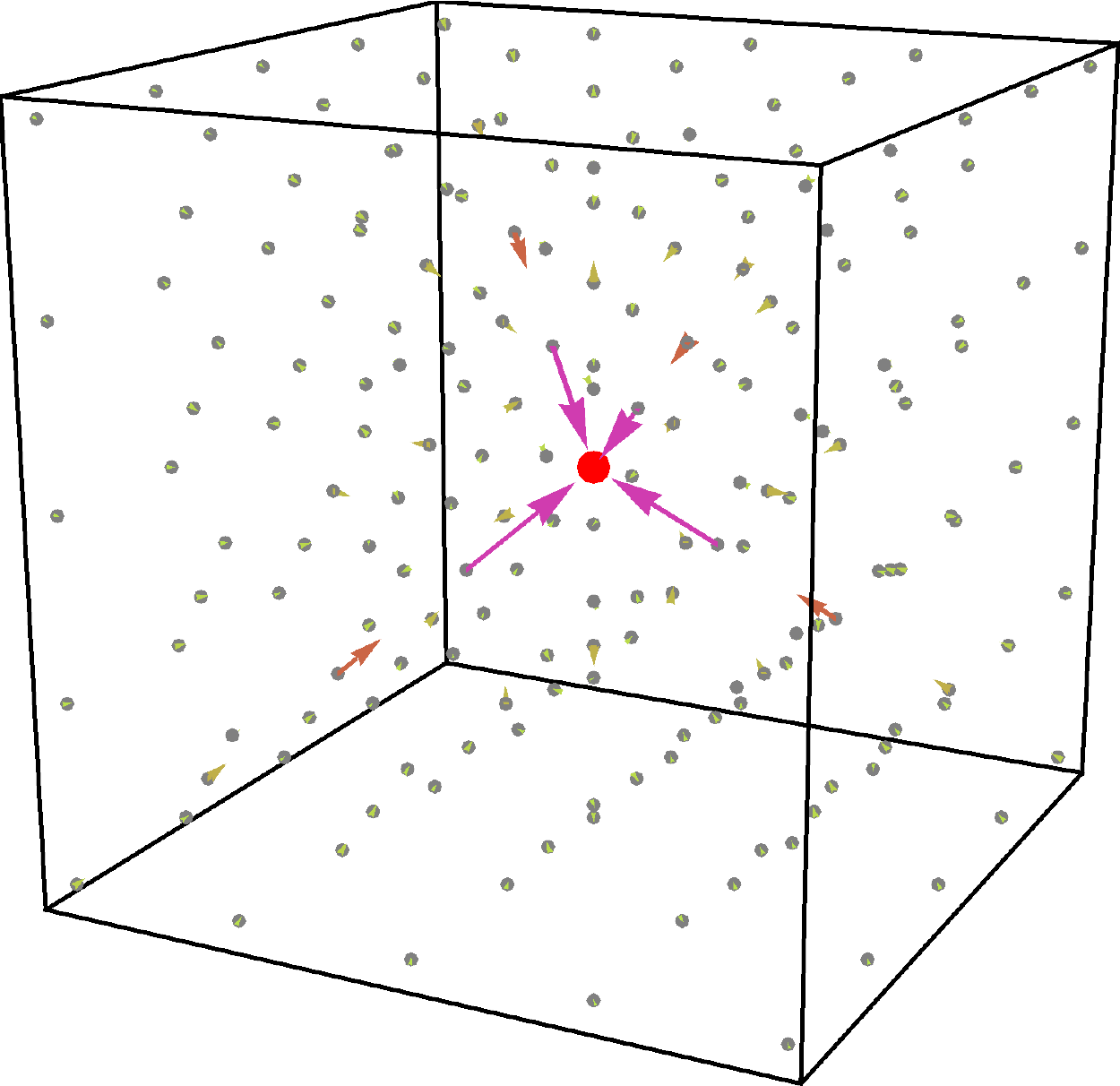} 
    & \includegraphics[width=\figwidthB, trim= 0cm -3.0cm 0.5cm 0.0cm,clip]{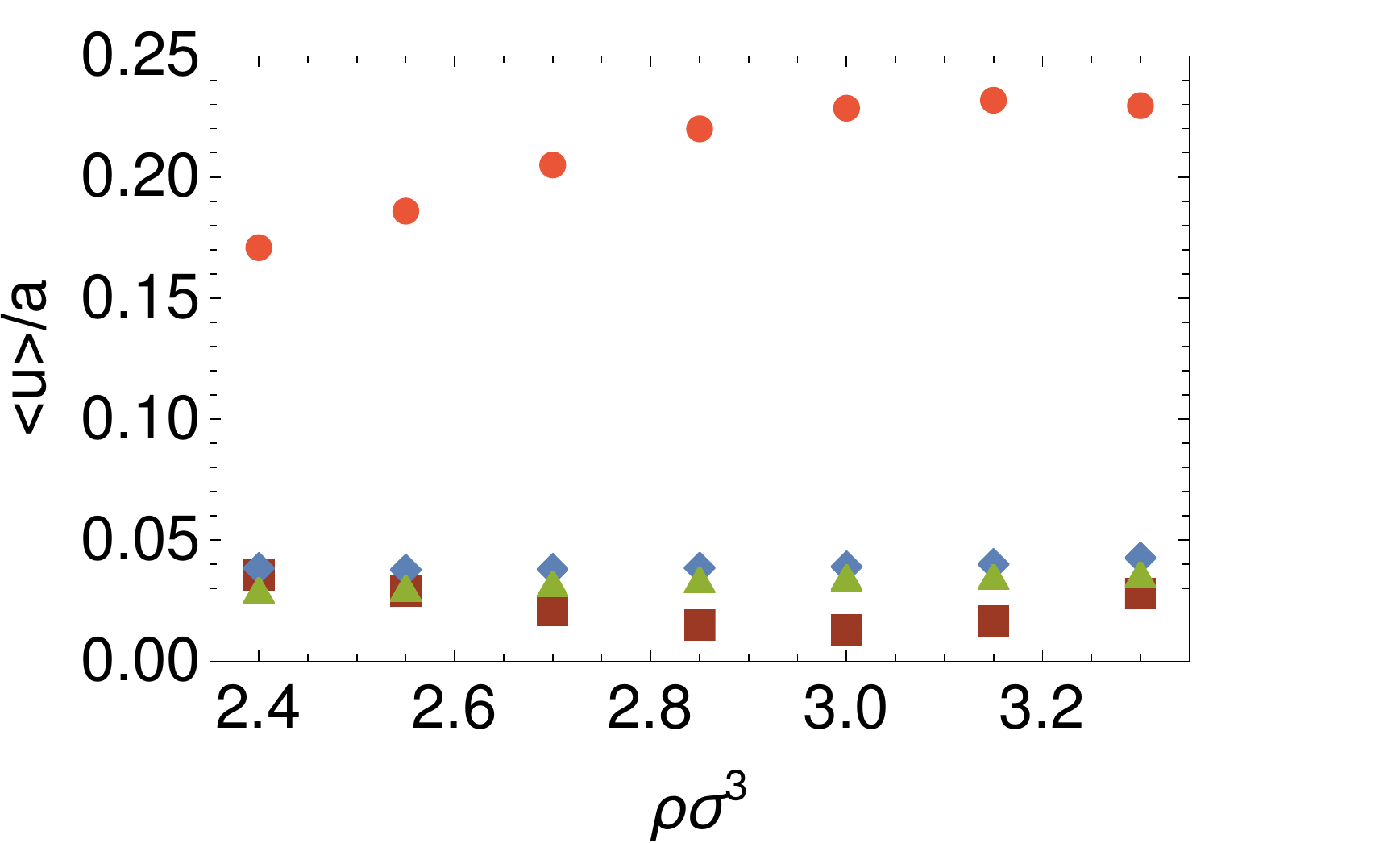}
    & \includegraphics[width=\figwidthB, trim= 0.5cm -2.1cm 0.0cm 0.0cm,clip]{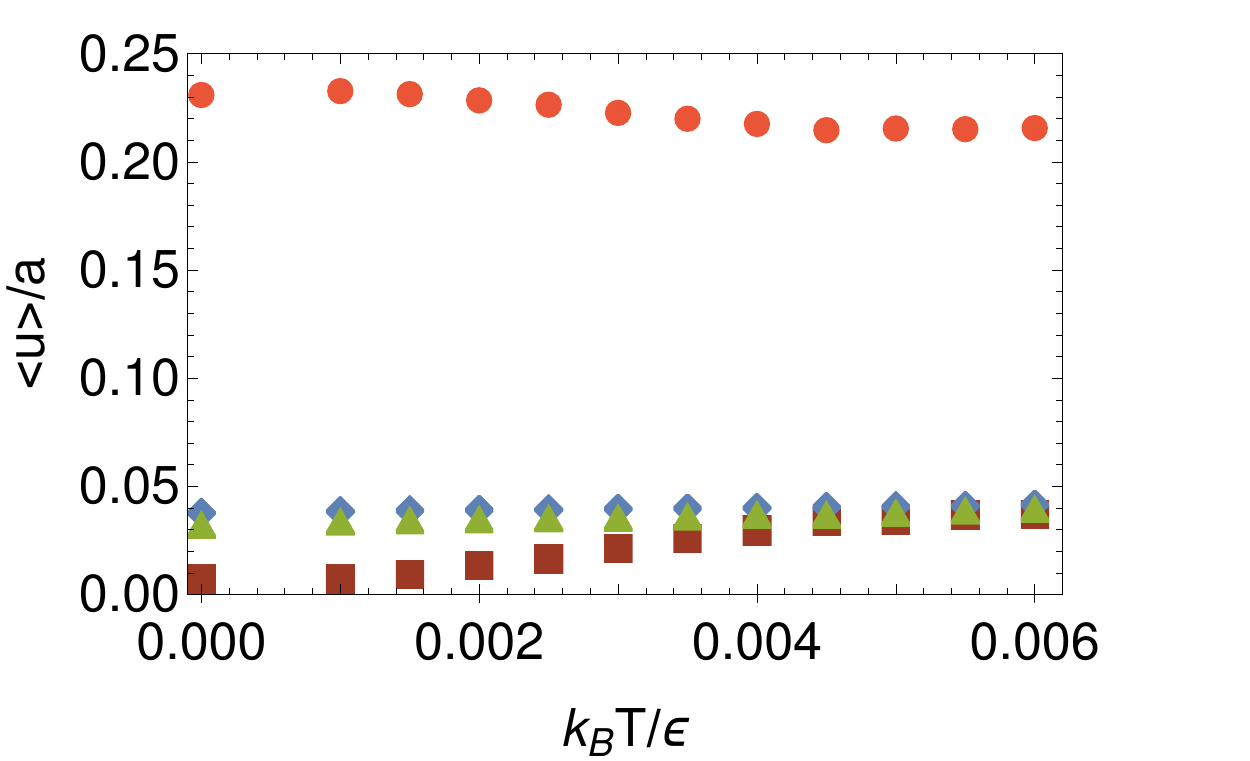}
    & \hspace{-0.2cm}\includegraphics[width=\legendsize,trim= 0.0cm -7.0cm 0.0cm 0.0cm]{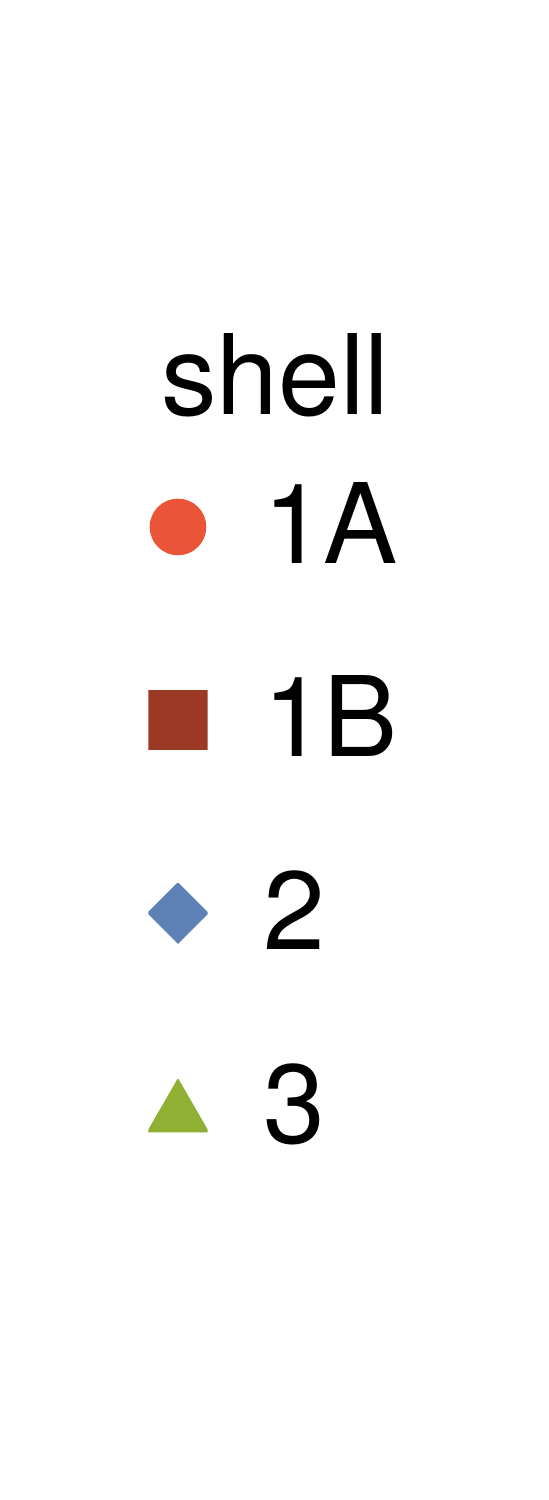} \\
    & d) & \,\, e) & f) &\\[-0.3cm]
    \includegraphics[width=\legendsizeD,trim= 0cm -0.5cm 0.0cm 0.0cm]{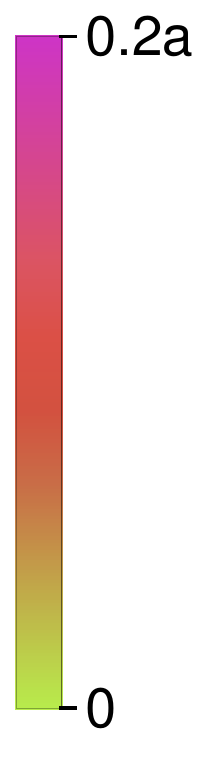} 
    & \includegraphics[width=\figwidth]{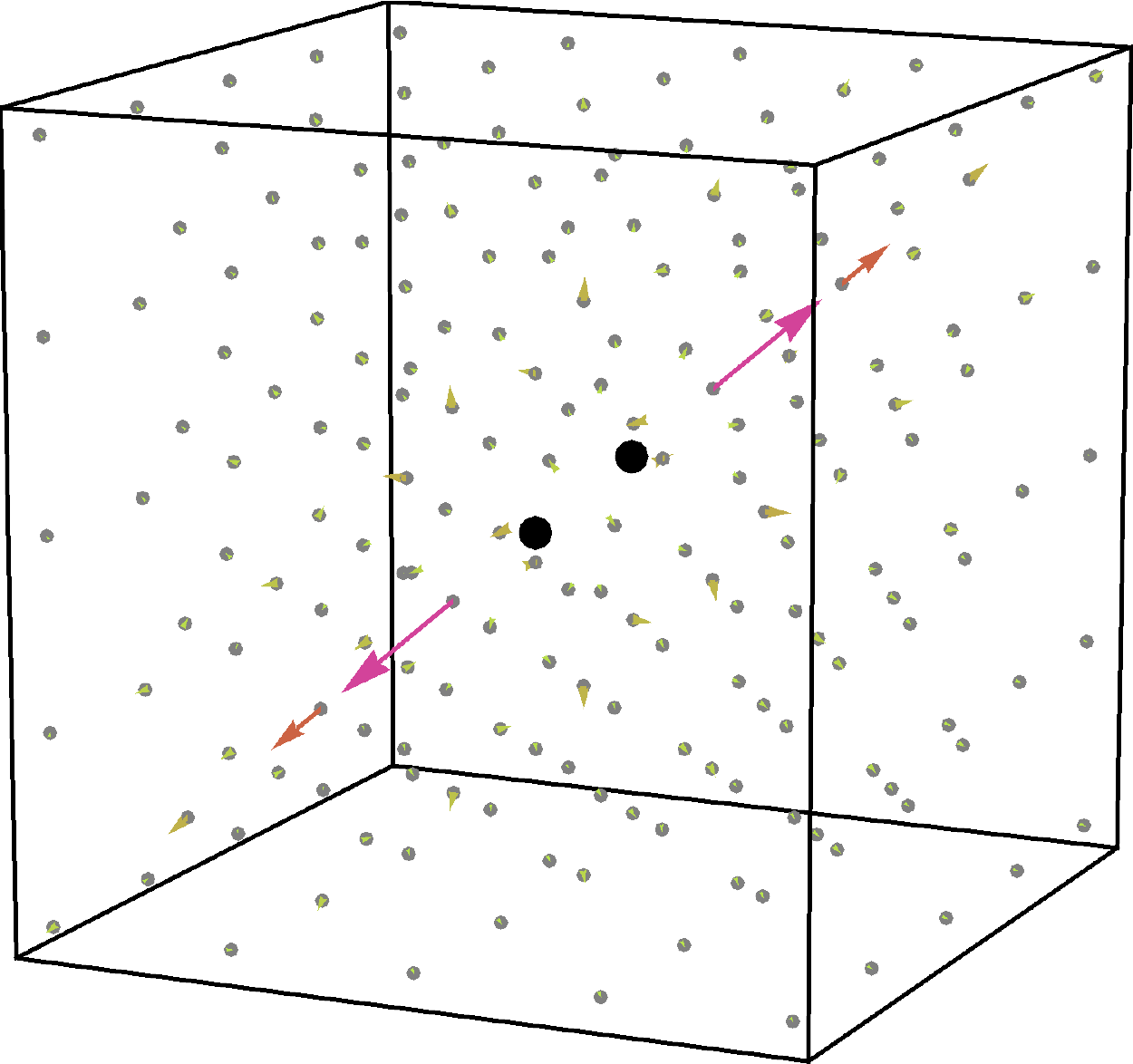} 
    & \includegraphics[width=\figwidthB, trim= 0cm -2.1cm 0.5cm 0.0cm,clip]{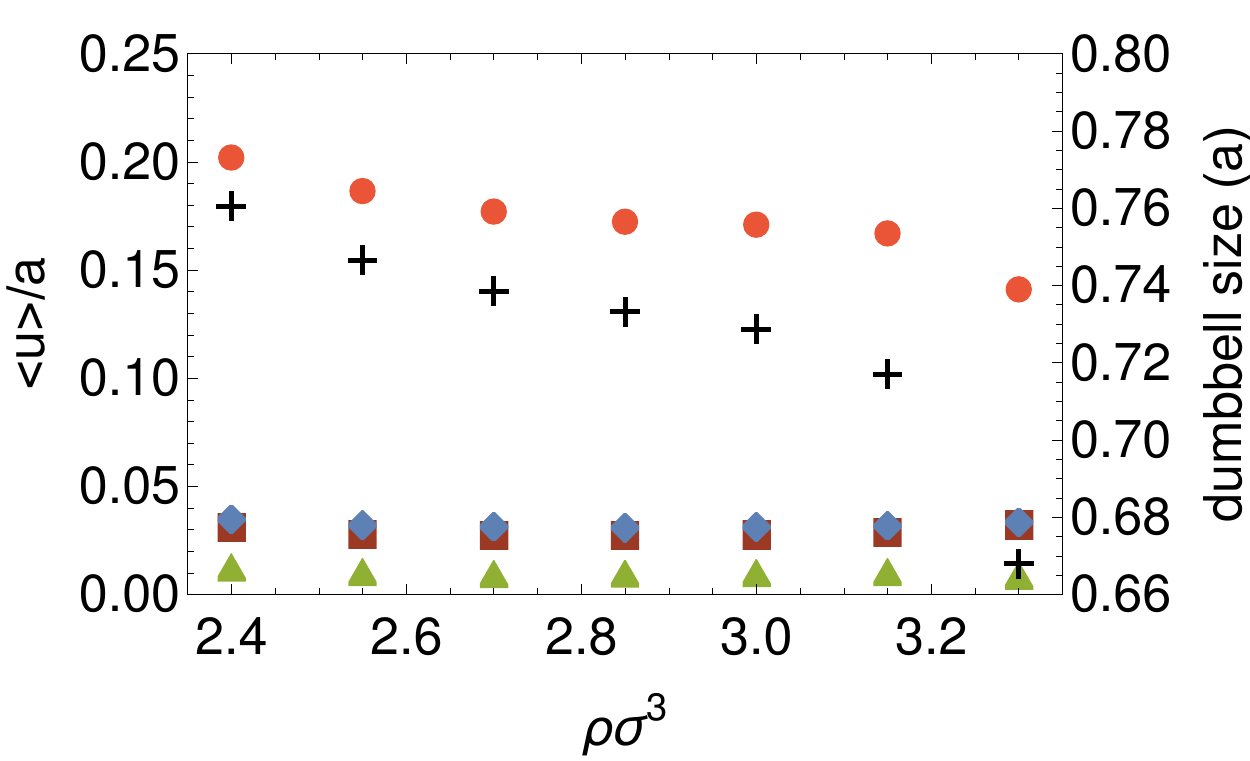}
    & \includegraphics[width=\figwidthB, trim= 0.5cm -2.1cm 0.0cm 0.0cm,clip]{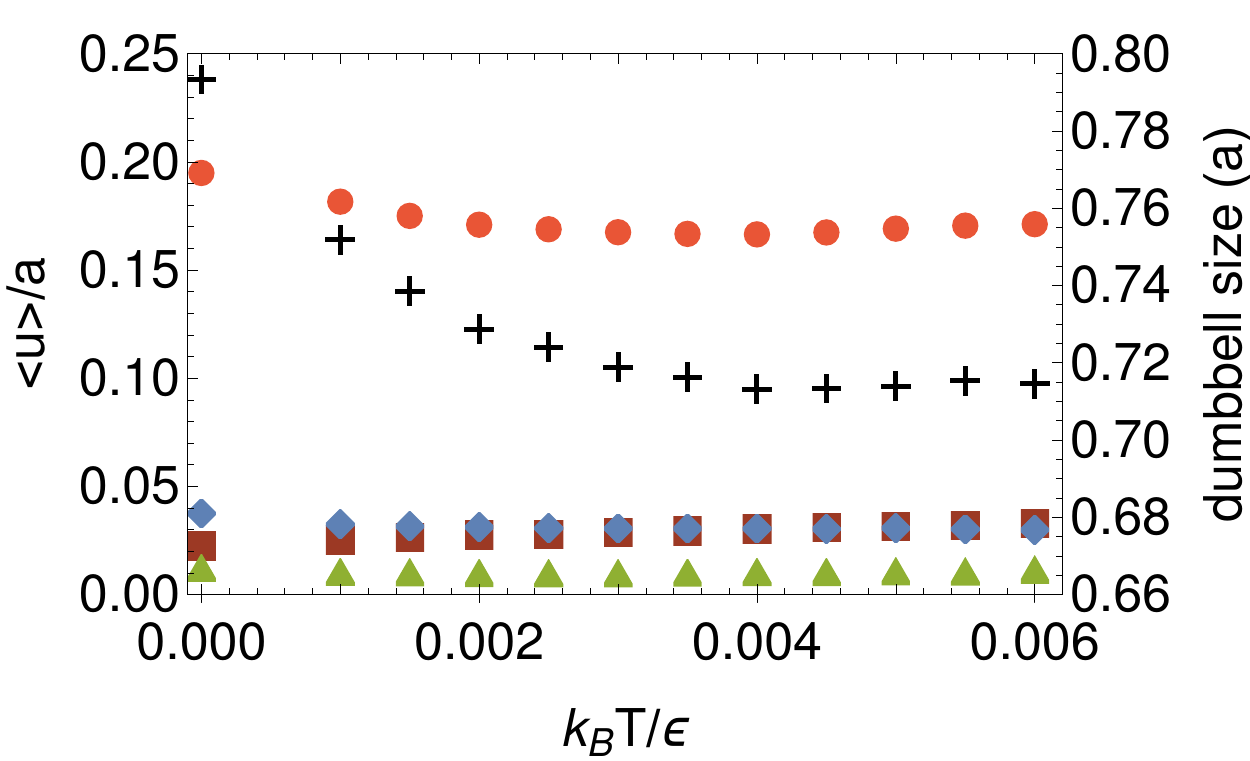}
    & \includegraphics[width=\legendsize,trim= 0.0cm -7.8cm 0.0cm 0.0cm]{images/legend-int-1ab23.pdf}
	\end{tabular}
	\caption[width=1\linewidth]{
	Average deformation of the BCC crystal of Hertzian spheres associated with a-c) a vacancy and d-f) an interstitial at density $\rho\sigma^3=3.00$ and temperature $k_BT/\epsilon=0.0020$. 
	Figures and legends as explained in the caption of Fig. \ref{fig:hertzfcc}.
	}
	\label{fig:hertzbcc}
\end{figure*}

\newcommand{\figwidthC}{0.27\linewidth}
\begin{figure*}
\begin{tabular}{lllll}
	& a) & \,\, b) & c) & \\[-0.3cm]
	\includegraphics[width=\legendsizeA,trim= 0cm -0.5cm 0.0cm 0.0cm]{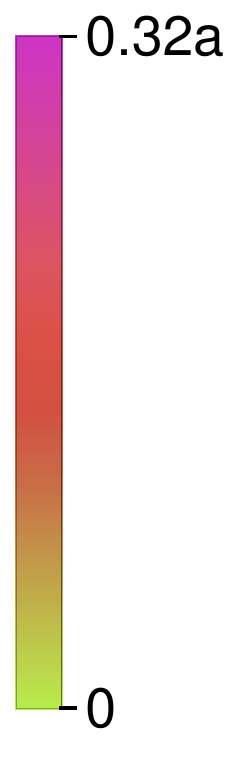} 
    & \includegraphics[width=\figwidth]{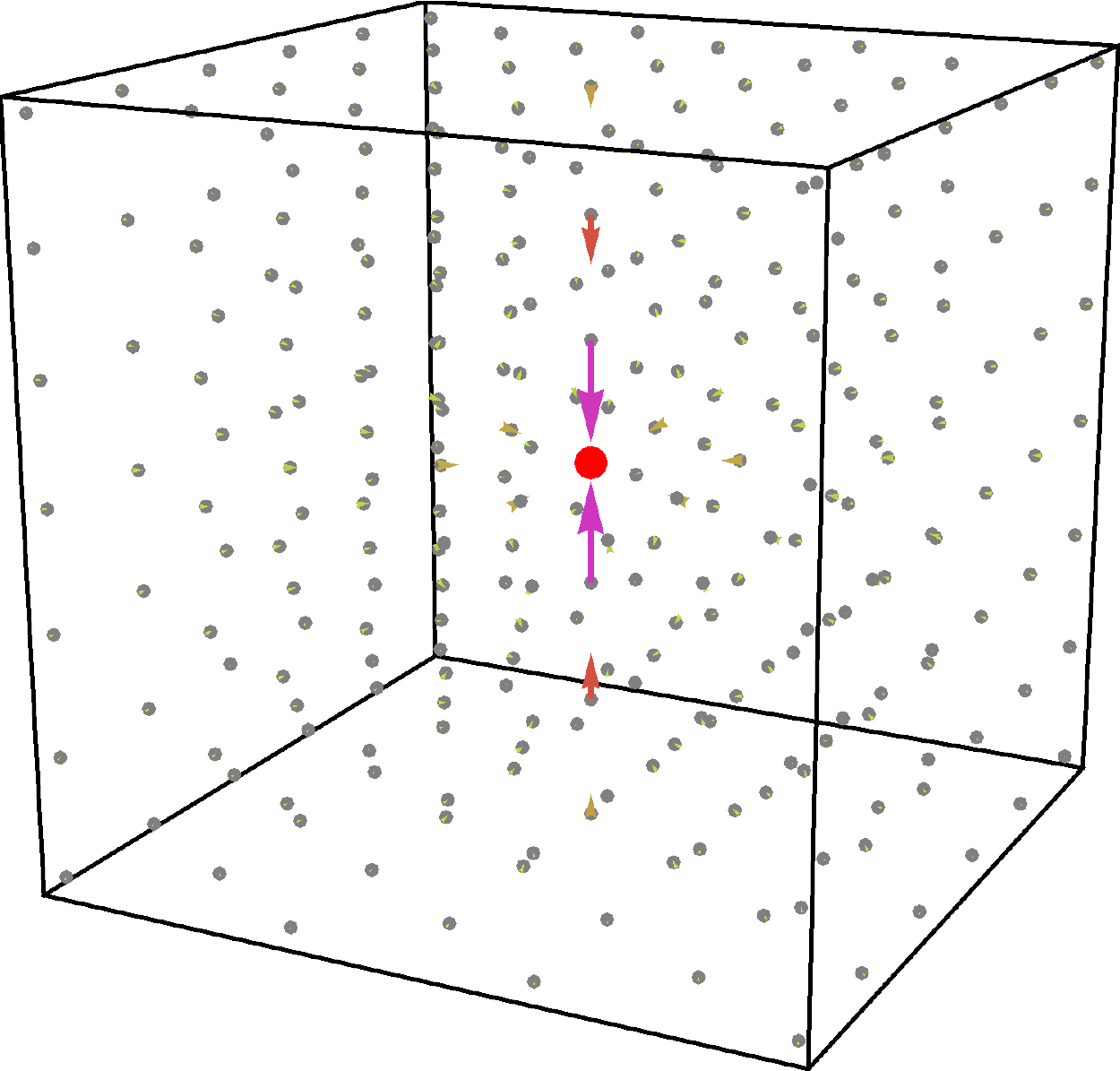} 
    & \includegraphics[width=\figwidthB, trim= 0cm -3.0cm 0.5cm 0.0cm,clip]{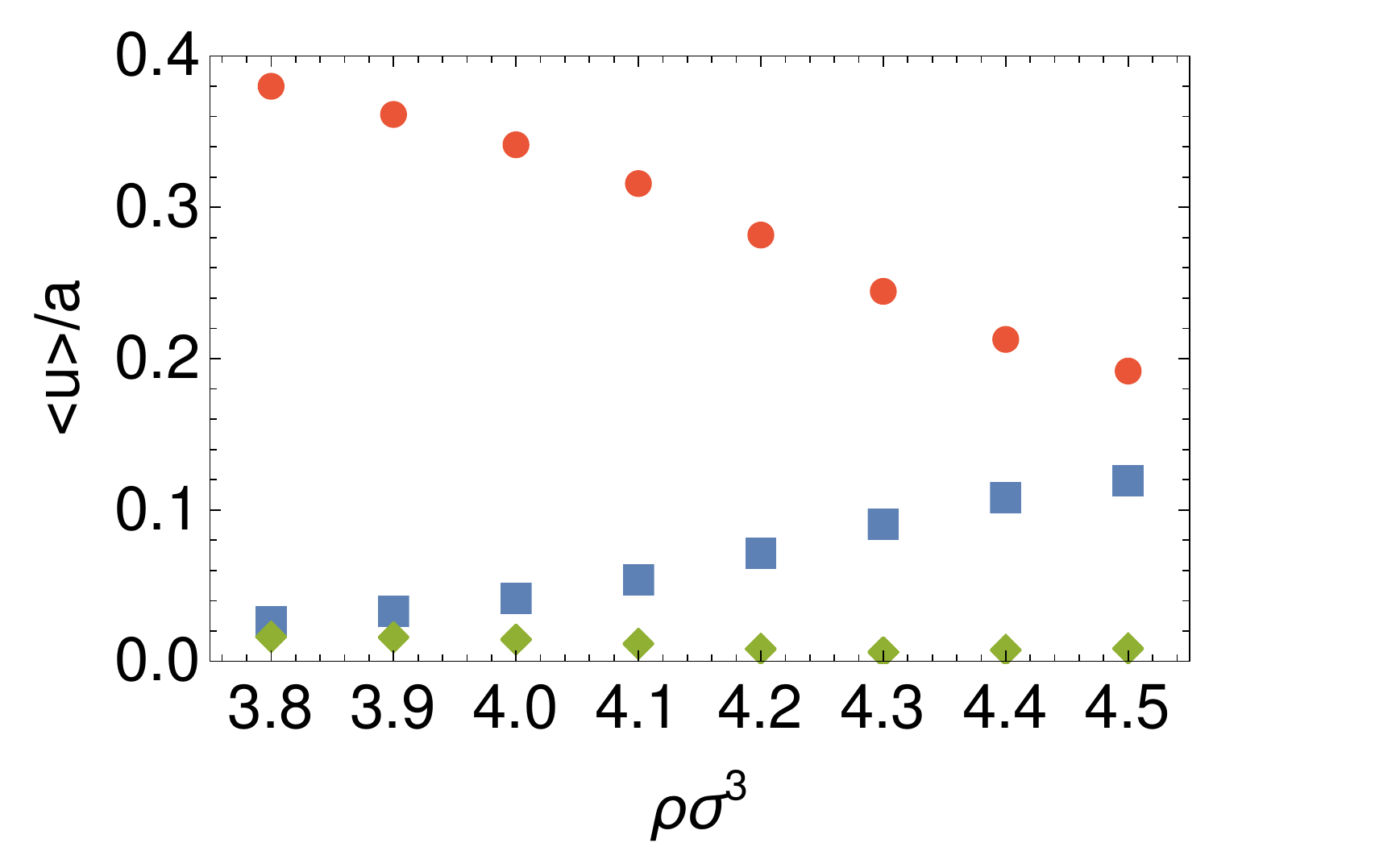}
    & \hspace{0.1cm}\includegraphics[width=\figwidthC, trim= 0.8cm -2.0cm 0.0cm 0.0cm,clip]{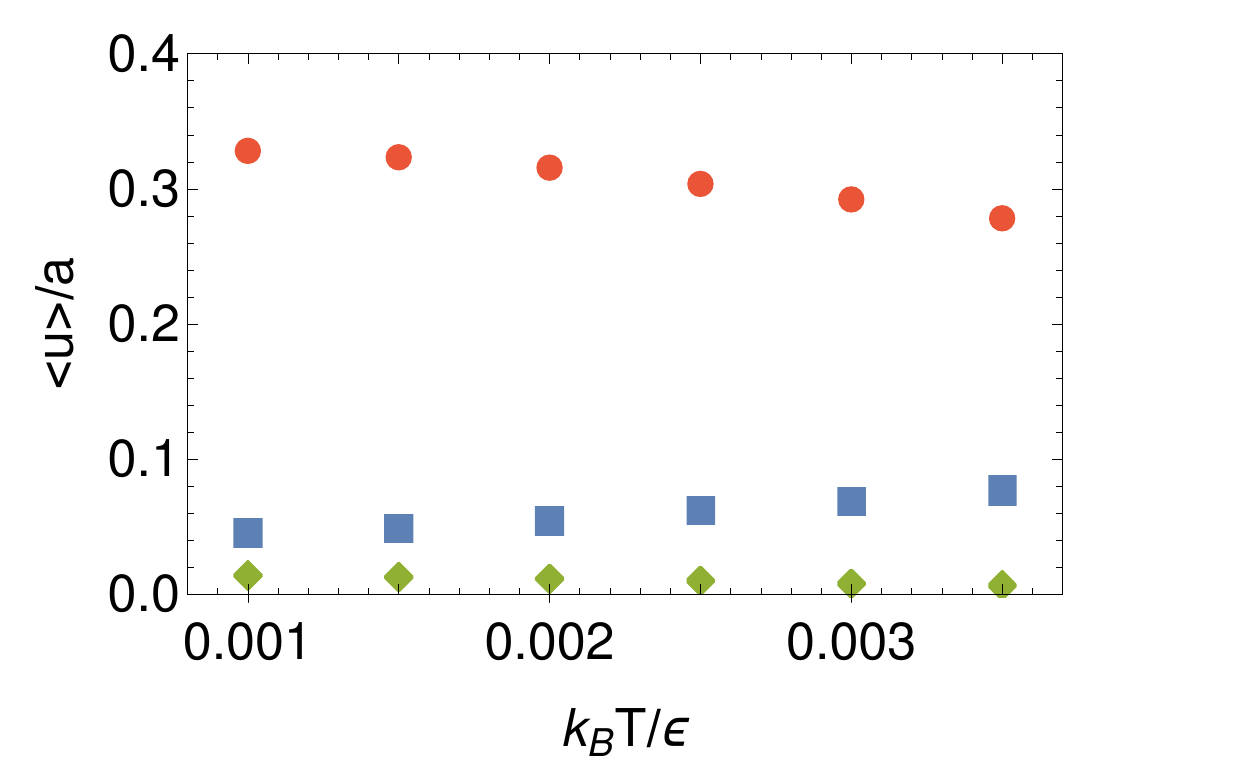}
    & \hspace{-0.2cm}\includegraphics[width=\legendsize,trim= 0.0cm -7.0cm 0.0cm 0.0cm]{images/legend-vac1.pdf} \\
    & d) & \,\, e) & f) &\\[-0.3cm]
    \includegraphics[width=\legendsizeA,trim= 0cm -0.5cm 0.0cm 0.0cm]{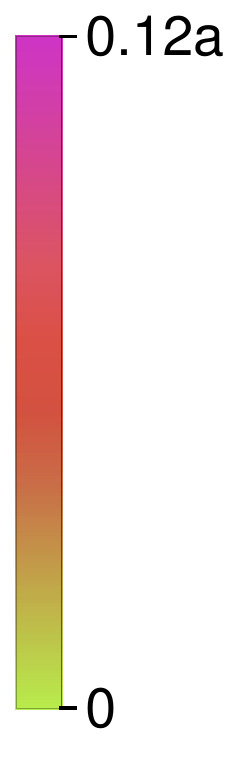} 
    & \includegraphics[width=\figwidth]{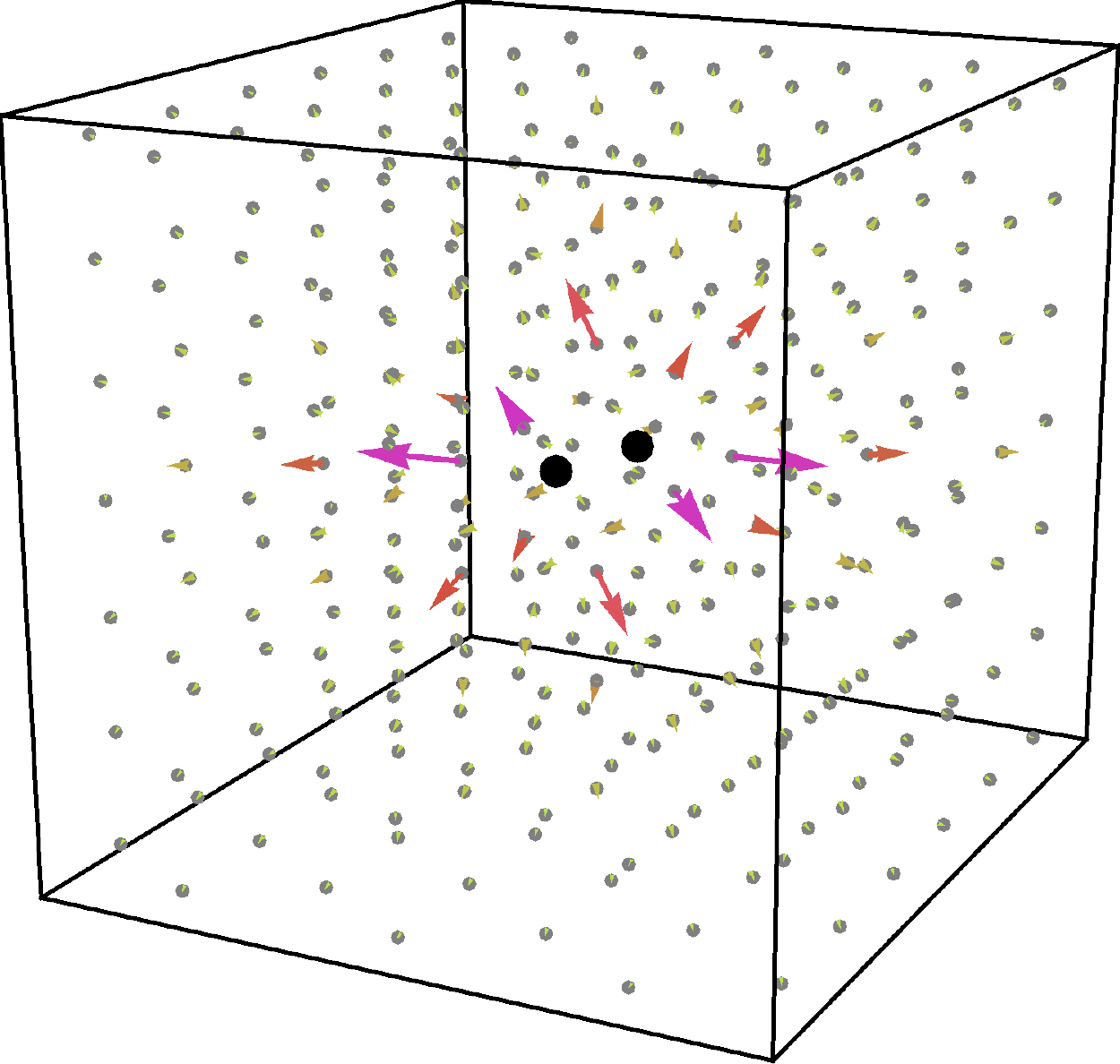}
    & \includegraphics[width=\figwidthB, trim= 0cm -2.1cm 0.5cm 0.0cm,clip]{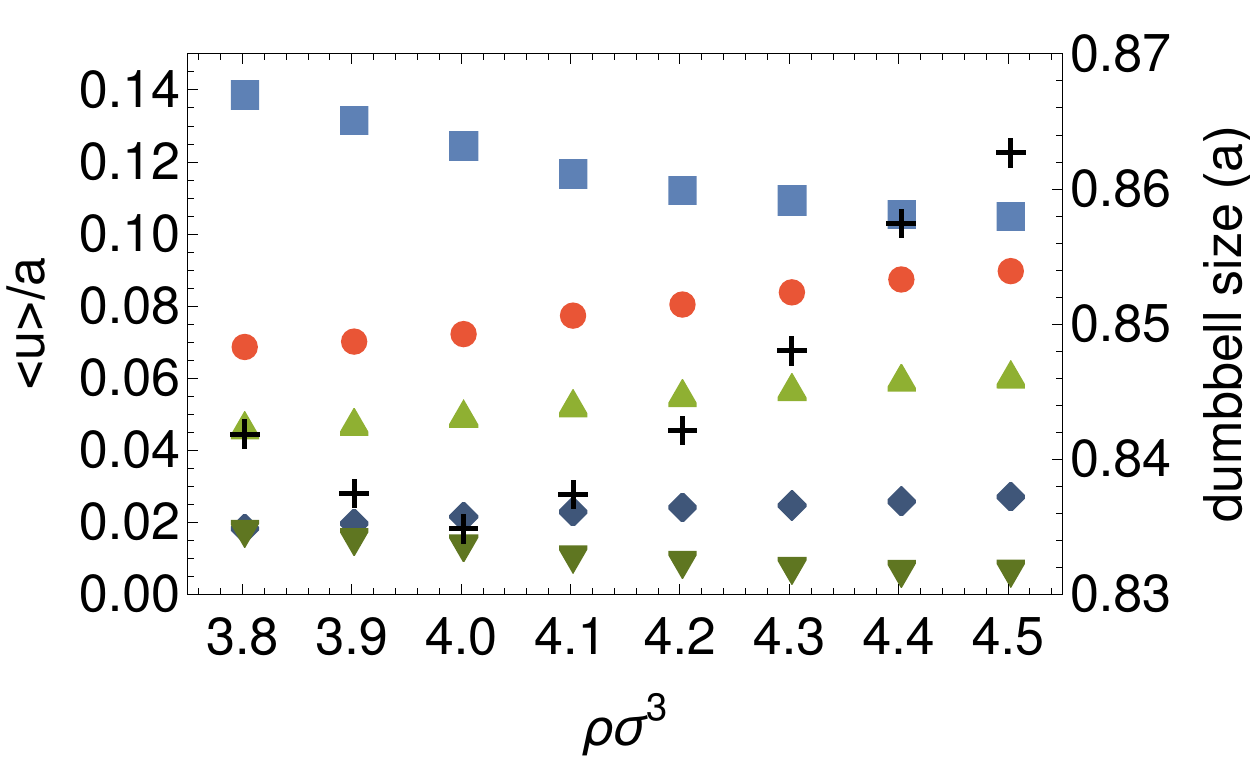}
    & \includegraphics[width=\figwidthB, trim= 0.5cm -2.1cm 0.0cm 0.0cm,clip]{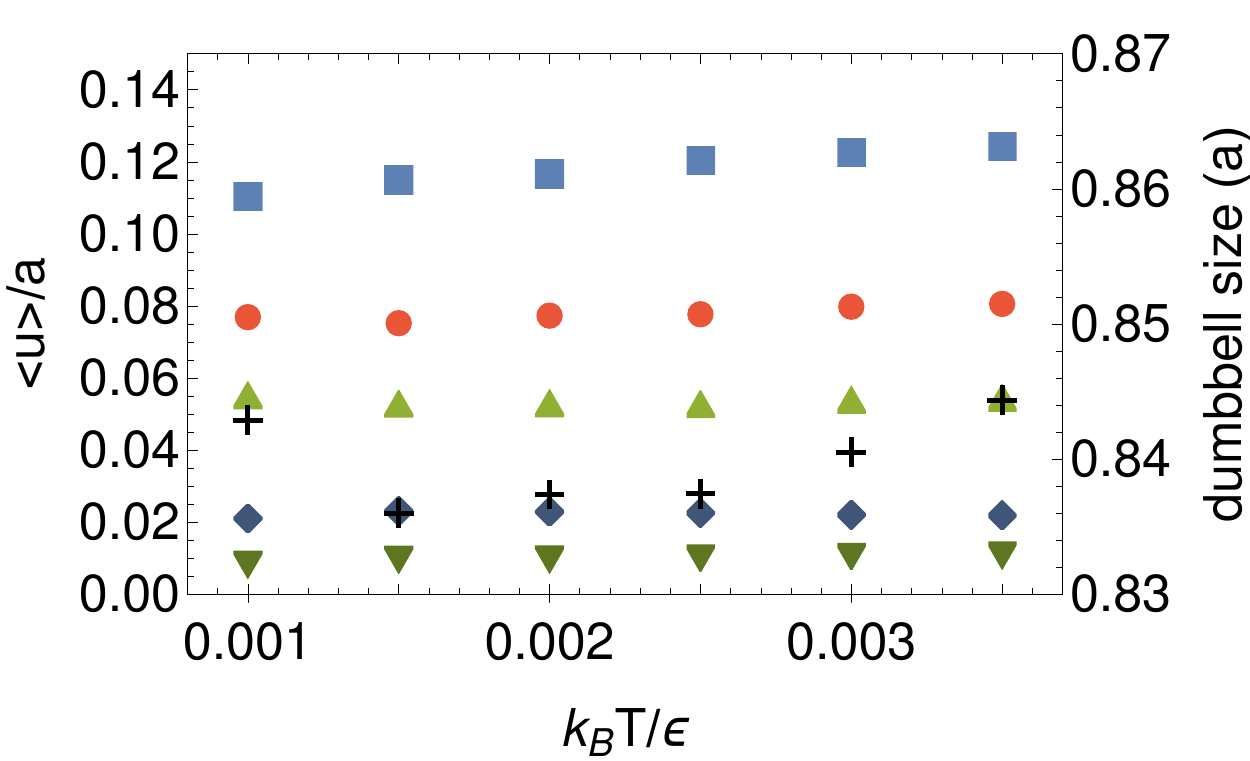}
    & \includegraphics[width=\legendsize,trim= 0.0cm -7.8cm 0.0cm 0.0cm]{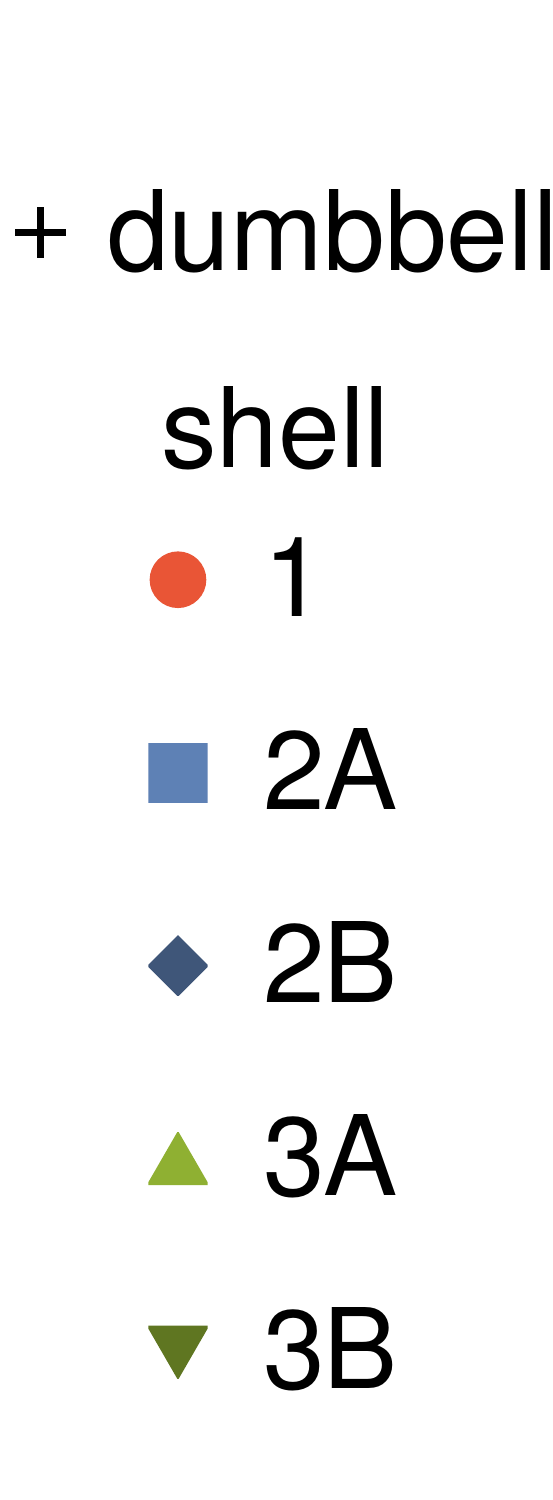}
	\end{tabular}
	\caption[width=1\linewidth]{
	Average deformation of the H crystal of Hertzian spheres associated with a-c) a vacancy and d-f) an interstitial at density $\rho\sigma^3=4.10$ and temperature $k_BT/\epsilon=0.0020$. 
	Figures and legends as explained in the caption of Fig. \ref{fig:hertzfcc}.
	}
	\label{fig:hertzh}
\end{figure*}

\begin{figure*}
\begin{tabular}{lllll}
	& a) & \,\, b) & c) & \\[-0.3cm]
	\includegraphics[width=\legendsizeA,trim= 0cm -0.5cm 0.0cm 0.0cm]{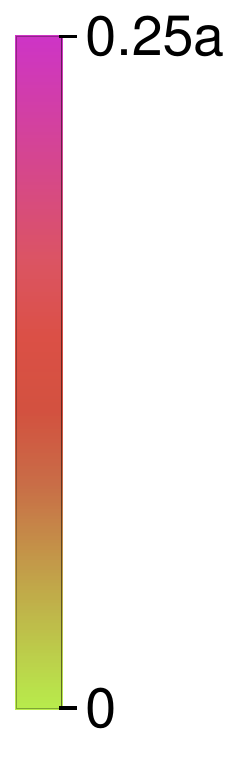} 
    & \includegraphics[width=\figwidth]{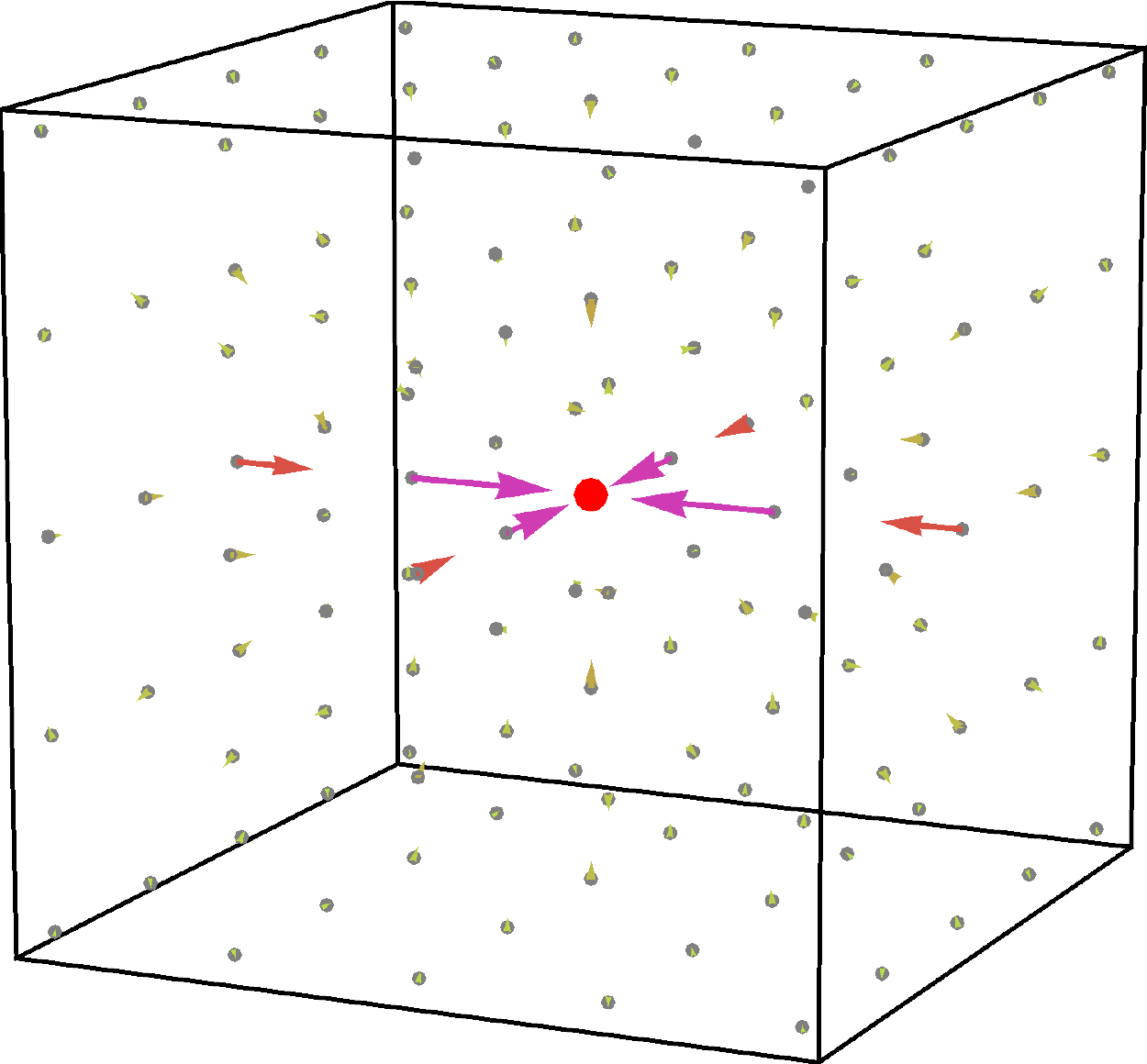} 
    & \includegraphics[width=\figwidthB, trim= 0cm -3.0cm 0.5cm 0.0cm,clip]{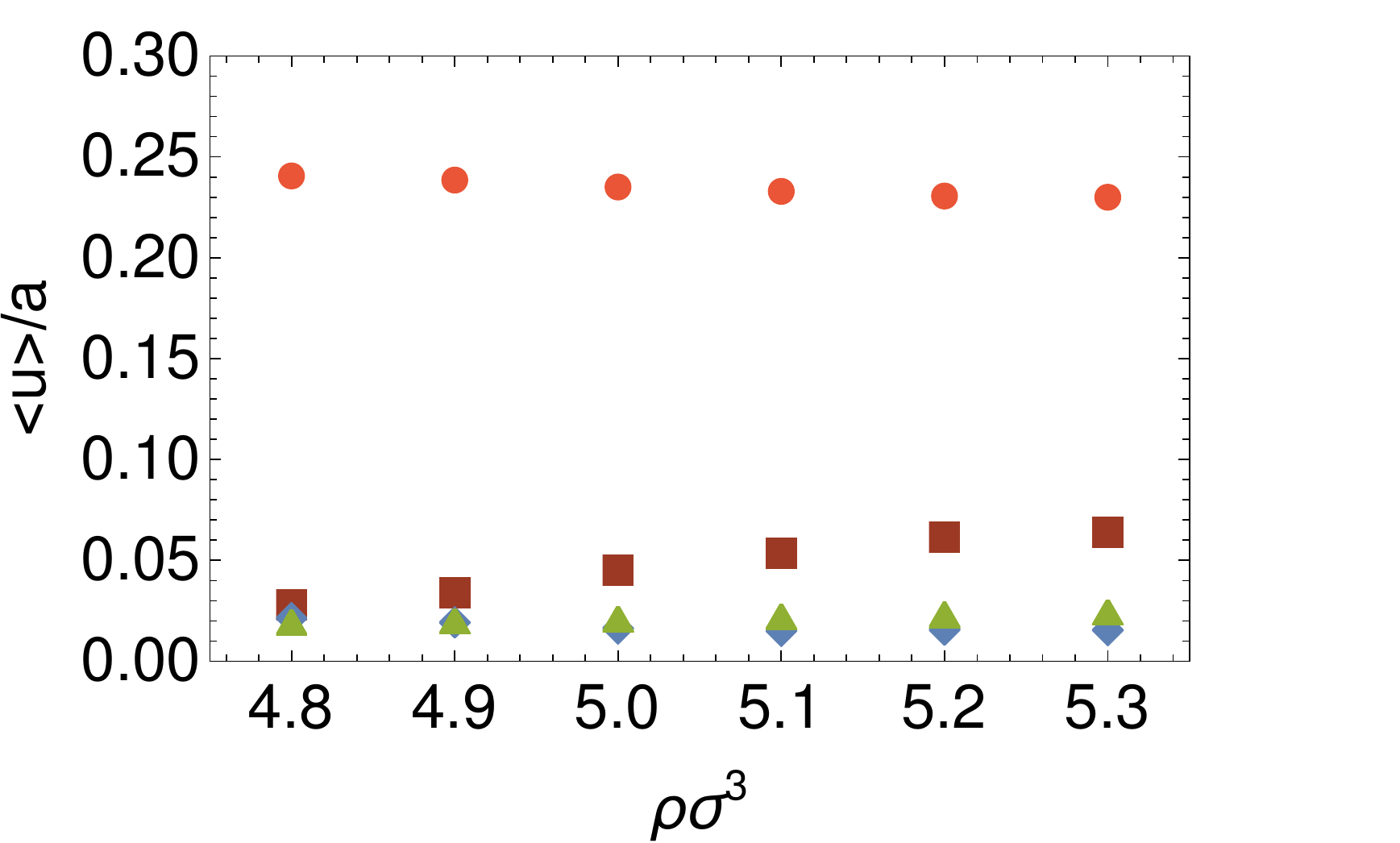}
    & \includegraphics[width=\figwidthB, trim= 0.5cm -2.1cm 0.0cm 0.0cm,clip]{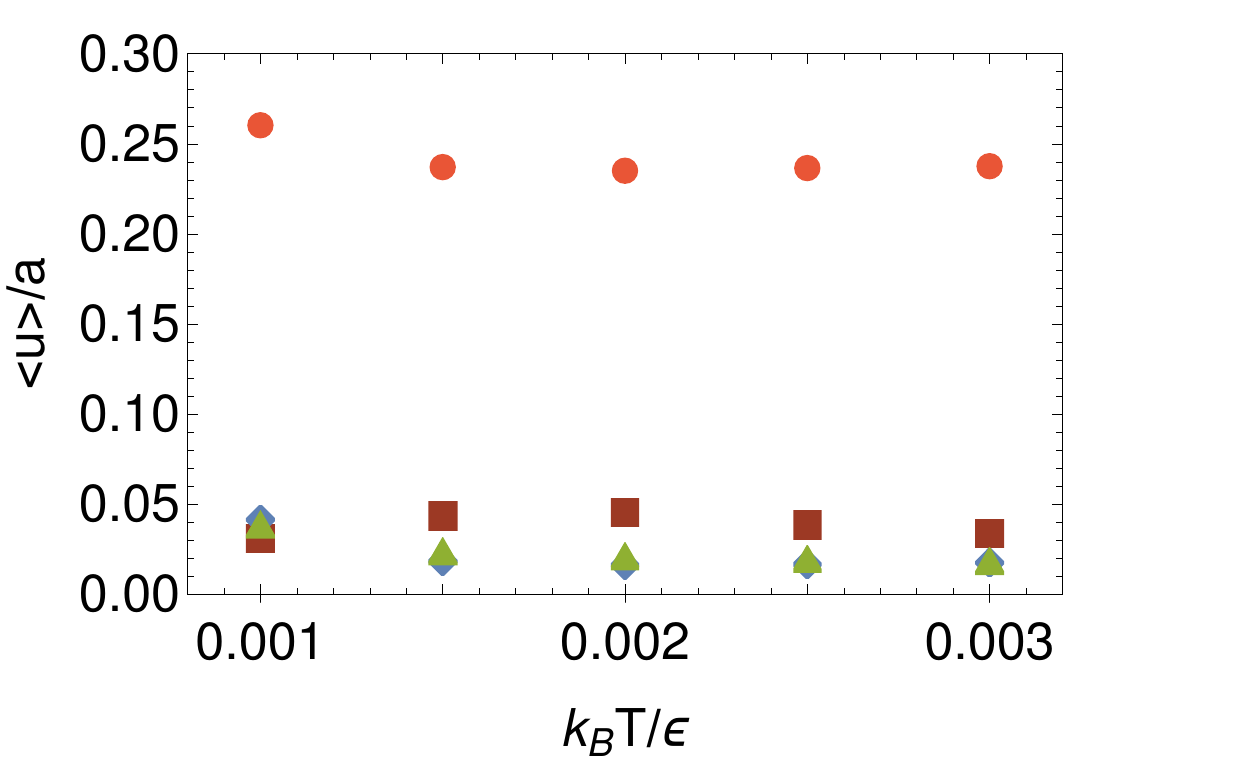}
    & \hspace{-0.2cm}\includegraphics[width=\legendsize,trim= 0.0cm -7.0cm 0.0cm 0.0cm]{images/legend-vac2.pdf} \\
    & d) & \,\, e) & f) &\\[-0.3cm]
    \includegraphics[width=\legendsizeD,trim= 0cm -0.5cm 0.0cm 0.0cm]{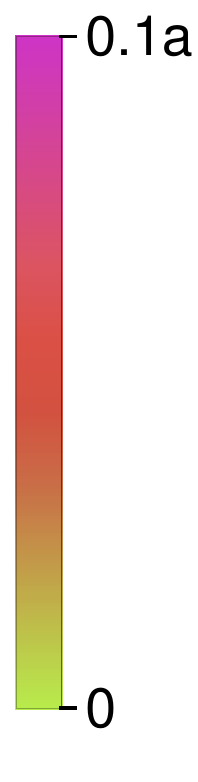} 
    & \includegraphics[width=\figwidth]{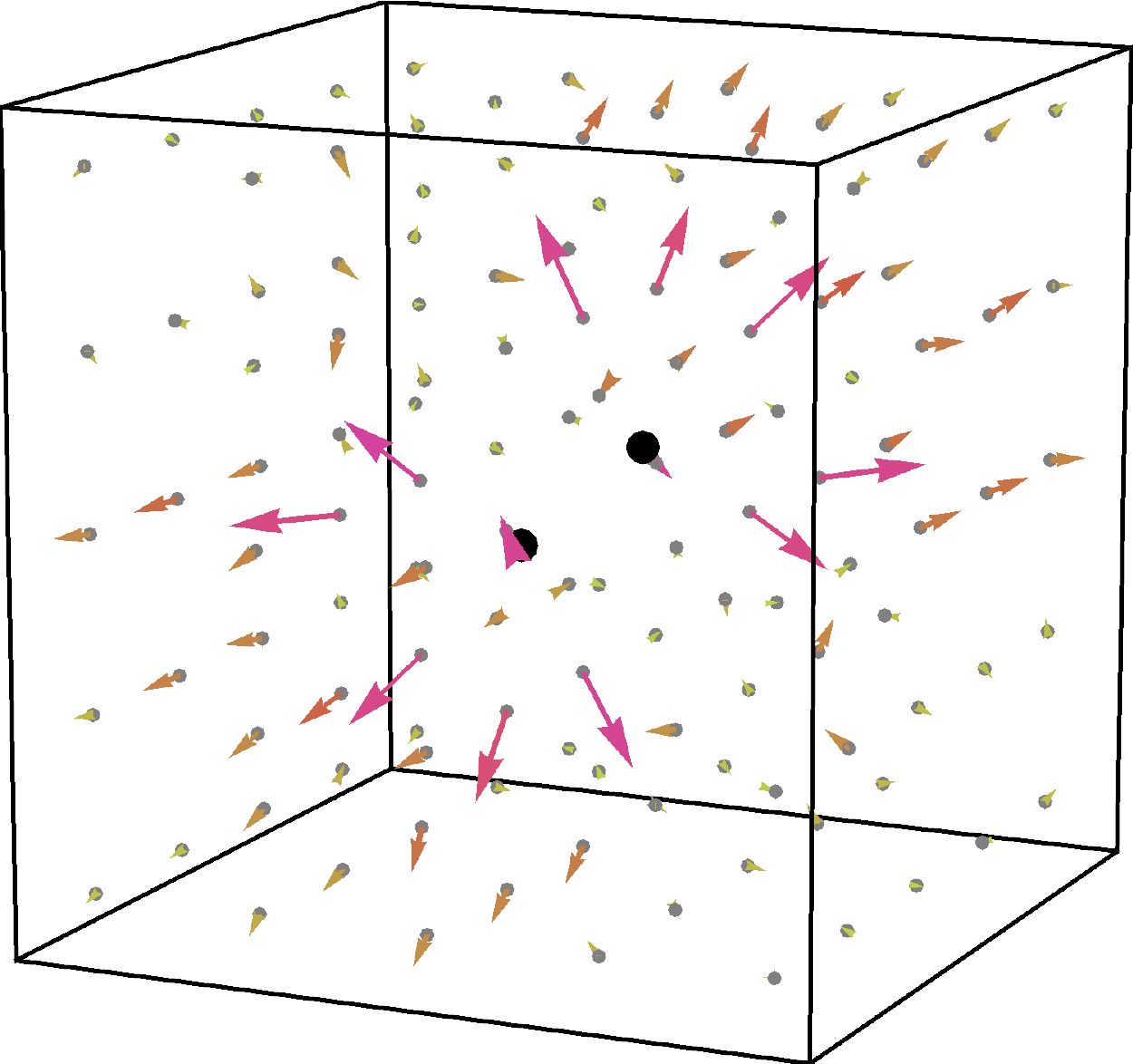} 
    & \includegraphics[width=\figwidthB, trim= 0cm -2.1cm 0.5cm 0.0cm,clip]{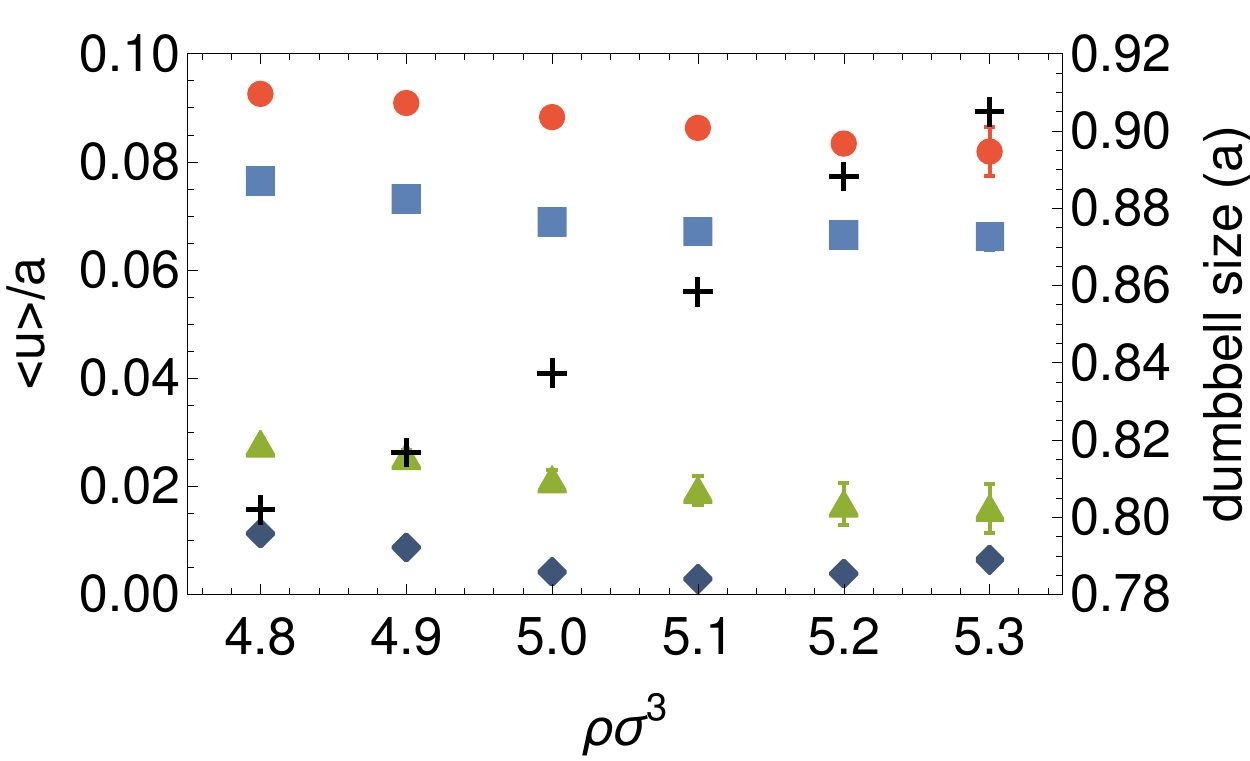}
    & \includegraphics[width=\figwidthB, trim= 0.5cm -2.1cm 0.0cm 0.0cm,clip]{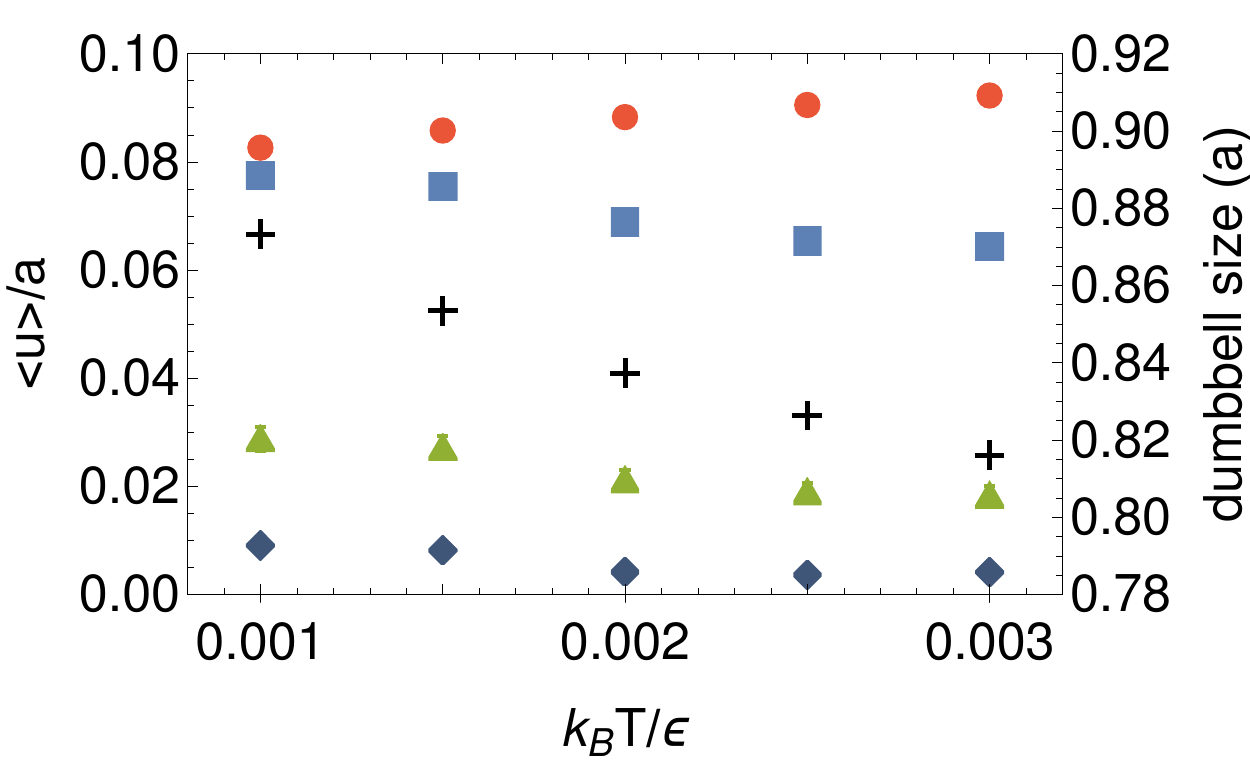}
    & \includegraphics[width=\legendsize,trim= 0.0cm -7.8cm 0.0cm 0.0cm]{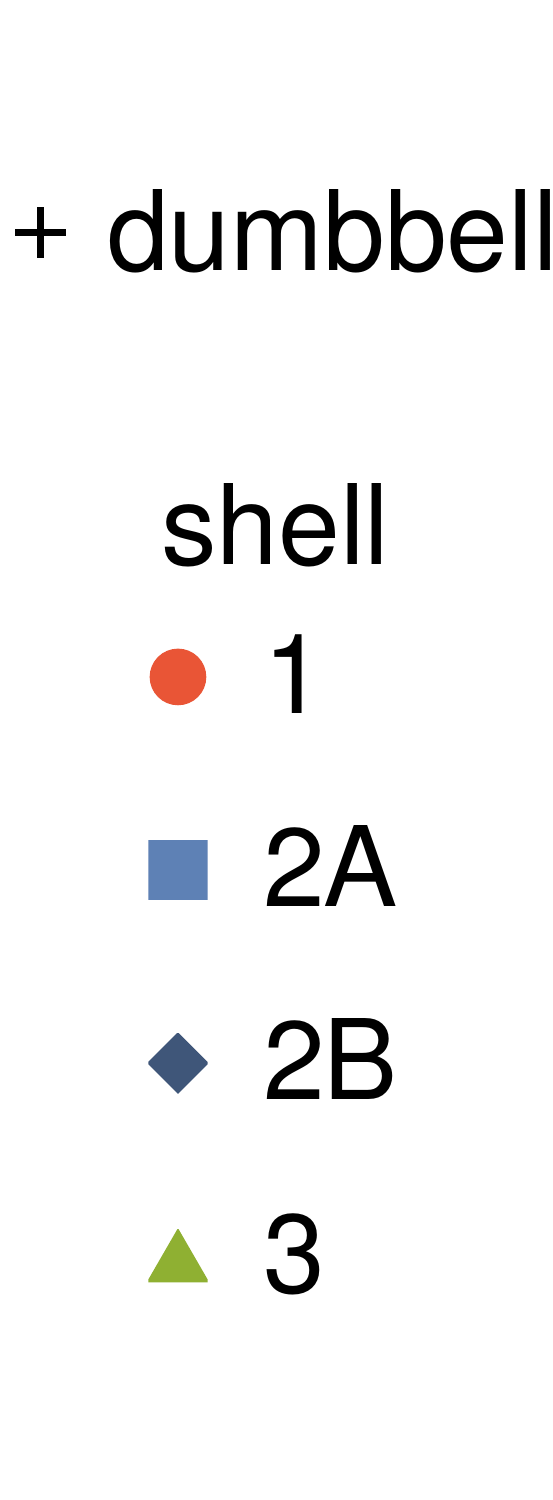}
	\end{tabular}
	\caption[width=1\linewidth]{
	Average deformation of the SC crystal of Hertzian spheres associated with a-c) a vacancy and d-f) an interstitial at density $\rho\sigma^3=5.00$ and temperature $k_BT/\epsilon=0.0020$. 
	Figures and legends as explained in the caption of Fig. \ref{fig:hertzfcc}.
	}
	\label{fig:hertzsc}
\end{figure*}

\begin{figure*}
\begin{tabular}{lllll}
	& a) & \,\, b) & c) & \\[-0.3cm]
	\includegraphics[width=\legendsizeD,trim= 0cm -0.5cm 0.0cm 0.0cm]{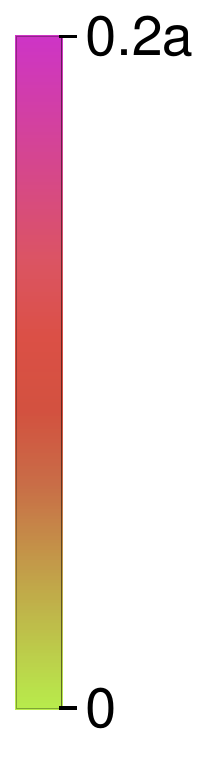} 
    & \includegraphics[width=\figwidth]{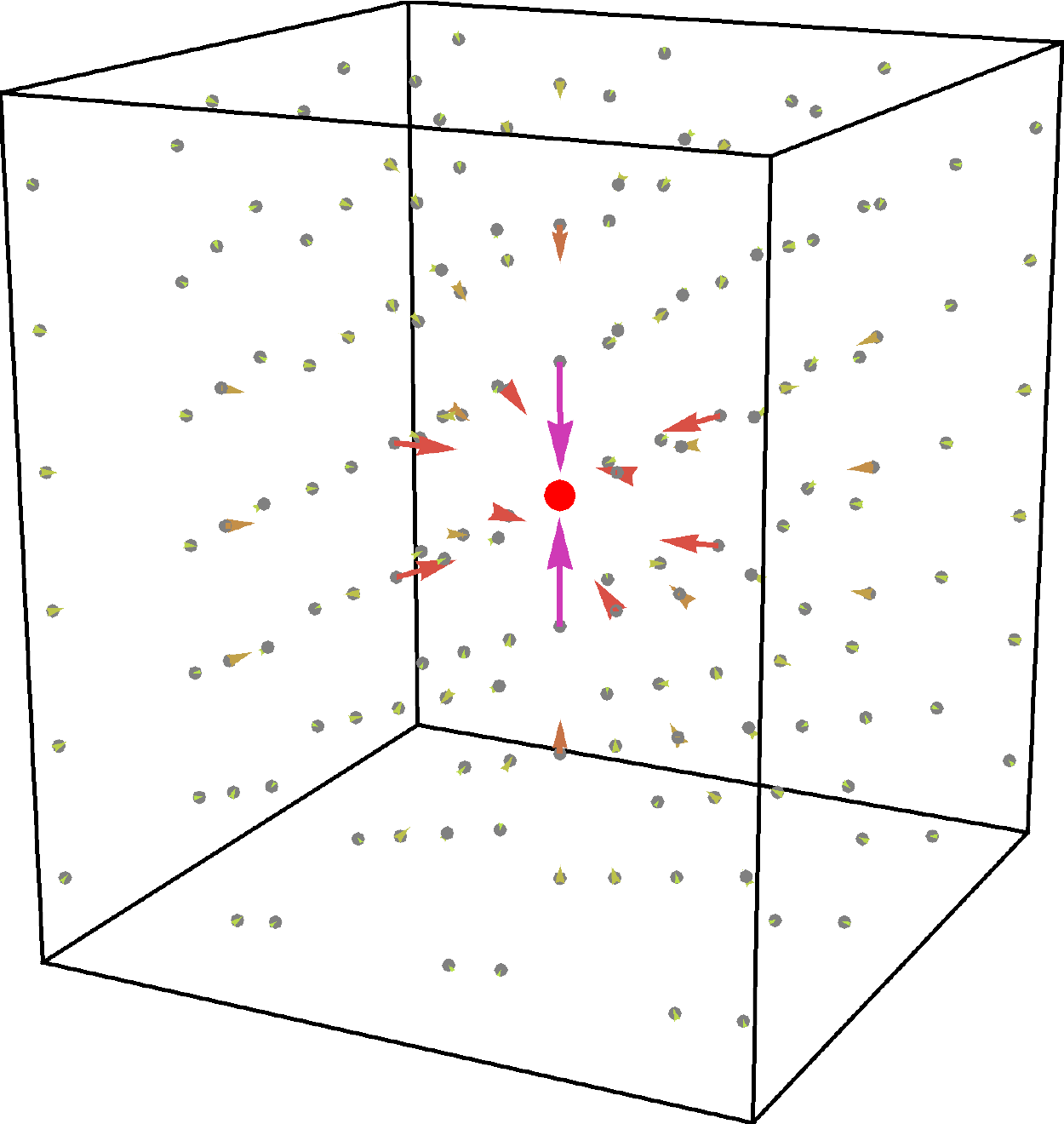} 
    & \includegraphics[width=\figwidthB, trim= 0cm -3.0cm 0.5cm 0.0cm,clip]{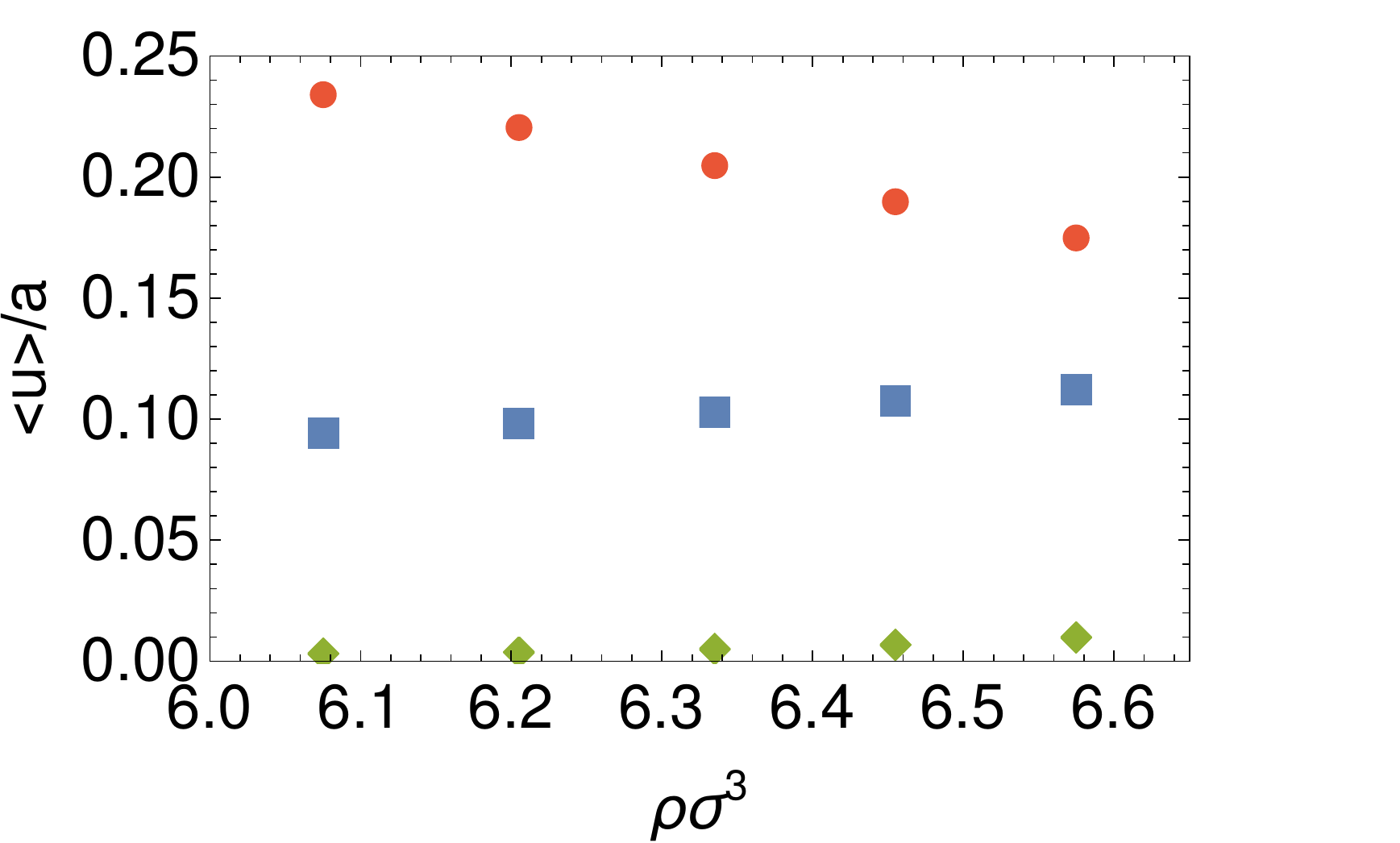}
    & \includegraphics[width=\figwidthB, trim= 0.5cm -2.1cm 0.0cm 0.0cm,clip]{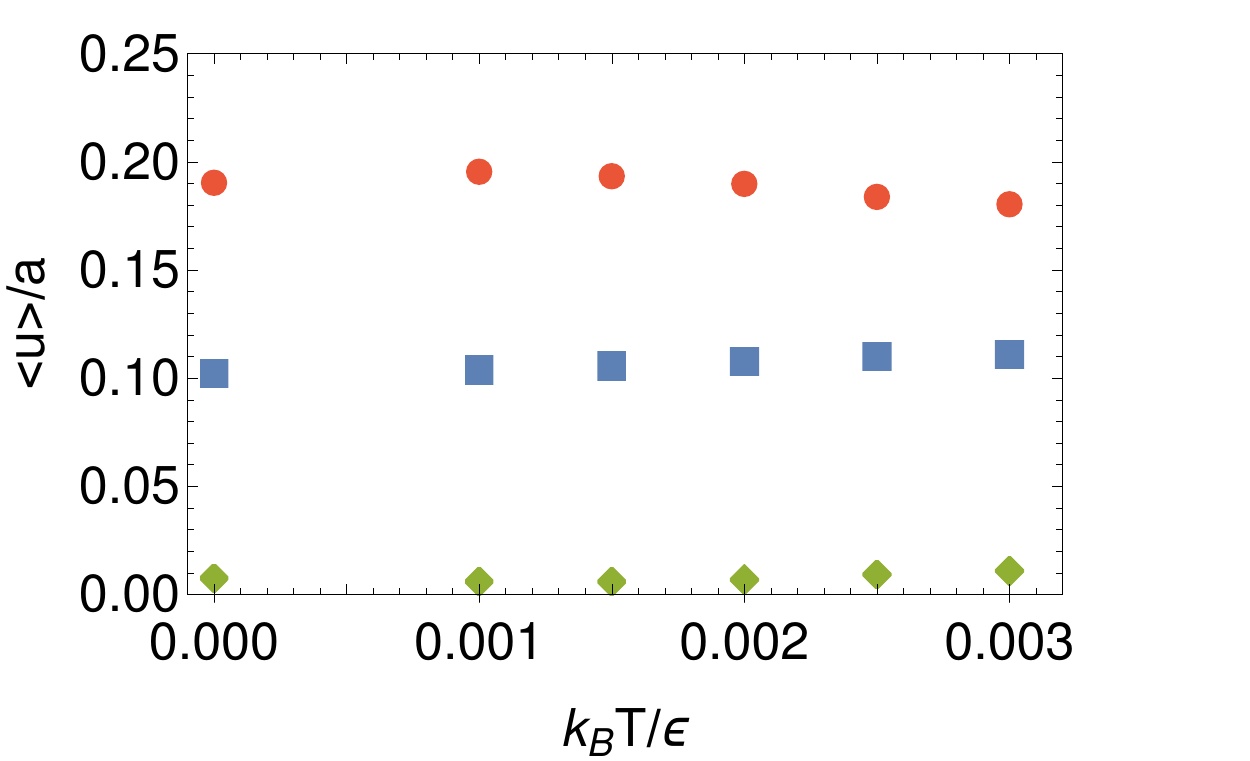}
    & \hspace{-0.2cm}\includegraphics[width=\legendsize,trim= 0.0cm -7.0cm 0.0cm 0.0cm]{images/legend-vac1.pdf} \\
    & d) & \,\, e) & f) &\\[-0.3cm]
    \includegraphics[width=\legendsizeA,trim= 0cm -0.5cm 0.0cm 0.0cm]{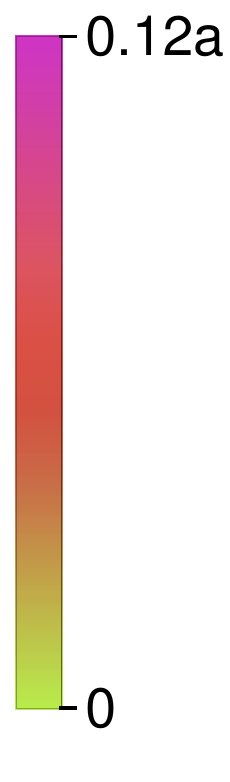} 
    & \includegraphics[width=\figwidth]{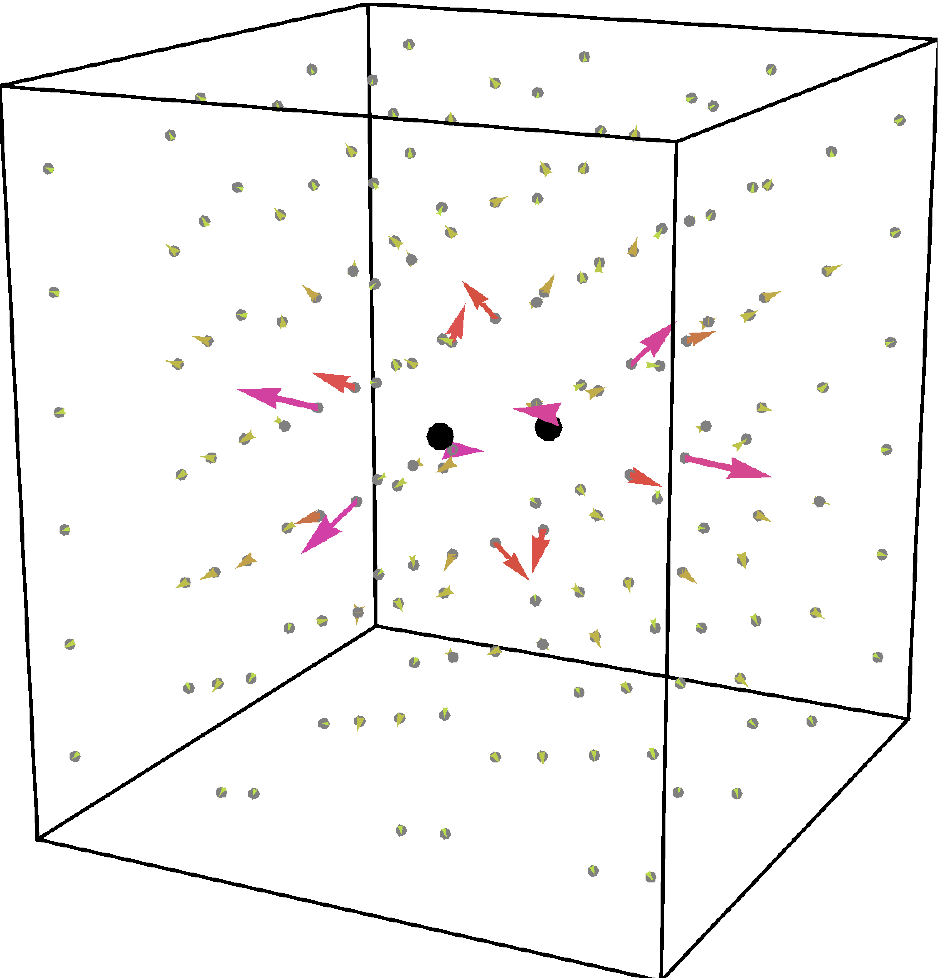} 
    & \includegraphics[width=\figwidthB, trim= 0cm -2.1cm 0.5cm 0.0cm,clip]{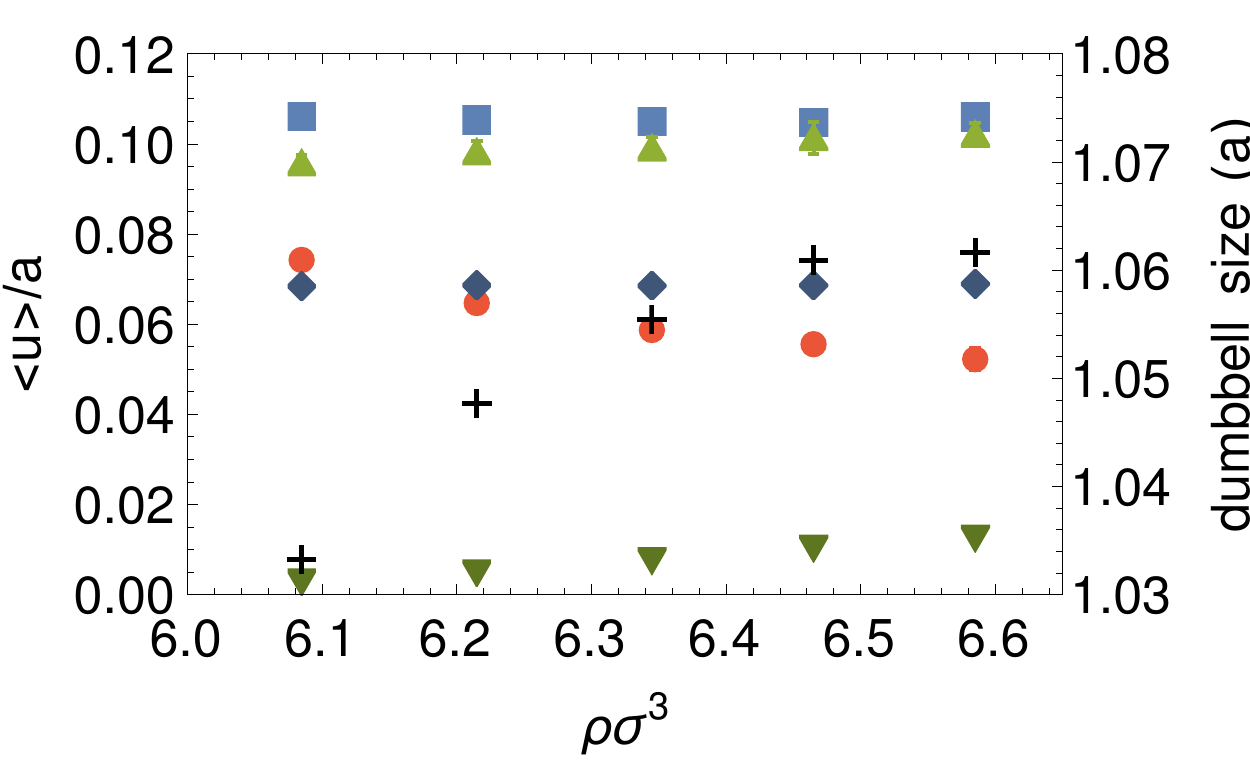}
    & \includegraphics[width=\figwidthB, trim= 0.5cm -2.1cm 0.0cm 0.0cm,clip]{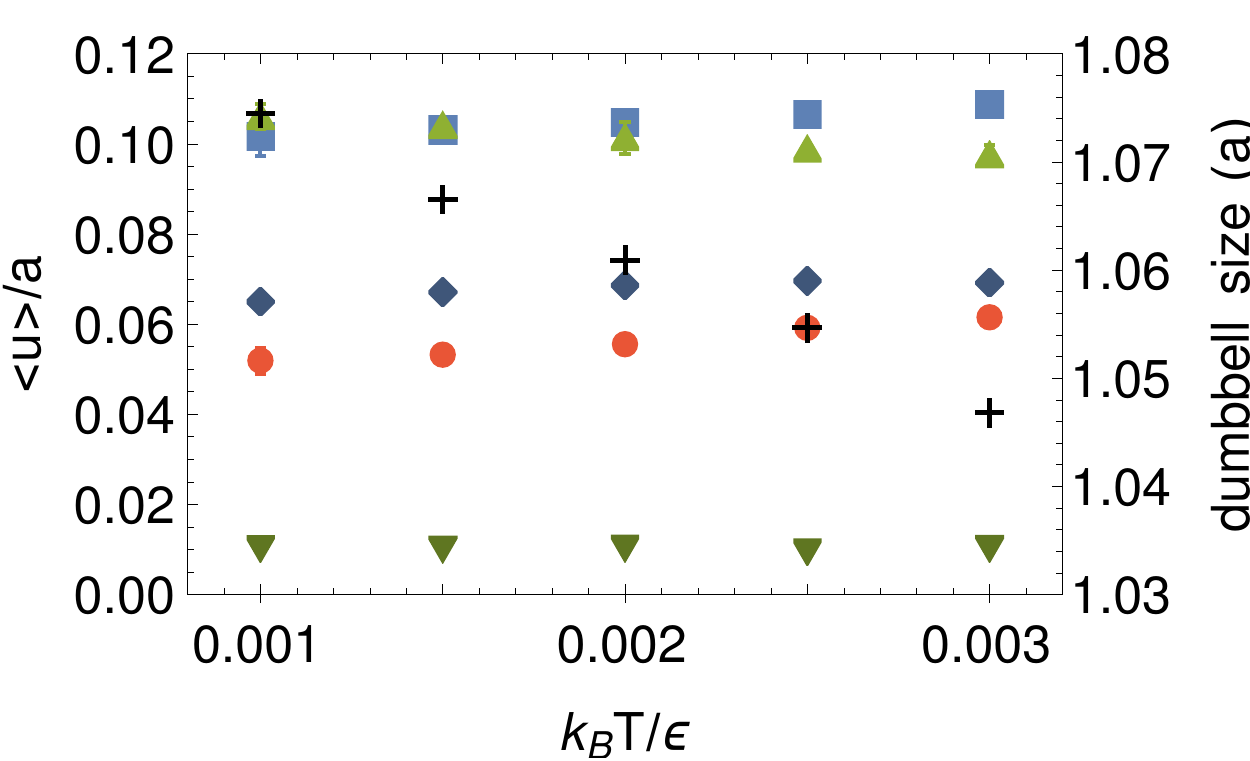}
    & \includegraphics[width=\legendsize,trim= 0.0cm -7.8cm 0.0cm 0.0cm]{images/legend-int-12ab3ab.pdf}
	\end{tabular}
	\caption[width=1\linewidth]{
	Average deformation of the BCT crystal of Hertzian spheres associated with a-c) a vacancy and d-f) an interstitial at density $\rho\sigma^3=6.46$ and temperature $k_BT/\epsilon=0.0020$. 
	Figures and legends as explained in the caption of Fig. \ref{fig:hertzfcc}.
	}
	\label{fig:hertzbct}
\end{figure*}

\newcommand{\figwidthvc}{0.42\linewidth}
\begin{figure*}
\begin{tabular}{lll}
	a) & \hspace{0.5cm} & b) \\[-0.3cm]
	\includegraphics[width=\figwidthvc]{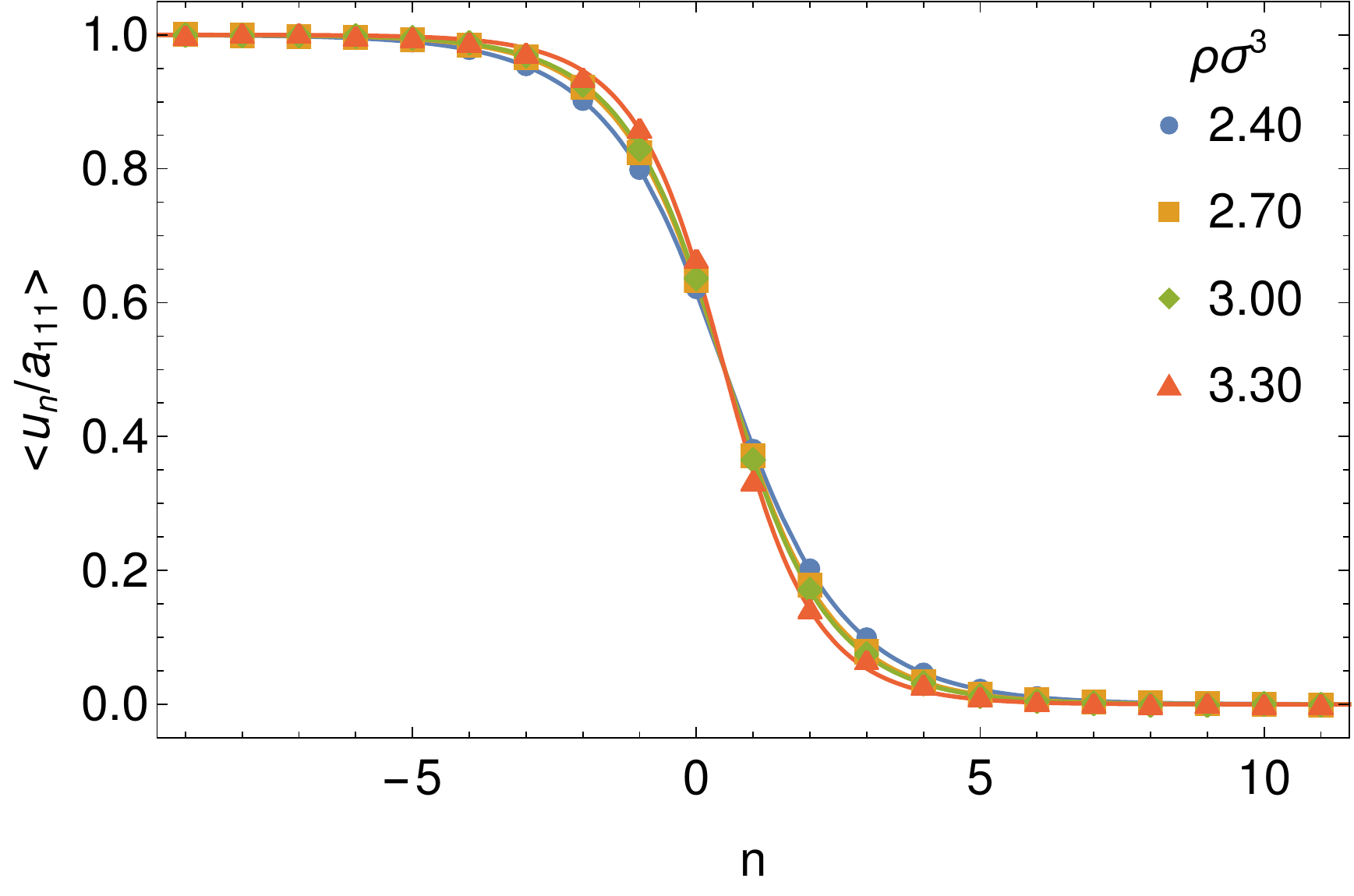}
	& & \includegraphics[width=\figwidthvc]{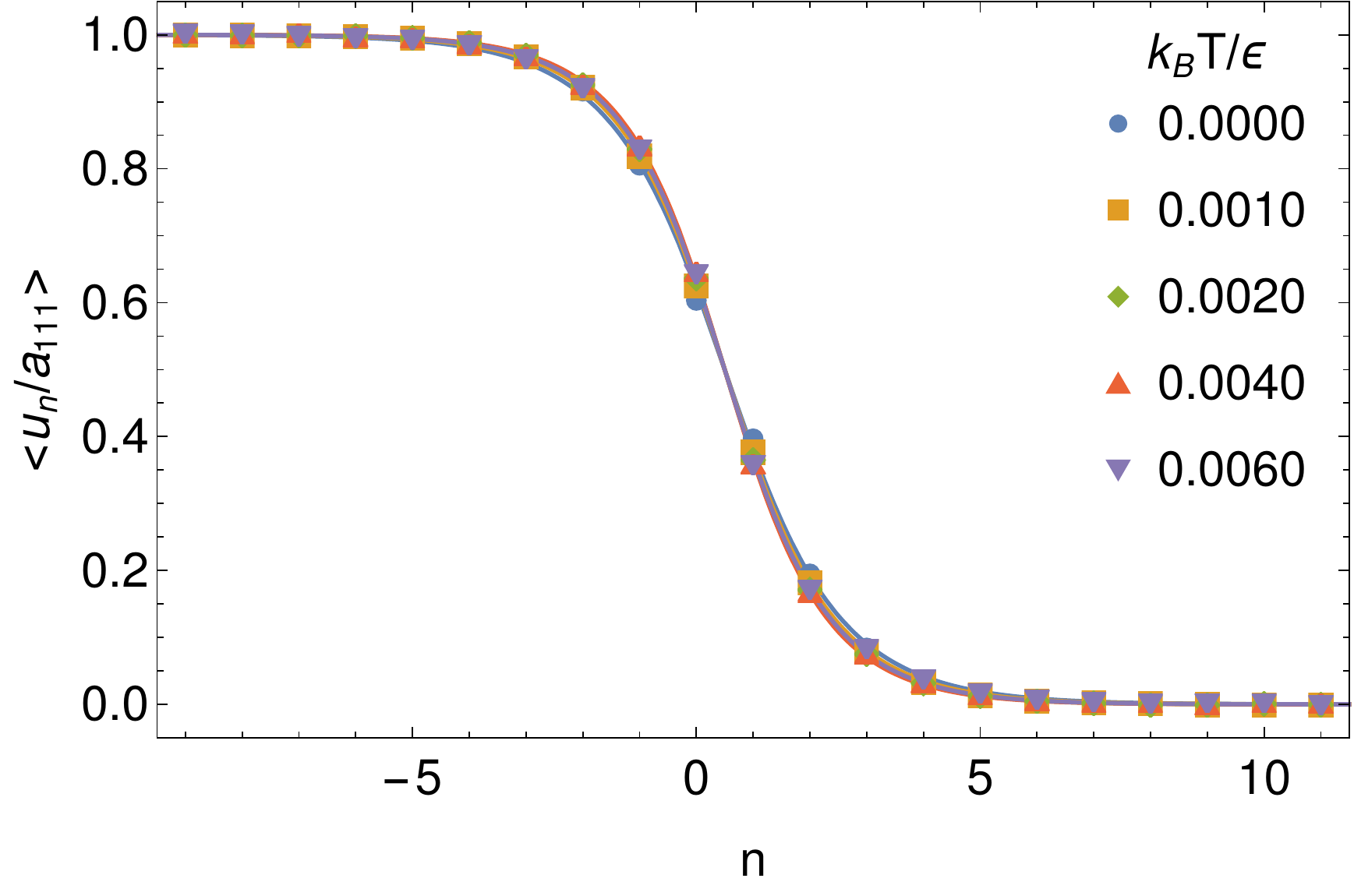} \\
	c) & & d) \\[-0.3cm]
	\includegraphics[width=\figwidthvc]{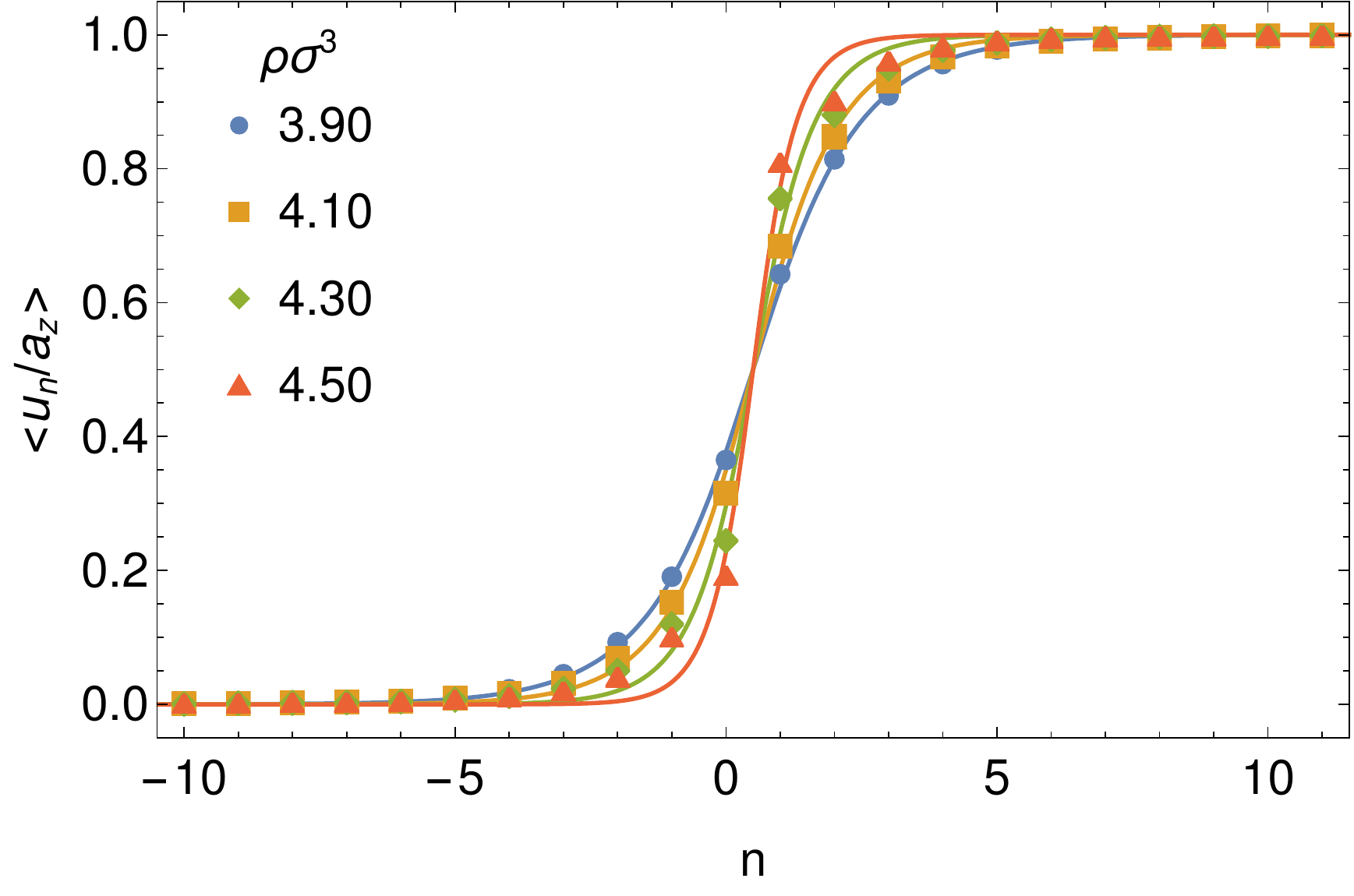}
	& & \includegraphics[width=\figwidthvc]{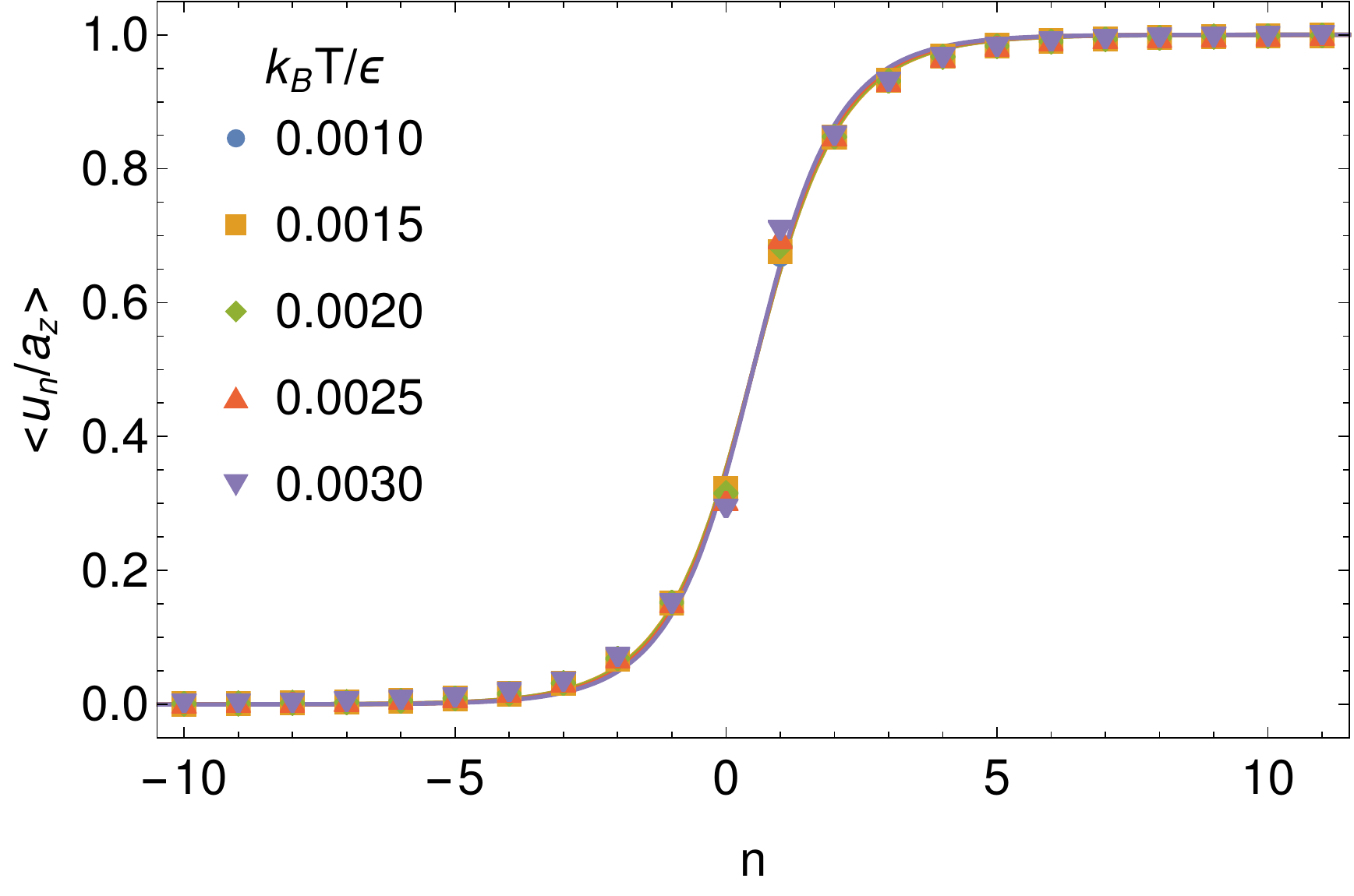}
	\end{tabular}
	\caption[width=1\linewidth]{
	Average displacement $\langle u_n\rangle$ along a-b) the $\langle 111\rangle$ direction of a crowdion in the BCC crystal and c-d) the z-direction of a voidion in the H crystal, both of Hertzian spheres. a,c) show the displacement for different densities at $k_BT/\epsilon=0.0020$ and b,d) for different temperatures at $\rho\sigma^3=3.00$ and $\rho\sigma^3=4.10$ for the BCC and H crystal, respectively.
	The lines represent the corresponding fitted soliton solutions. 
	Note that for the H crystal the simulation box was extended in the z-direction to prevent any effects of its own periodic image on the voidion.
	}
	\label{fig:hertzcrowdionvoidion}
\end{figure*}

\newcommand{\widthpdb}{0.32\linewidth}
\begin{figure*}
\begin{tabular}{lll}
     a) & b) & c) \\[-0.25cm]
     \includegraphics[width=\widthpdb]{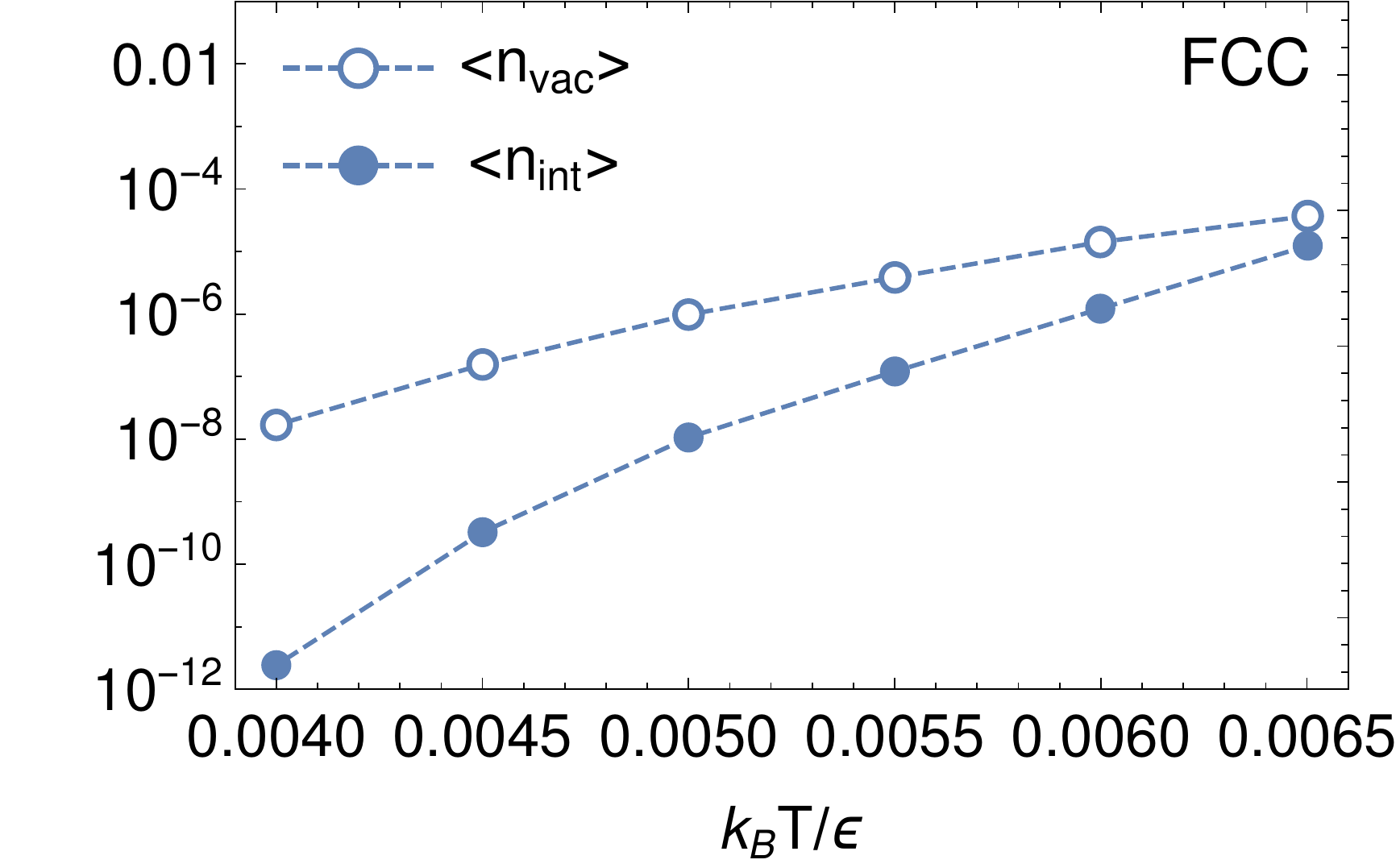}
     & \includegraphics[width=\widthpdb]{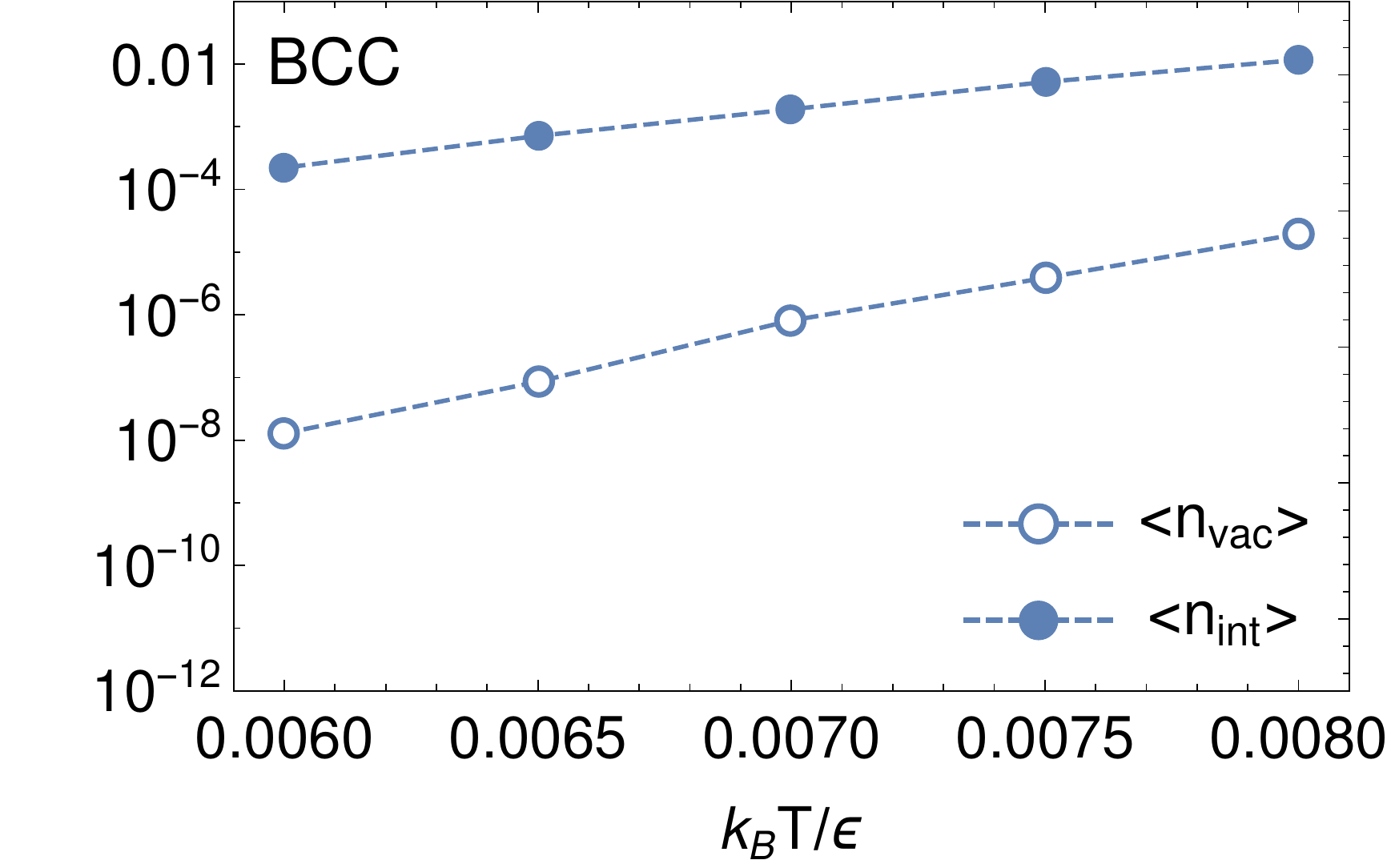}
     & \includegraphics[width=\widthpdb]{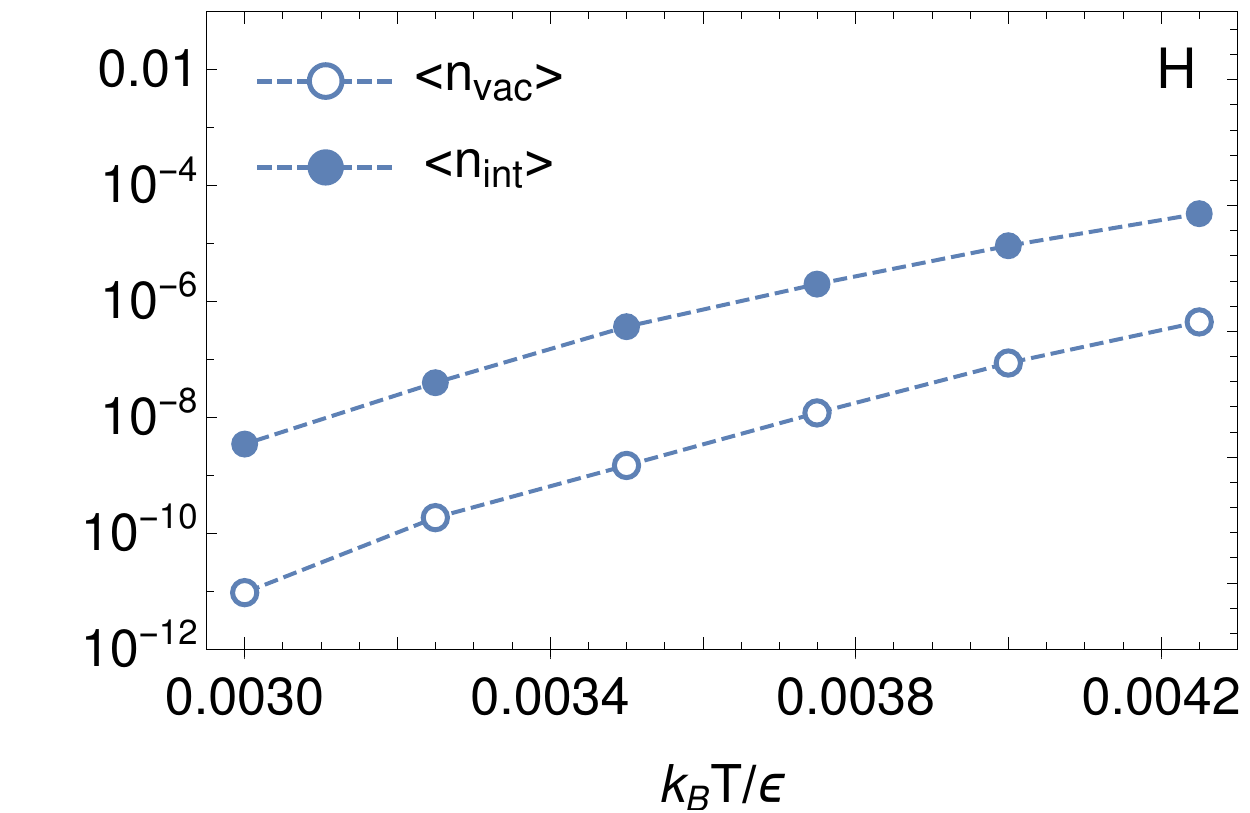}\\
\end{tabular} \\
\vspace{0.1cm}
\begin{tabular}{ll}
     c) & d) \\[-0.25cm]
     \includegraphics[width=\widthpdb]{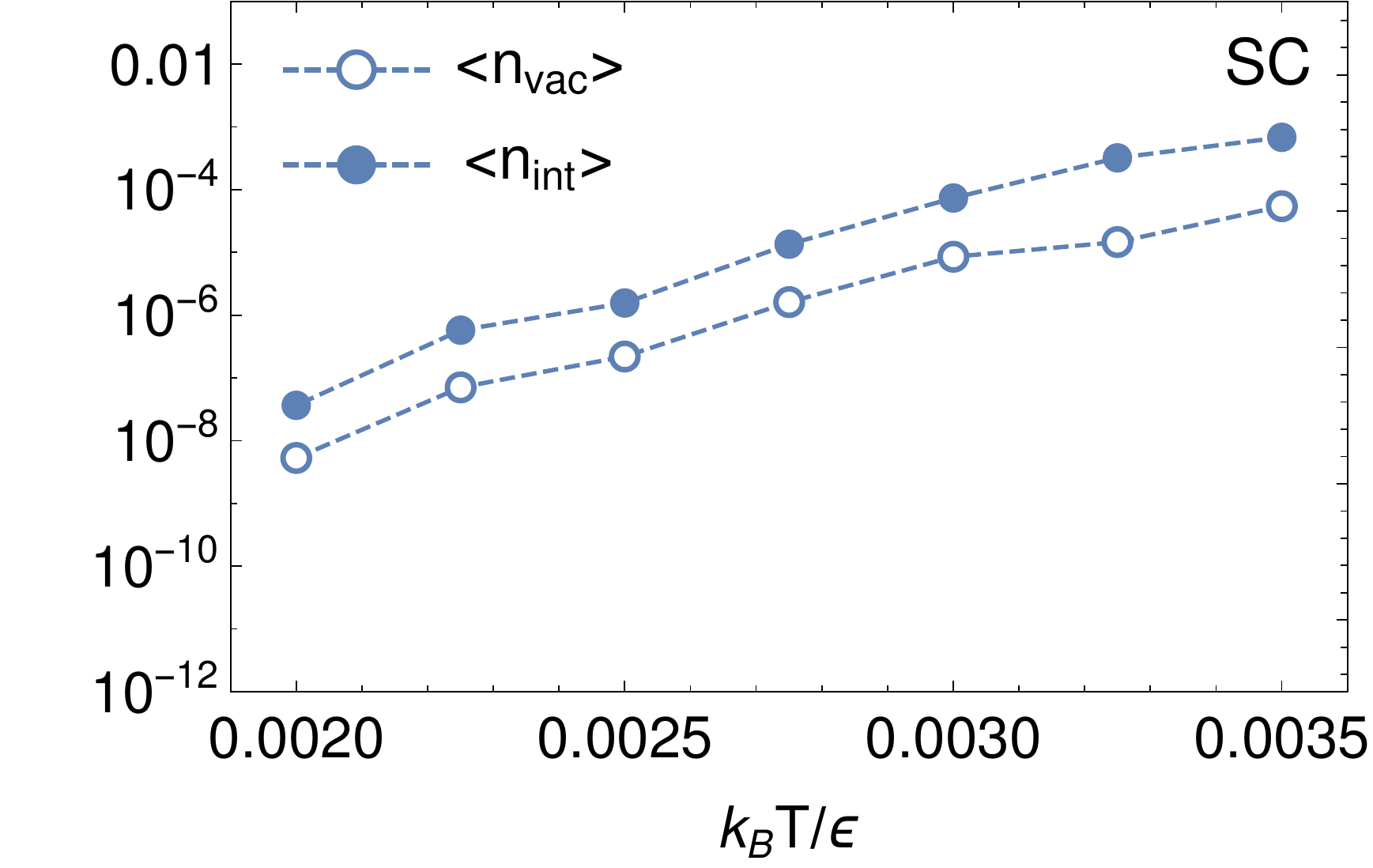}
     & \includegraphics[width=\widthpdb]{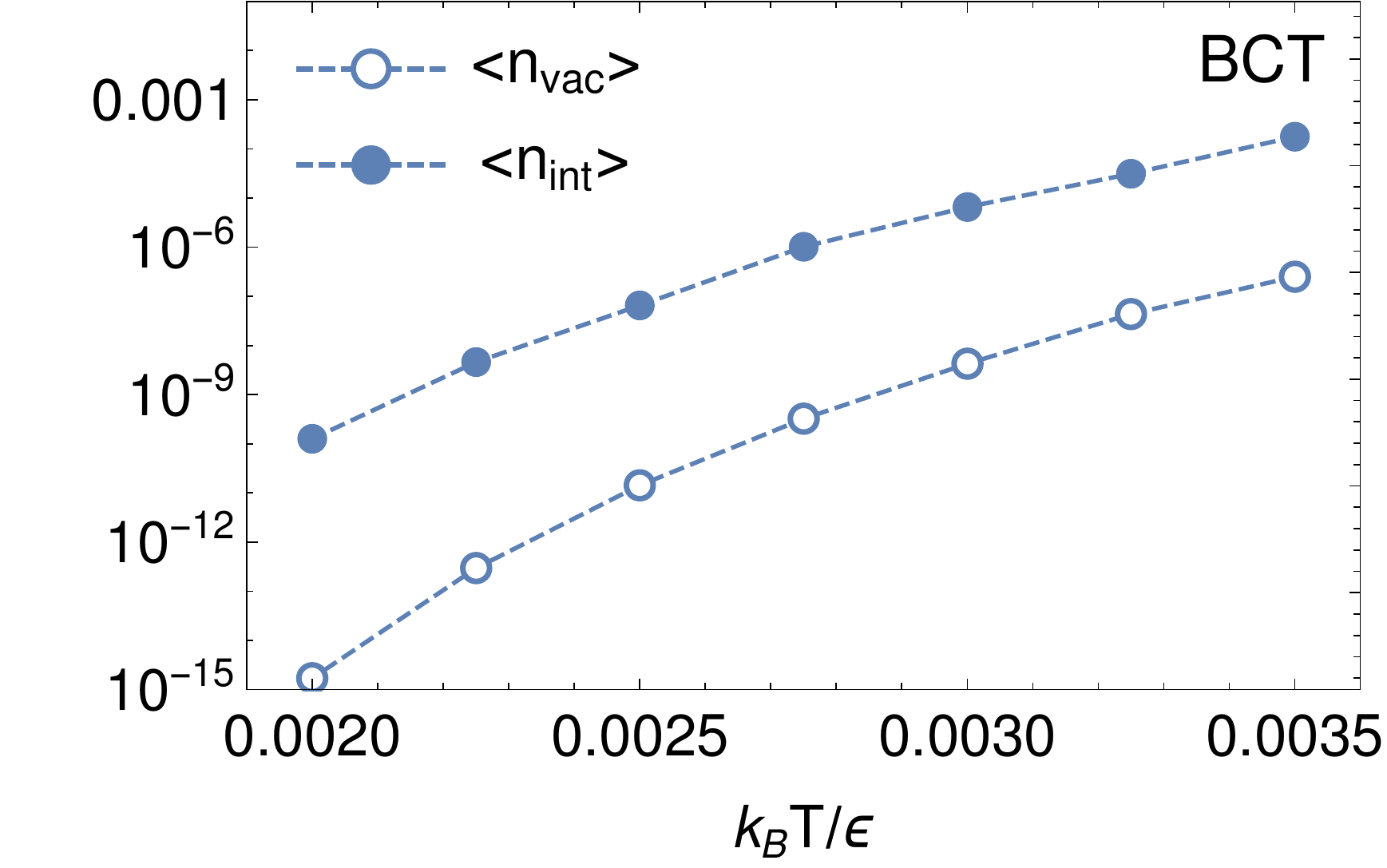}\\
\end{tabular}
    \caption[width=1\linewidth]{
	Vacancy and interstitial concentrations as a function of the temperature $k_BT/\epsilon$ for the different crystal phases of Hertzian spheres.
	The density of each crystal is given in Tab. \ref{tab:hertzstatepoints}.
	Note that the temperatures on the right side of the figures are near melting.
	}
	\label{fig:hertzconcentrations}
\end{figure*}

Figures \ref{fig:hertzfcc}-\ref{fig:hertzbct} show the average deformations of the FCC, BCC, H, SC, and BCT crystals of Hertzian spheres. 
In these figures, a) and d) show the deformation for a typical vacancy and interstitial, respectively, while b-c,e-f) give the displacement for the first three neighbor shells as a function of the density and temperature. Additionally, e-f) give the distance between the interstitial and its companion, i.e. the particle that it is sharing a Wigner-Seitz cell with. 
For convenience we will use the term ``dumbbell'' when referring to these two particles.

As expected, we observe that the largest deformations caused by a vacancy are associated with the first shell of neighbors. Notice, however, that the displacement in the FCC crystal is 6 to 10 times as small as in the other crystals. The reason for this difference is most likely the high number of nearest neighbors in the FCC crystal, i.e. 12 in comparison to 8 for BCC, 6 for SC, and 2 for H and BCT. The 12 nearest neighbors in the FCC crystal constrain each particle's displacement. 

Furthermore, we find that the deformation of the FCC and BCT crystals associated with a vacancy is largely three-dimensional, symmetric, and local. This was also found for the FCC crystal of hard spheres \cite{van2017diffusion} and repulsive point Yukawa particles \cite{alkemade2021point}. More interesting are the deformations of the other crystals. 
In the BCC crystal, the vacancy causes a symmetry-broken, three-dimensional deformation in which only 4 of the 8 nearest neighbors are displaced, each leaving a trail of inwards moving particles behind.
Similarly, in the SC crystal, only 4 of the 6 nearest neighbors are displaced, again each leaving a trail of inwards moving particles behind. As a result, the deformation of the BCC and SC crystals has a longer range than that of the FCC and BCT crystals.
The two-dimensional deformation of the SC crystal is suprising, as vacancies in the SC crystal of (slanted) hard cubes and several other repulsive potentials cause one-dimensional deformations \cite{smallenburg2012vacancy,van2017phase,van2018revealing}.
The softness combined with the shape of the Hertzian spheres make it apparently more favorable for 4 of the nearest neighbors to move into the vacancy instead of 2.
For the H crystal, we do find that the deformation associated with a vacancy is essentially one-dimensional, with the main distortion oriented along the z-axis through the vacancy. Notice that the size and range of this vacancy structure is greater than in any of the other crystal structures. This further indicates that decreasing the dimensionality of the lattice deformation increases its range. 

Contrary to the vacancy cases, the deformation caused by an interstitial has a broken symmetry by default, as it depends on the preferred orientation of the dumbbell. 
For the FCC crystal, we find that the dumbbell orients itself along one of the three $\langle100\rangle$ directions, pushing 8 of the 12 nearest neighbors away (Fig. \ref{fig:hertzfcc}d). Interestingly, the displacement caused by an interstitial is larger than the displacement caused by a vacancy, which was seen for hard spheres as well \cite{van2017diffusion}, whereas it is the other way around for the other crystals.
Nevertheless, the deformation of the FCC crystal associated with an interstitial is still local. 

In general, we find that the deformations associated with an interstitial are three-dimensional and local. The only exception is the BCC crystal, in which the dumbbell orients itself along one of the $\langle111\rangle$ directions, causing an extended, one-dimensional deformation. 
This one-dimensional deformation was also found in the BCC crystal of repulsive point Yukawa particles \cite{alkemade2021point}.

Moreover, looking at the displacement of the first three neighbor shells as a function of density, 
we find that the deformation caused by a vacancy generally increases as the density increases, while the deformation caused by an interstitial decreases. Note that this is directly opposite what is observed for defects in crystals of hard spheres, where the deformation from a vacancy decreases with density, and the deformation of an interstitial grows with density \cite{van2017diffusion}. However, in the case of soft, repulsive particles like we examine here, the dominant feature controlling the density dependence of the defect deformation is likely the  ``push'' from the surrounding particles on the particles nearest to the defect. For the case of a vacancy, this means that the neighboring particles are pushed more into the vacant lattice site, whereas the outward displacement of the particles neighboring an interstitial is restricted by this ``push''.
A noteworthy exception to this general behavior is the vacancy in the H crystal, for which the displacement of the two nearest neighbors drastically decreases with increasing density. The reason for this is that the particles of the second neighbor shell, which lie in the xy-plane of the vacancy, are pushed more into the vacancy, hindering the displacement of the two nearest neighbors positioned above and below the vacancy.

Likewise, we find that increasing the temperature generally increases the deformation caused by a vacancy and decreases that caused by an interstitial. This can be explained by observing that increasing the temperature makes it easier for Hertzian spheres to overlap. 
For neighbors of a vacancy, this means that their displacement towards the vacancy is less constrained by the displacement of other neighbors. For the interstitial, this means that the distance between the dumbbell particles and its effect on the surrounding particles decreases.
Note, however, that the effect of the temperature is generally significantly smaller than the effect of the density.

The extended, one-dimensional deformations of the H crystal associated with a vacancy and the BCC crystal associated with an interstitial strongly resemble a voidion and crowdion. 
The classic characteristic of crowdions and voidions is that their shape is well captured by the Frenkel-Kontorova model \cite{kontorova1938theory,braun1998nonlinear,landau1993model,dudarev2003coherent}. Hence, in order to characterize the structure of these one-dimensional defects, we measure the average particle displacements $u_n=x_n-a_d n$ around the vacancy and interstitial, along the defect direction. Here $x_n$ is the position of particle $n$ along the defect and $a_d$ is the crystal lattice spacing in this direction. We choose $n=0$ to correspond to the particle just before the defect center and use the standard boundary conditions: $u_{n=-\infty}=a_d$, $u_{n=+\infty}=0$ for the interstitial and $u_{n=-\infty}=0$, $u_{n=+\infty}=a_d$ for the vacancy \cite{kovalev1993generalized}. 

Figure \ref{fig:hertzcrowdionvoidion} shows the average displacement along the $\langle111\rangle$ direction for an interstitial in the BCC crystal and along the z-direction for a vacancy in the H crystal for different densities and temperatures. We compare these displacements to the soliton solution of the sine-Gordon equation, i.e. the continuum limit of the Frenkel-Kontorova model, using a single fitting parameter to match the extension of the defect. We observe excellent agreement for the interstitial in the BCC crystal (Fig. \ref{fig:hertzcrowdionvoidion}a-b) and good agreement for $\rho\sigma^3\lesssim4.1$ for the vacancy in the H crystal (Fig. \ref{fig:hertzcrowdionvoidion}c-d). 
Recall that increasing the density results in a less one-dimensional and more three-dimensional deformation of the H crystal associated with a vacancy. For this reason, the deformation cannot be fully characterized as a voidion for higher densities.

Aside from this, notice that the structure and range of the crowdion and voidion are essentially independent of the density and temperature. This was seen in other systems as well \cite{van2018revealing,alkemade2021point}.

Next, to fully understand the impact of the defects on the overall structure of the different crystals, we also predicted the equilibrium concentration of vacancies and interstitials in the different crystals, as explained in Methods.
For each crystal phase we choose one density (see Tab. \ref{tab:hertzstatepoints}) and varied the temperature to get some insight  in the concentrations as a function of the proximity of the solid-fluid phase transition. The resulting concentrations are shown in Fig. \ref{fig:hertzconcentrations}.
We observe that all defect concentrations increase with increasing temperature; thus, similar to what was seen for hard (slanted) cubes \cite{van2017phase,smallenburg2012vacancy}, the concentrations are the highest closest to melting.

\begin{figure*}
\begin{tabular}{lllll}
	& a) & \,\, b) & c) & \\[-0.3cm]
	\includegraphics[width=\legendsizeA,trim= 0cm -0.5cm 0.0cm 0.0cm]{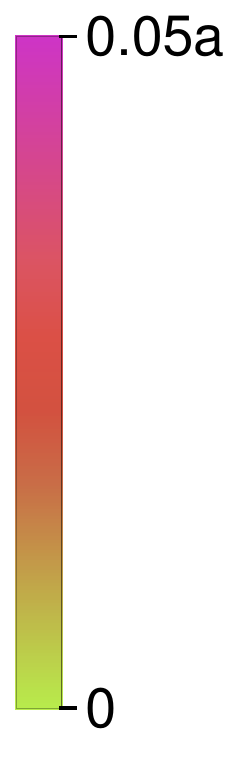} 
    & \includegraphics[width=\figwidth]{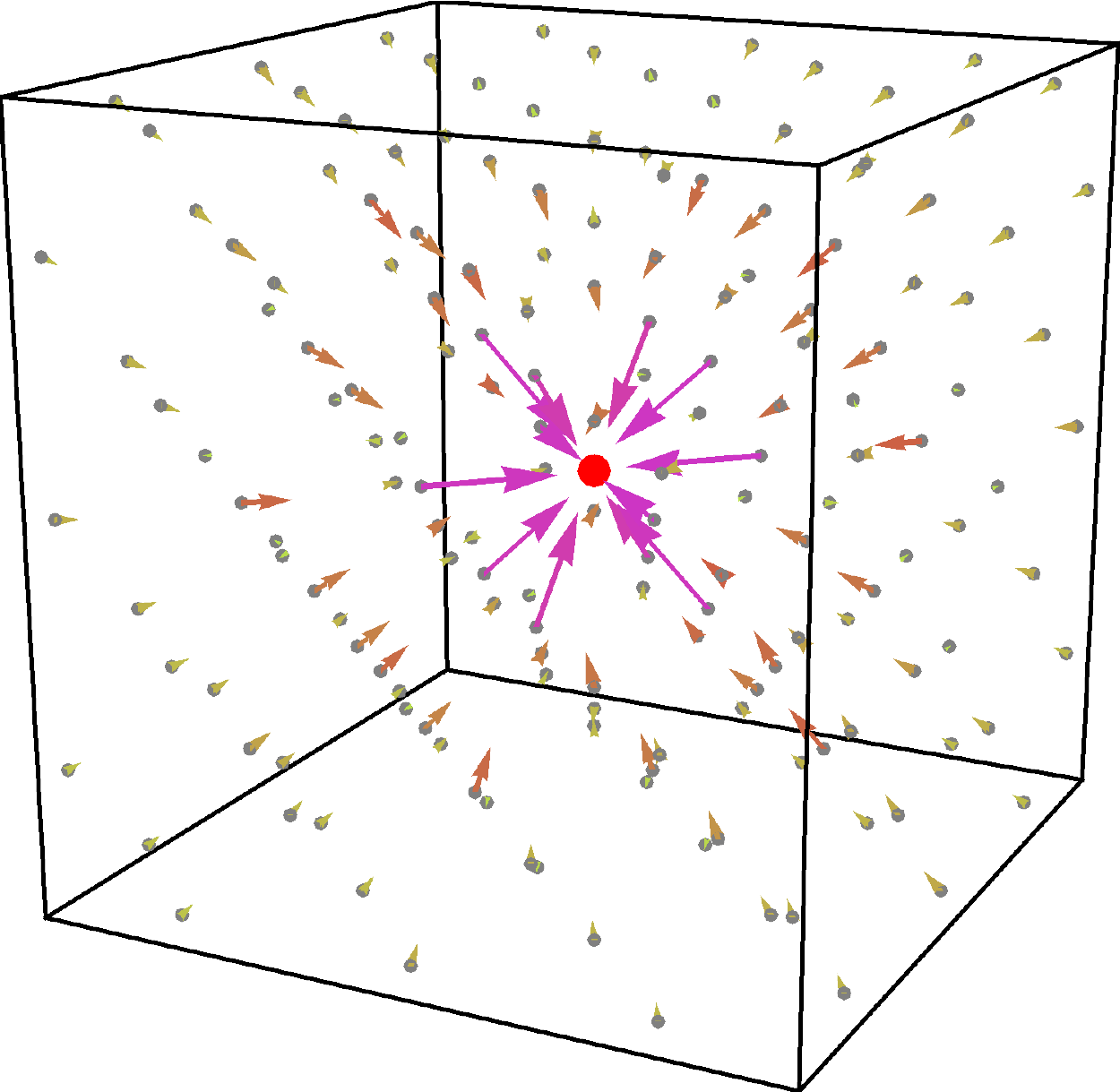} 
    & \includegraphics[width=\figwidthB, trim= 0cm -2.1cm 0.8cm 0.0cm,clip]{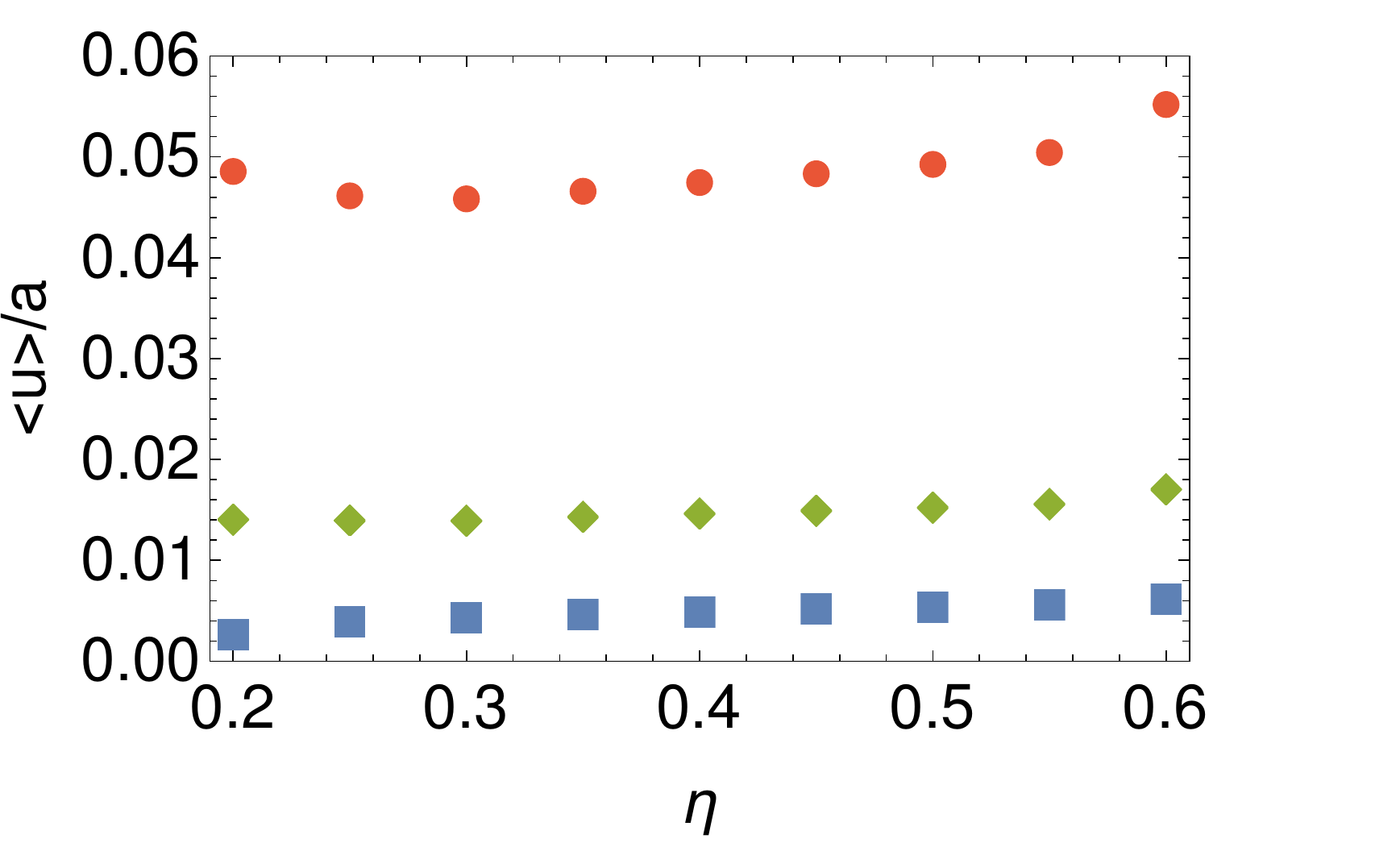}
    & \includegraphics[width=\figwidthB, trim= 0.8cm -2.1cm 0.0cm 0.0cm,clip]{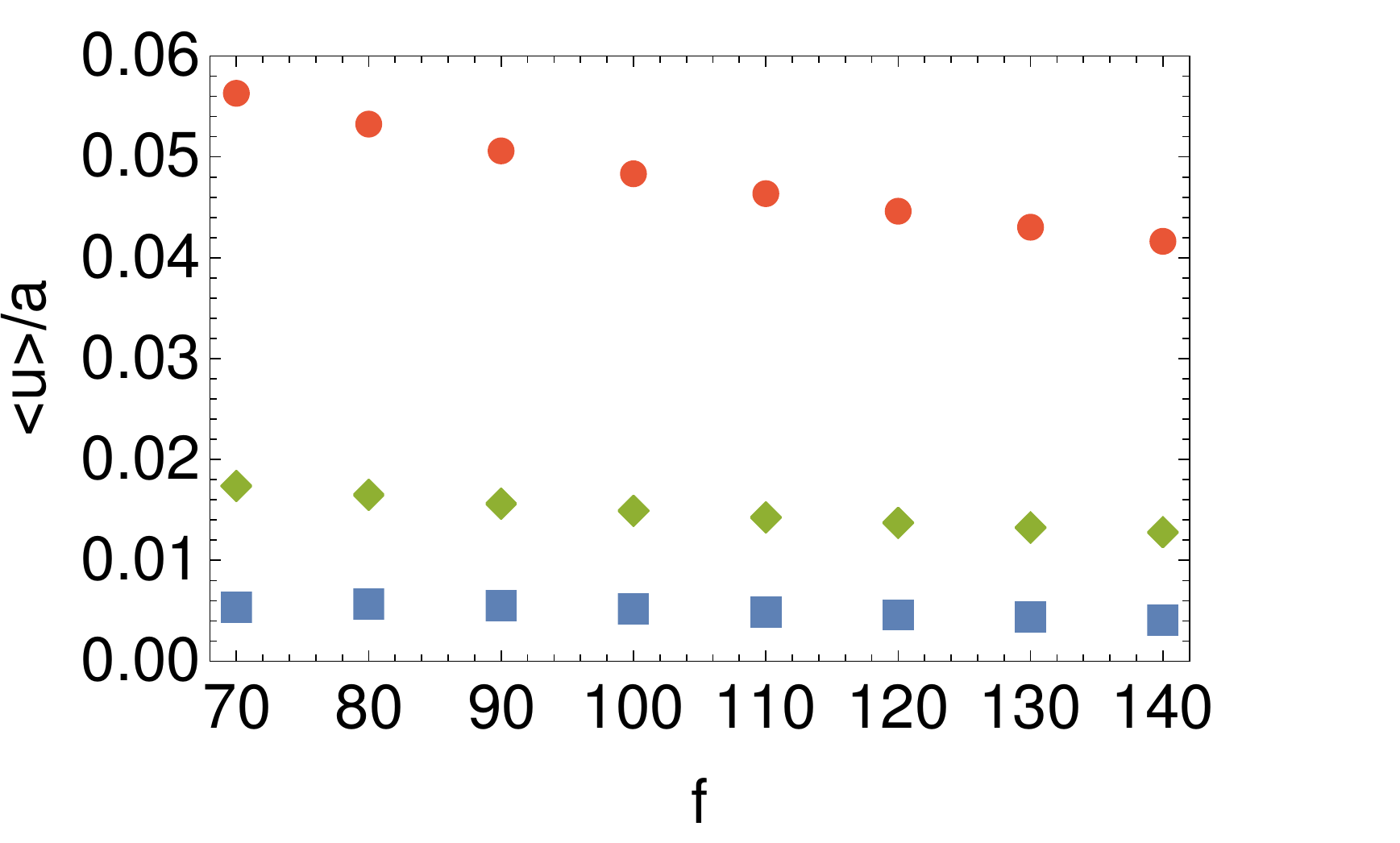}
    & \hspace{-0.2cm}\includegraphics[width=\legendsize,trim= 0.0cm -7.0cm 0.0cm 0.0cm]{images/legend-vac1.pdf} \\
    & d) & \,\, e) & f) &\\[-0.3cm]
    \includegraphics[width=\legendsizeD,trim= 0cm -0.5cm 0.0cm 0.0cm]{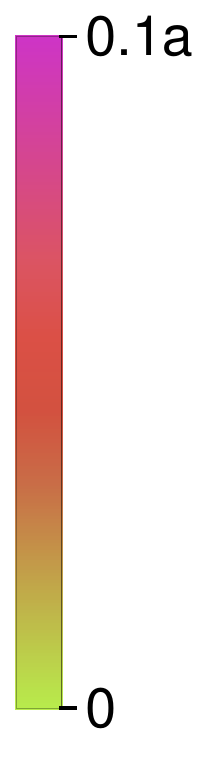} 
    & \includegraphics[width=\figwidth]{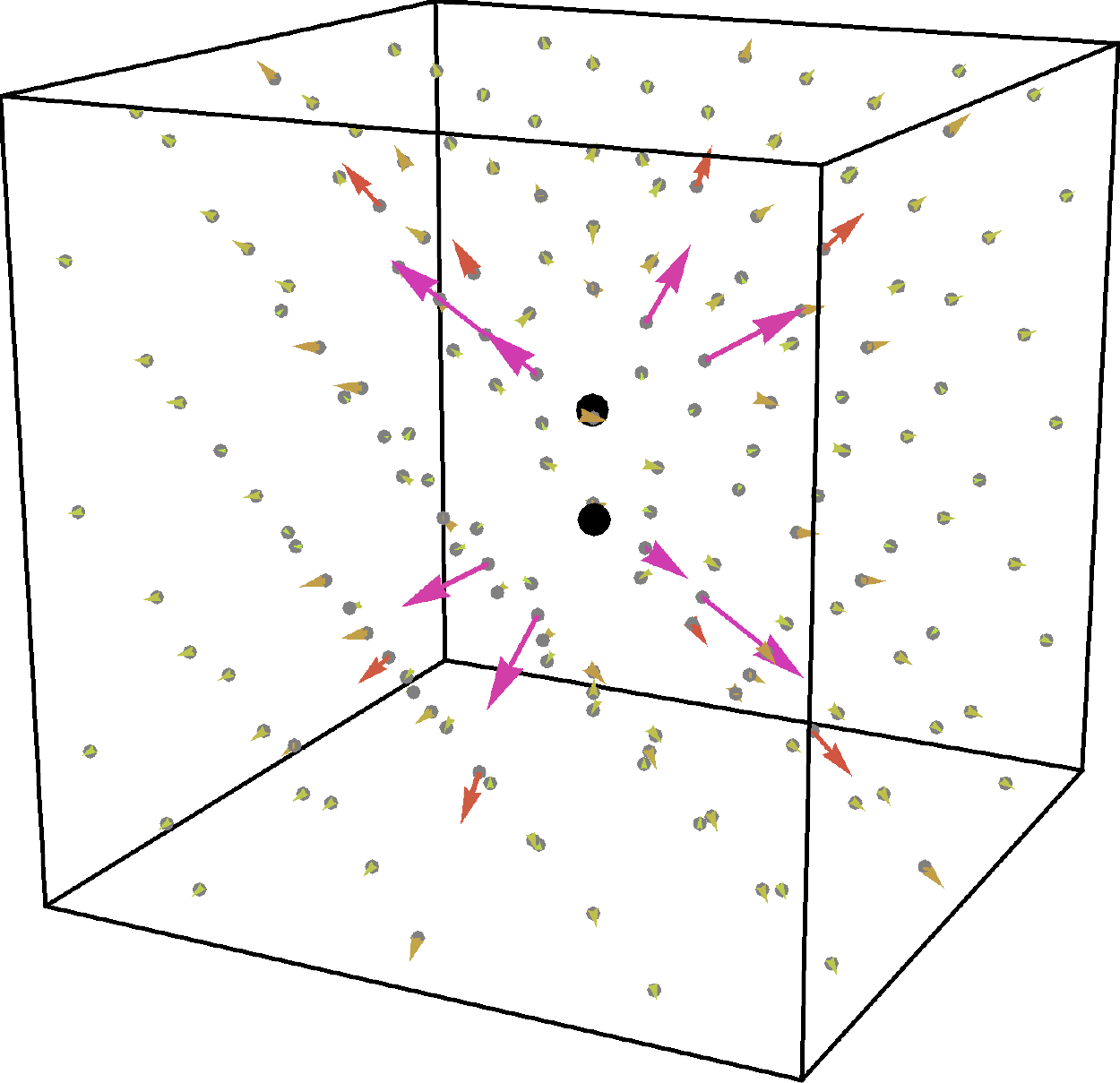} 
    & \includegraphics[width=\figwidthB, trim= 0cm -2.1cm 0.8cm 0.0cm,clip]{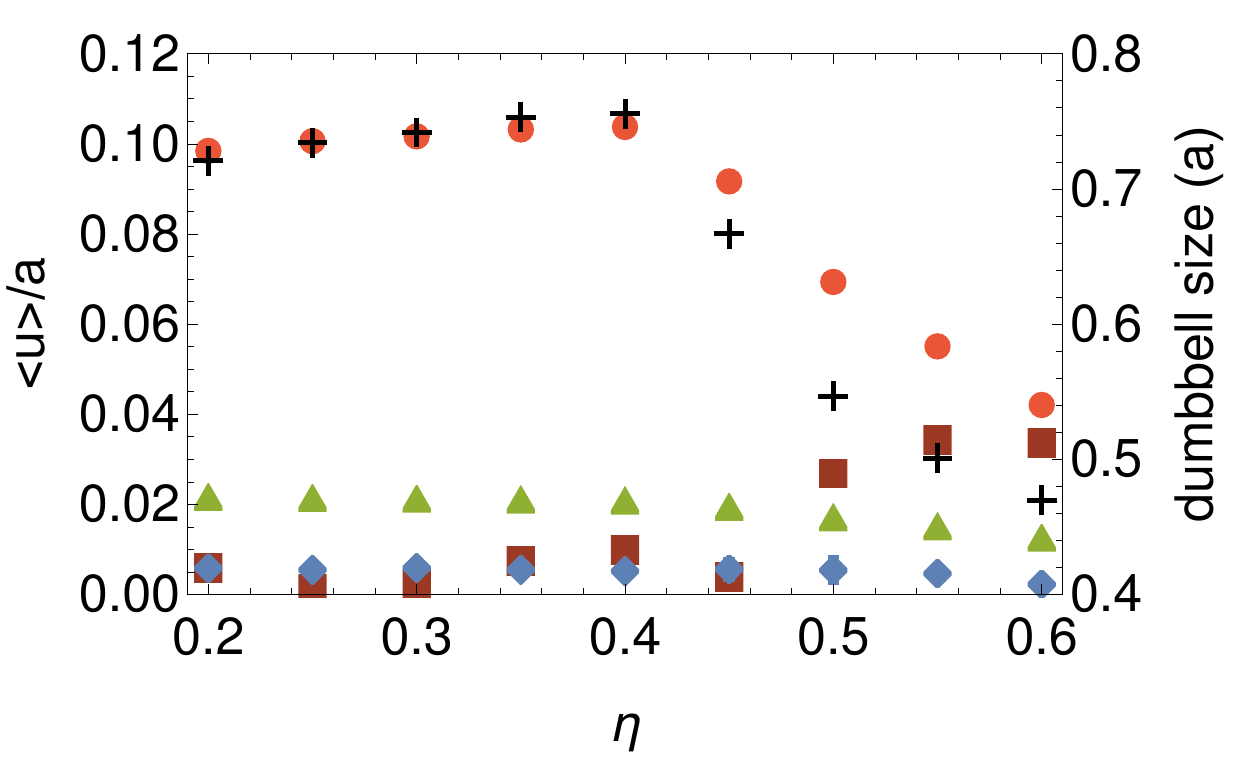}
    & \includegraphics[width=\figwidthB, trim= 0.8cm -2.1cm 0.0cm 0.0cm,clip]{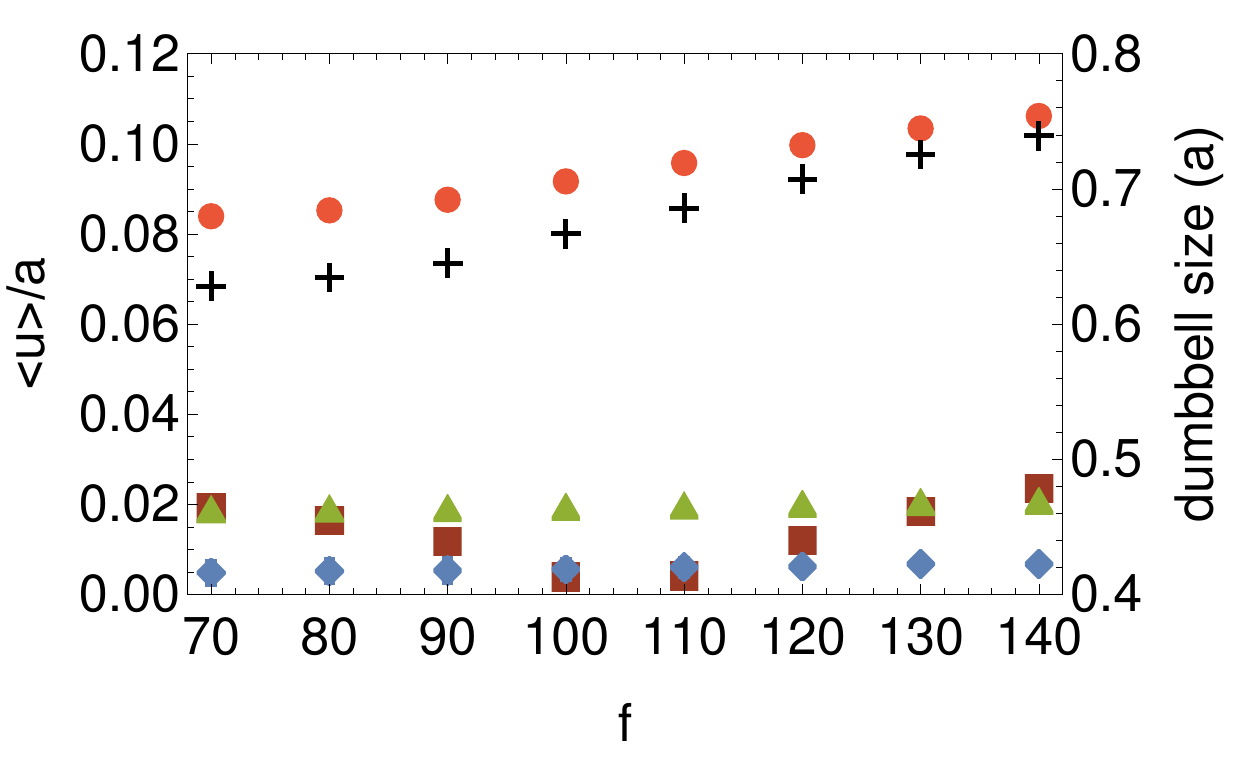}
    & \includegraphics[width=\legendsize,trim= 0.0cm -7.8cm 0.0cm 0.0cm]{images/legend-int-1ab23.pdf}
	\end{tabular}
	\caption[width=1\linewidth]{
	Average deformation of the FCC crystal of star polymers associated with a-c) a vacancy and d-f) an interstitial. In a,d) the deformation at packing fraction $\eta=0.45$ and arm number $f=100$ is shown. The vacancy is indicated by the red dot, and the interstitial and its companion by the two black dots. The gray points represent the lattice sites and the arrows represent the deformation. The size of the arrows is exaggerated, but the color indicates the deformation in terms of the nearest-neighbor distance $a$. 
	b-c,e-f) Average deformation $\langle u\rangle/a$ for the first three neighbor shells (1=nn, 2=nnn, and 3=nnnn) as a function of the packing fraction (at $f=100$) and arm number (at $\eta=0.45$). 
	When the displacement of the particles in a neighbor shell has a broken symmetry, the label ``A'' indicates the most displaced particles and ``B'' the others.
	The distance between the interstitial and its companion - the ``dumbbell'' - is given in e-f) as well (right axis). 
	}
	\label{fig:starpolfcc}
\end{figure*}

Furthermore, we find that the vacancy concentration is higher than the interstitial concentration in the FCC crystal, which was also found for the FCC crystal of hard spheres near melting \cite{bennett1971studies,pronk2001point}. Surprisingly, in the other crystals, the interstitial concentrations are higher than the vacancy concentrations, with the difference ranging from 4 orders magnitude in the BCC and BCT crystals to 3 orders in the H crystal and 1 order in the SC crystal.
This goes against the general expectation that vacancies are easier to accommodate than interstitials, leading to higher vacancy concentrations.
The highest concentrations are found for interstitials in the BCC crystal. Specifically, near melting the concentration is on the order of 1\%, which is 6 orders of magnitude higher than the interstitial concentration for hard spheres near melting \cite{pronk2001point}.


\subsection{Star Polymers}
We continue our investigation by exploring the manifestation of defects in the different crystal phases of star polymers in a similar fashion as we did for Hertzian spheres. First, we take a look at the deformations.
For each crystal phase, we focus on one state point where the crystal is stable, and examine the deformation as a function of the arm number and packing fraction starting from this state point. 
Based on Ref. \onlinecite{watzlawek1999phase}, the arm number and packing fraction at these state points are given in Tab. \ref{tab:starpolstatepoints} together with the number of particles simulated. Note that Ref. \onlinecite{watzlawek1999phase} also predicts a BCO crystal; however, we did not find it to be stable in our simulations.

\begin{table}
	\centering
	\begin{tabular}{|p{0.18\linewidth}<{\raggedright\arraybackslash}||p{0.13\linewidth}<{\centering} p{0.13\linewidth}<{\centering} p{0.13\linewidth}<{\centering}|}
		\hline
		Crystal  & $\eta$ & $f$ & $N$\\
		\hline \hline
		FCC  & $0.45$ & 100 & 2048\\
		BCC  & $0.45$ & 45 & 2000\\ 
		diamond  & $1.25$ & 100 & 2744\\ 
		\hline
	\end{tabular}
	\caption{The number of star polymers $N$ simulated for each crystal phase together with the packing fraction $\eta$.}
	\label{tab:starpolstatepoints}
\end{table}

\begin{figure*}
\begin{tabular}{lllll}
	& a) & \,\, b) & c) & \\[-0.3cm]
	\includegraphics[width=\legendsizeA,trim= 0cm -0.5cm 0.0cm 0.0cm]{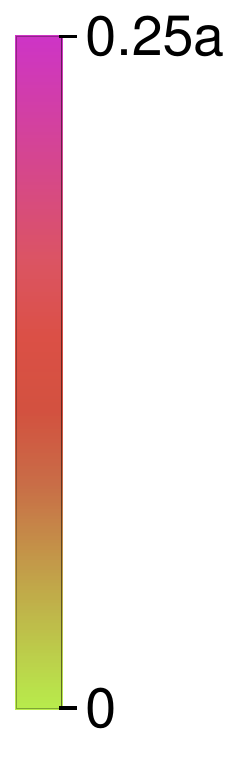} 
    & \includegraphics[width=\figwidth]{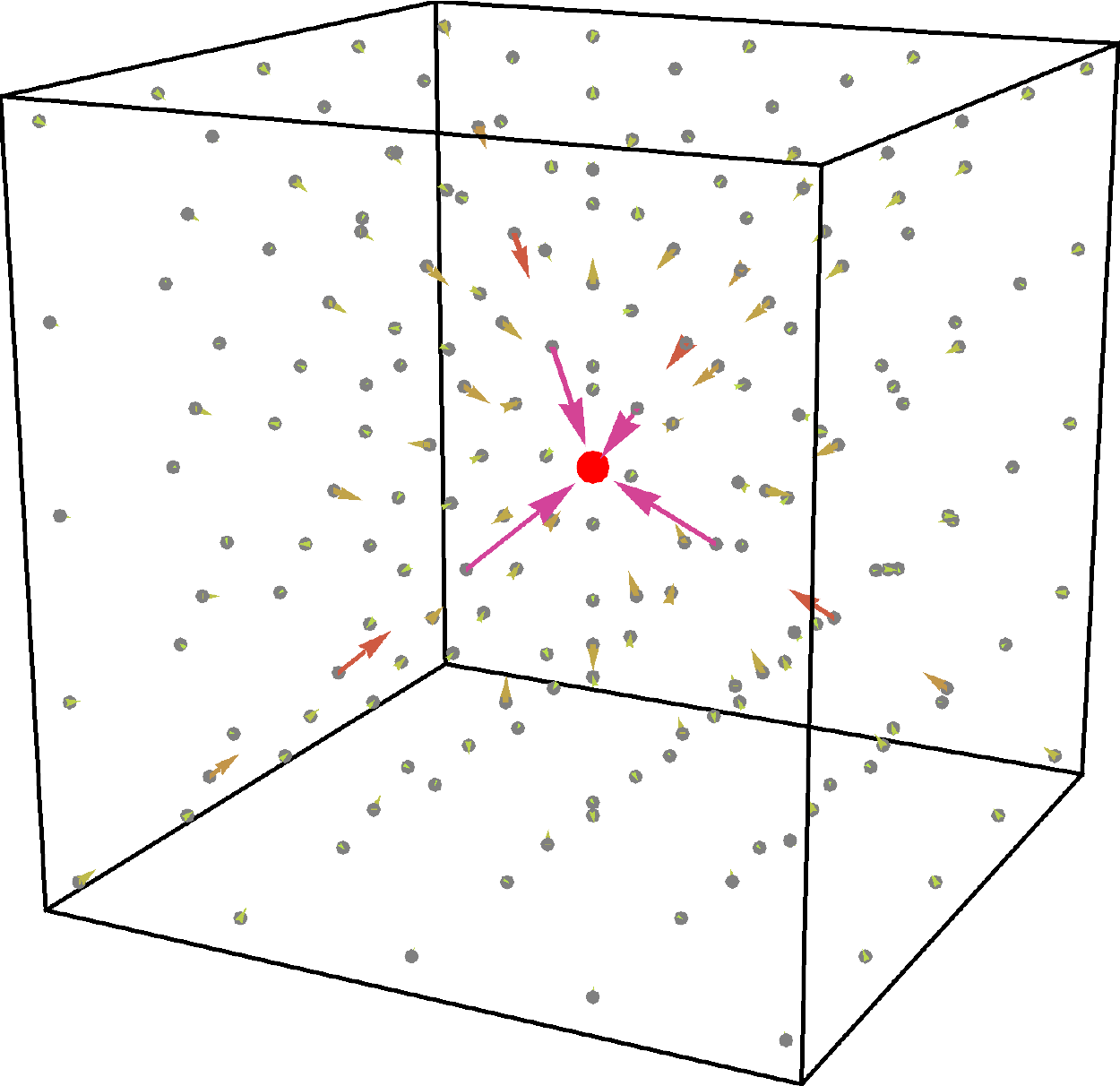} 
    & \includegraphics[width=\figwidthB, trim= 0cm -3.0cm 0.6cm 0.0cm,clip]{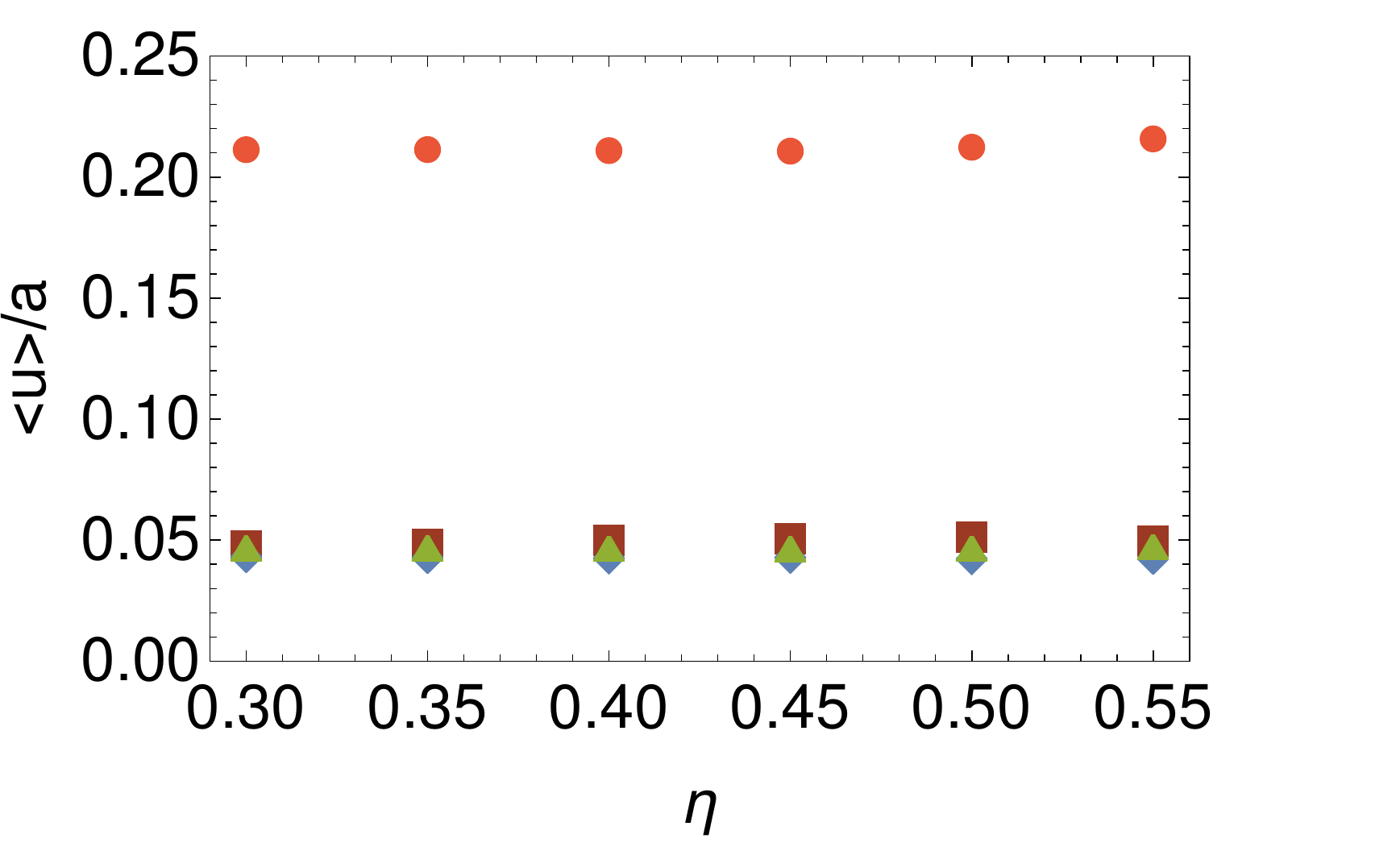}
    & \includegraphics[width=\figwidthB, trim= 0.6cm -2.9cm 0.0cm 0.0cm,clip]{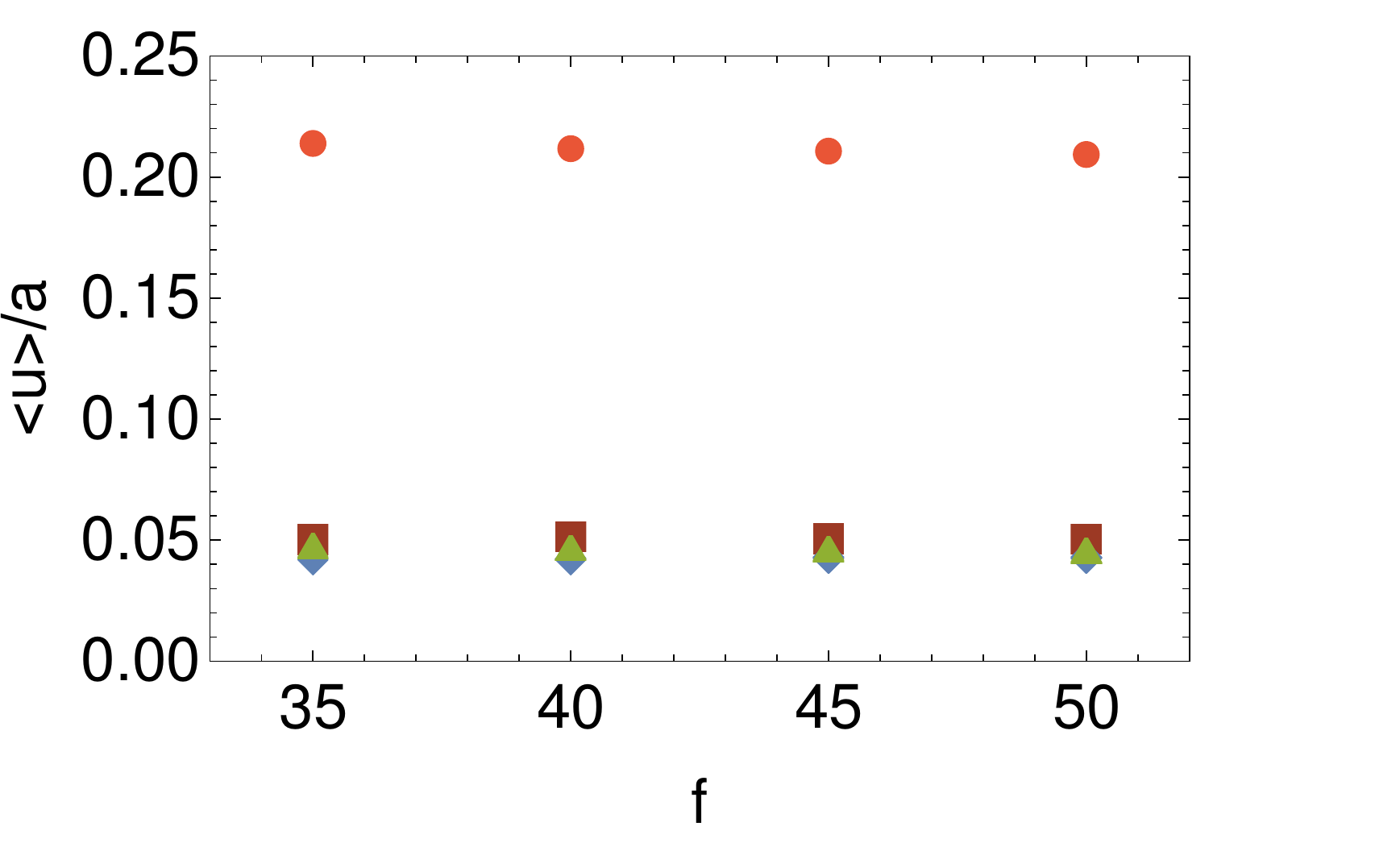}
    & \hspace{-0.2cm}\includegraphics[width=\legendsize,trim= 0.0cm -7.0cm 0.0cm 0.0cm]{images/legend-vac2.pdf} \\
    & d) & \,\, e) & f) &\\[-0.3cm]
    \includegraphics[width=\legendsizeA,trim= 0cm -0.5cm 0.0cm 0.0cm]{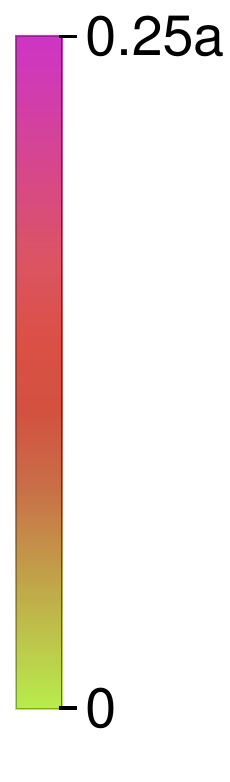} 
    & \includegraphics[width=\figwidth]{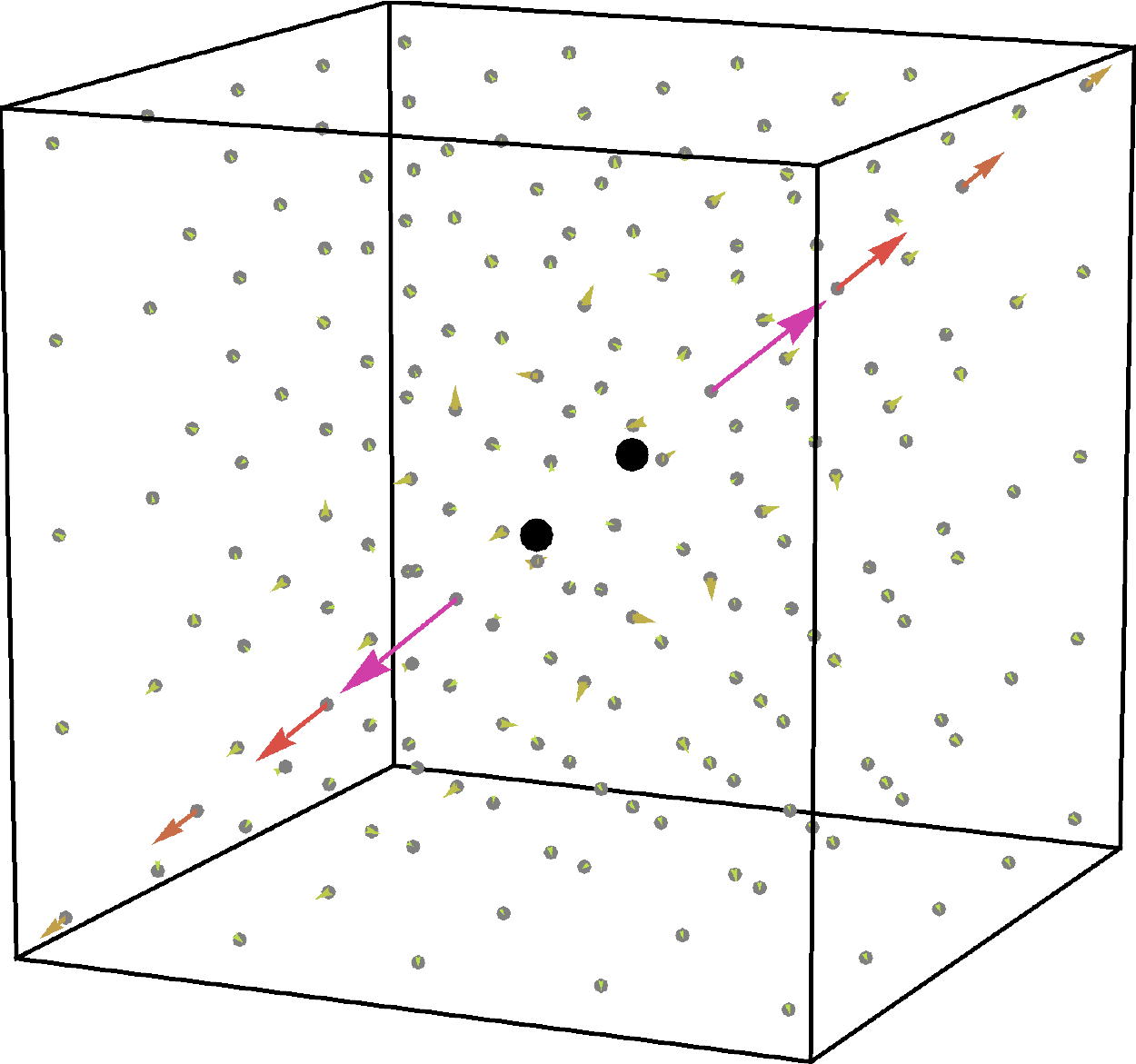} 
    & \includegraphics[width=\figwidthB, trim= 0cm -2.1cm 0.6cm 0.0cm,clip]{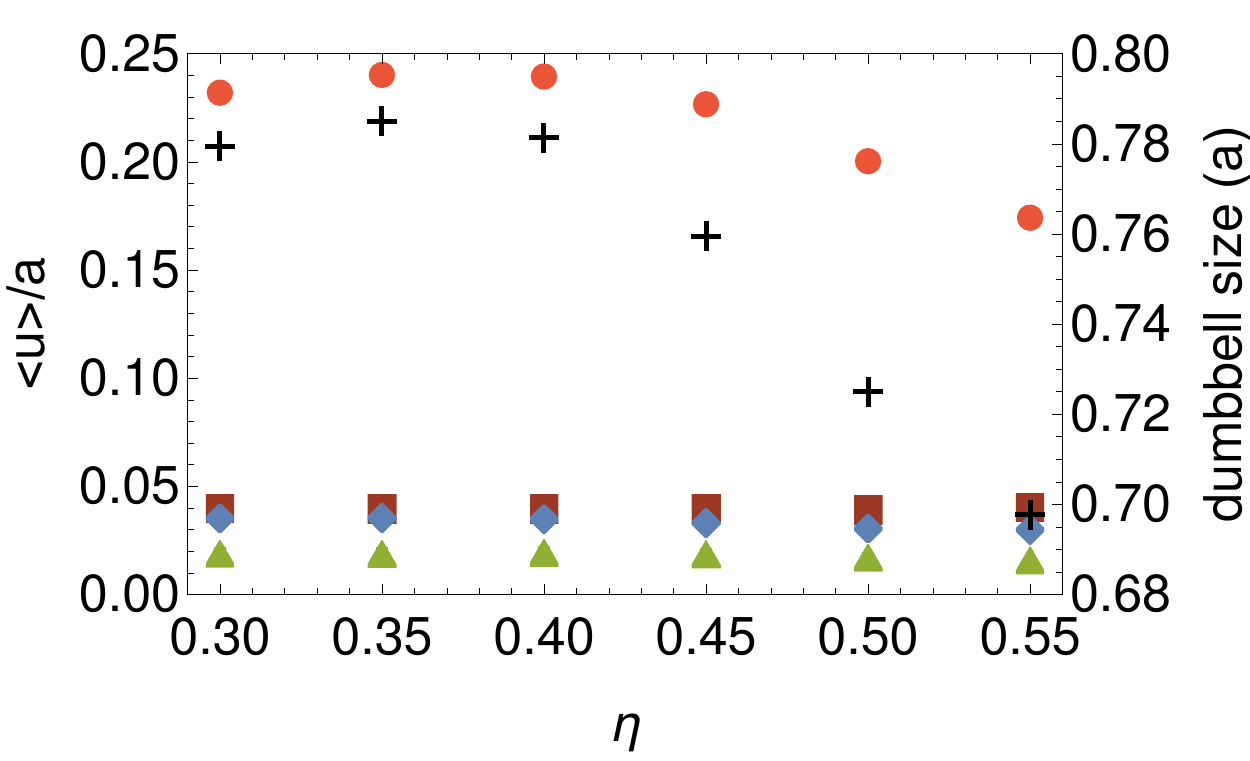}
    & \includegraphics[width=\figwidthB, trim= 0.6cm -2.1cm 0.0cm 0.0cm,clip]{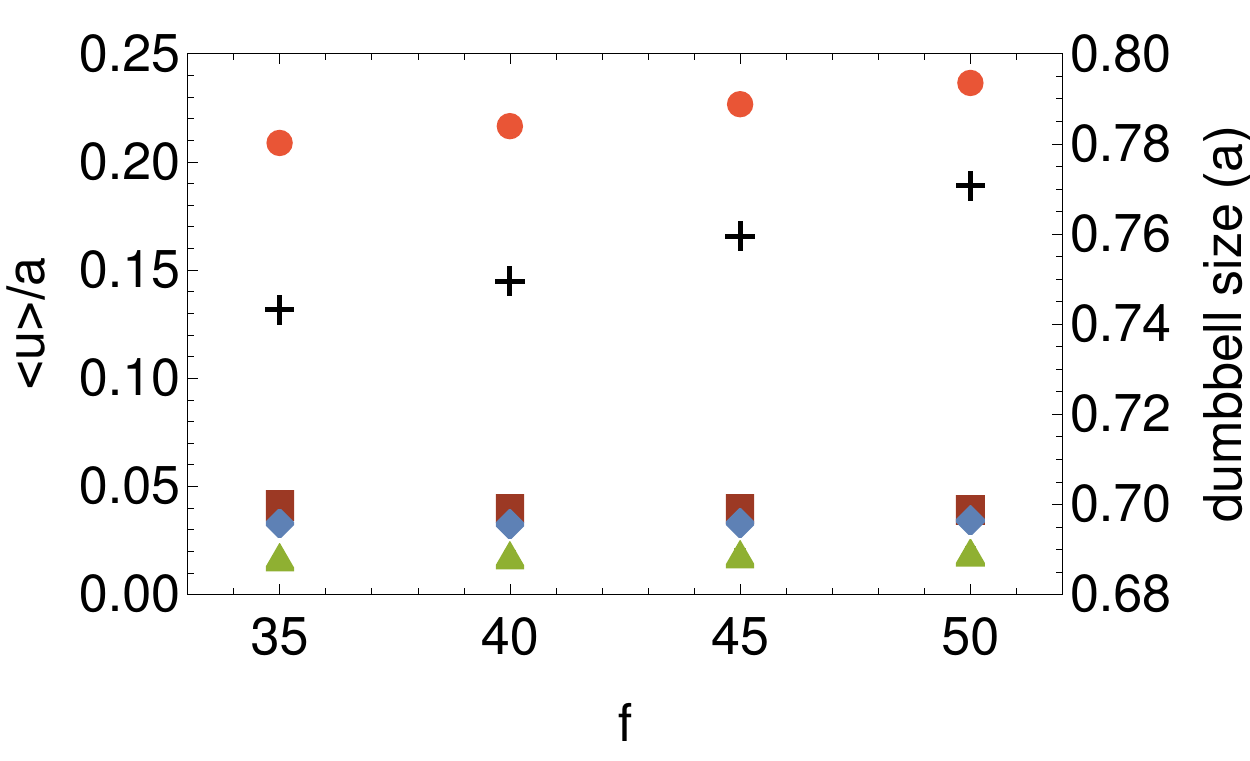}
    & \includegraphics[width=\legendsize,trim= 0.0cm -7.8cm 0.0cm 0.0cm]{images/legend-int-1ab23.pdf}
	\end{tabular}
	\caption[width=1\linewidth]{
	Average deformation of the BCC crystal of star polymers associated with a-c) a vacancy and d-f) an interstitial at packing fraction $\eta=0.45$ and arm number $f=45$. 
	Figures and legends as explained in the caption of Fig. \ref{fig:starpolfcc}.
	}
	\label{fig:starpolbcc}
\end{figure*}

\begin{figure*}
\begin{tabular}{lllll}
	& a) & \,\, b) & c) & \\[-0.3cm]
	\includegraphics[width=\legendsizeA,trim= 0cm -0.5cm 0.0cm 0.0cm]{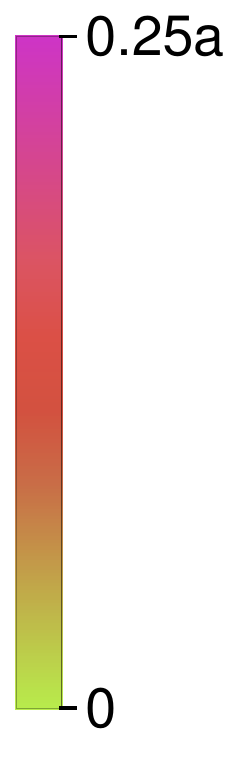} 
    & \includegraphics[width=\figwidth]{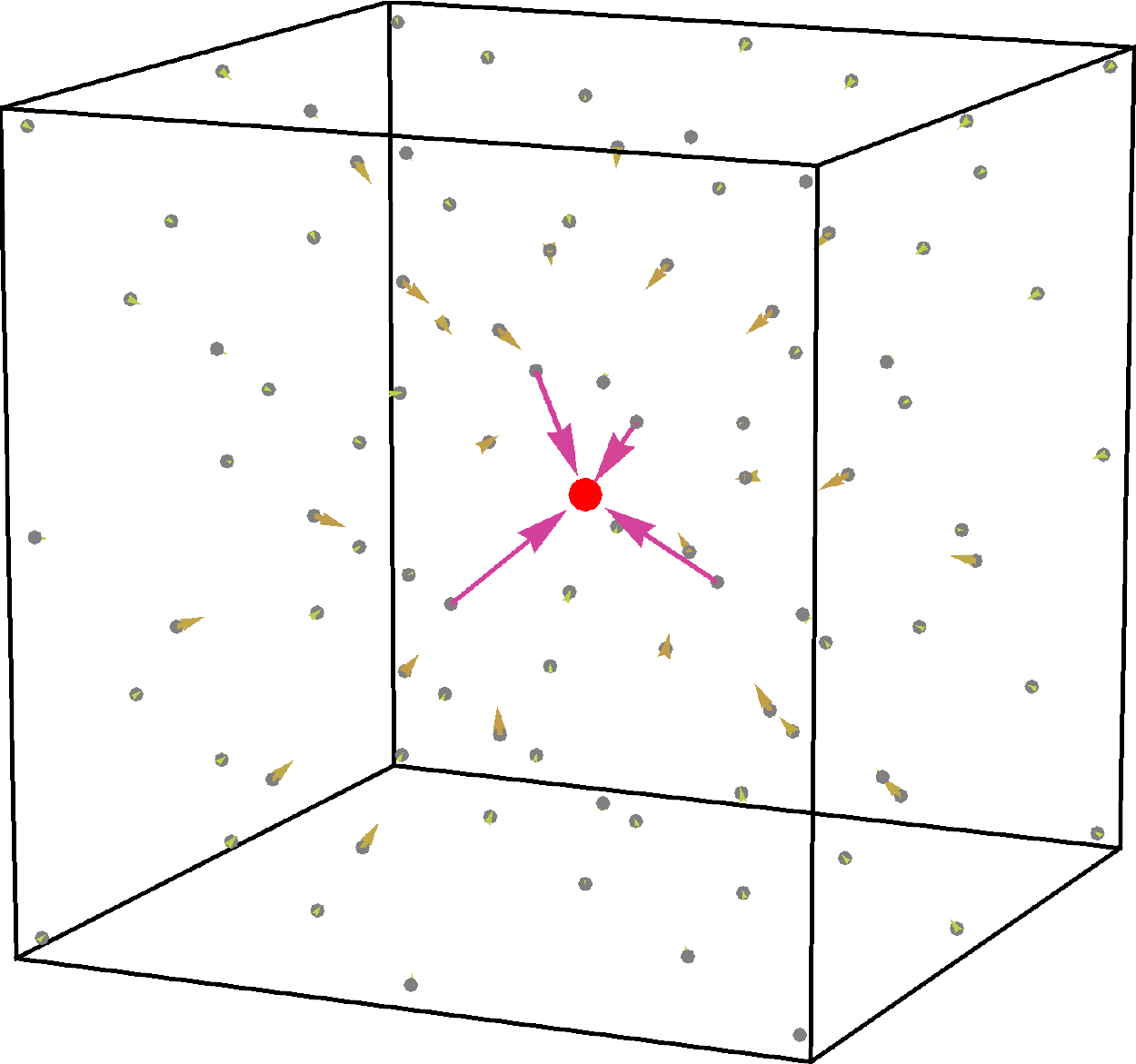} 
    & \includegraphics[width=\figwidthB, trim= 0cm -1.9cm 0.6cm 0.0cm,clip]{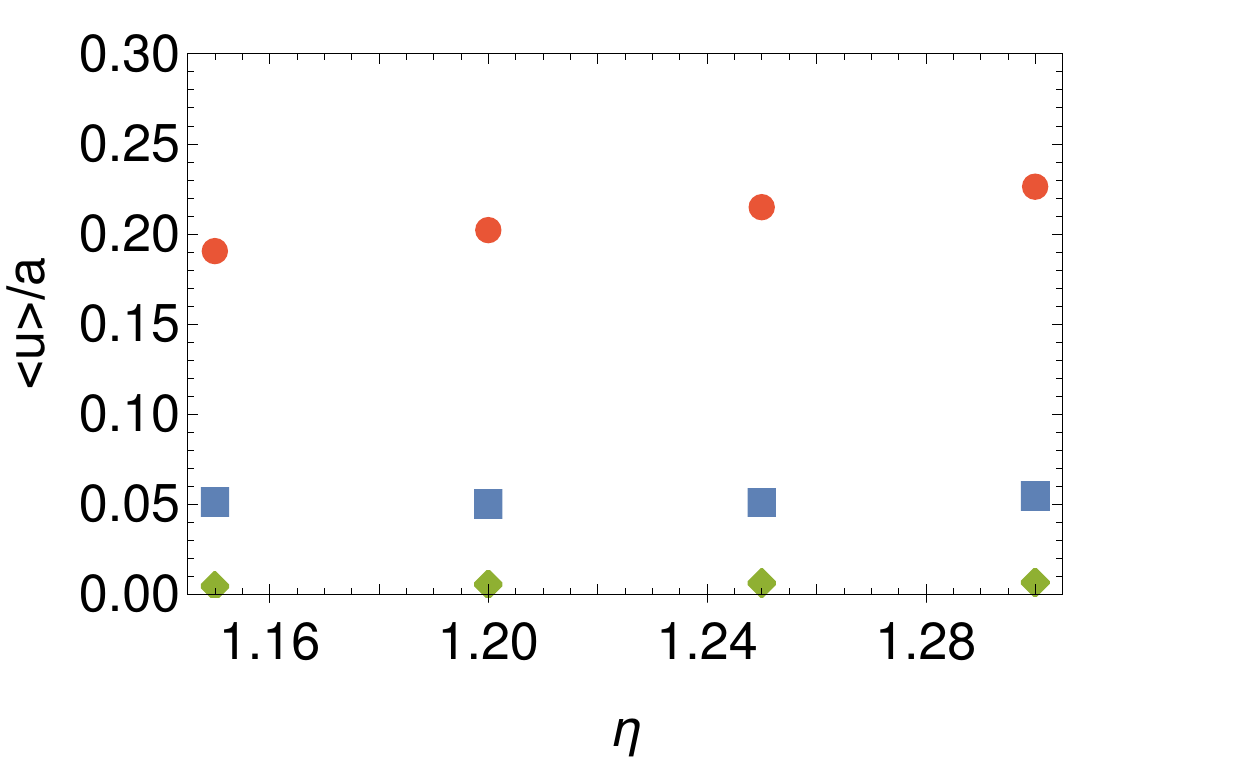}
    & \includegraphics[width=\figwidthB, trim= 0.6cm -2.9cm 0.0cm 0.0cm,clip]{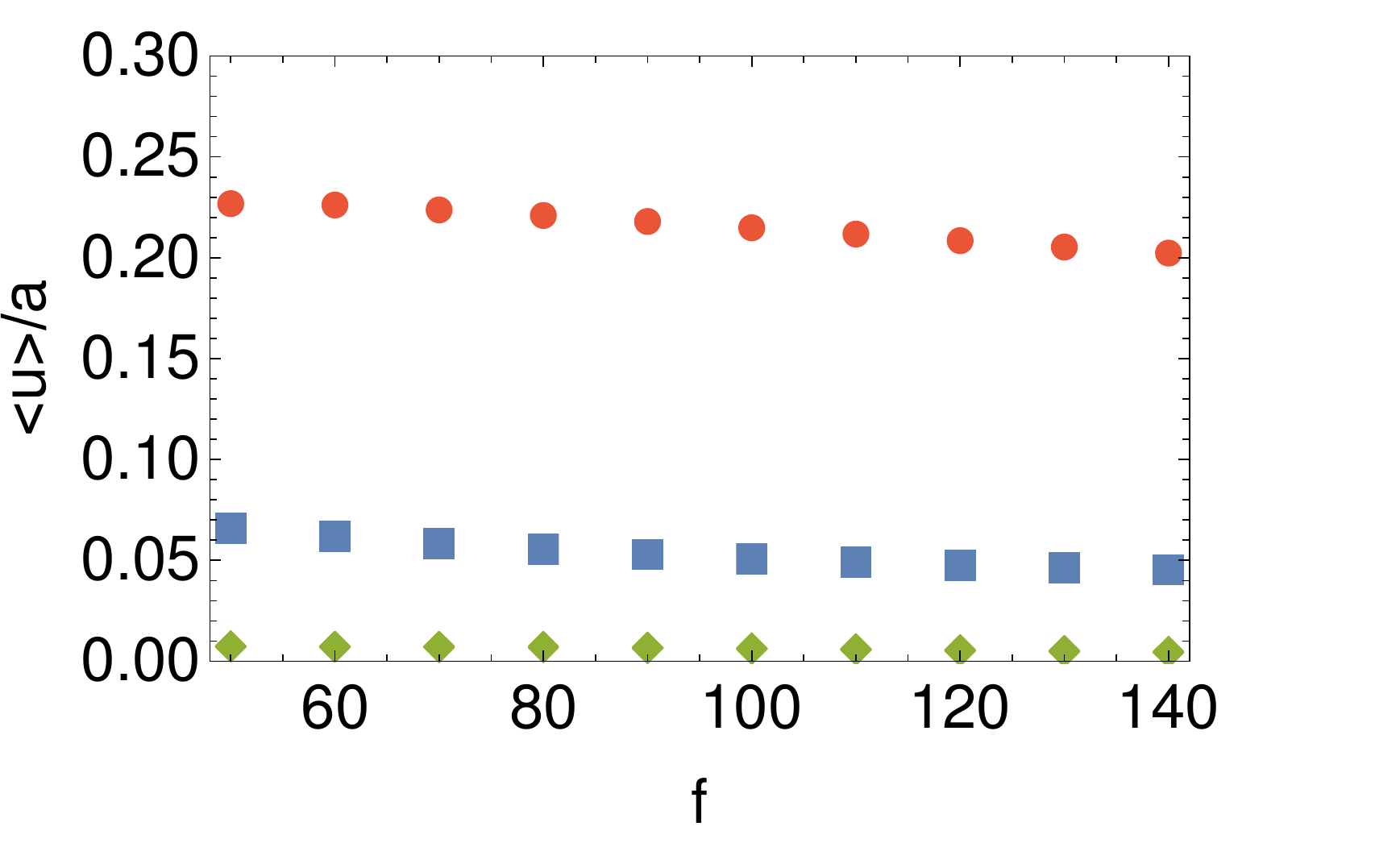}
    & \hspace{-0.2cm}\includegraphics[width=\legendsize,trim= 0.0cm -7.0cm 0.0cm 0.0cm]{images/legend-vac1.pdf} \\
    & d) & \,\, e) & f) &\\[-0.3cm]
    \includegraphics[width=\legendsizeA,trim= 0cm -0.5cm 0.0cm 0.0cm]{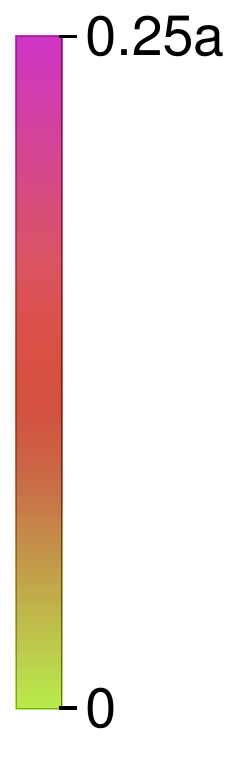} 
    & \includegraphics[width=\figwidth]{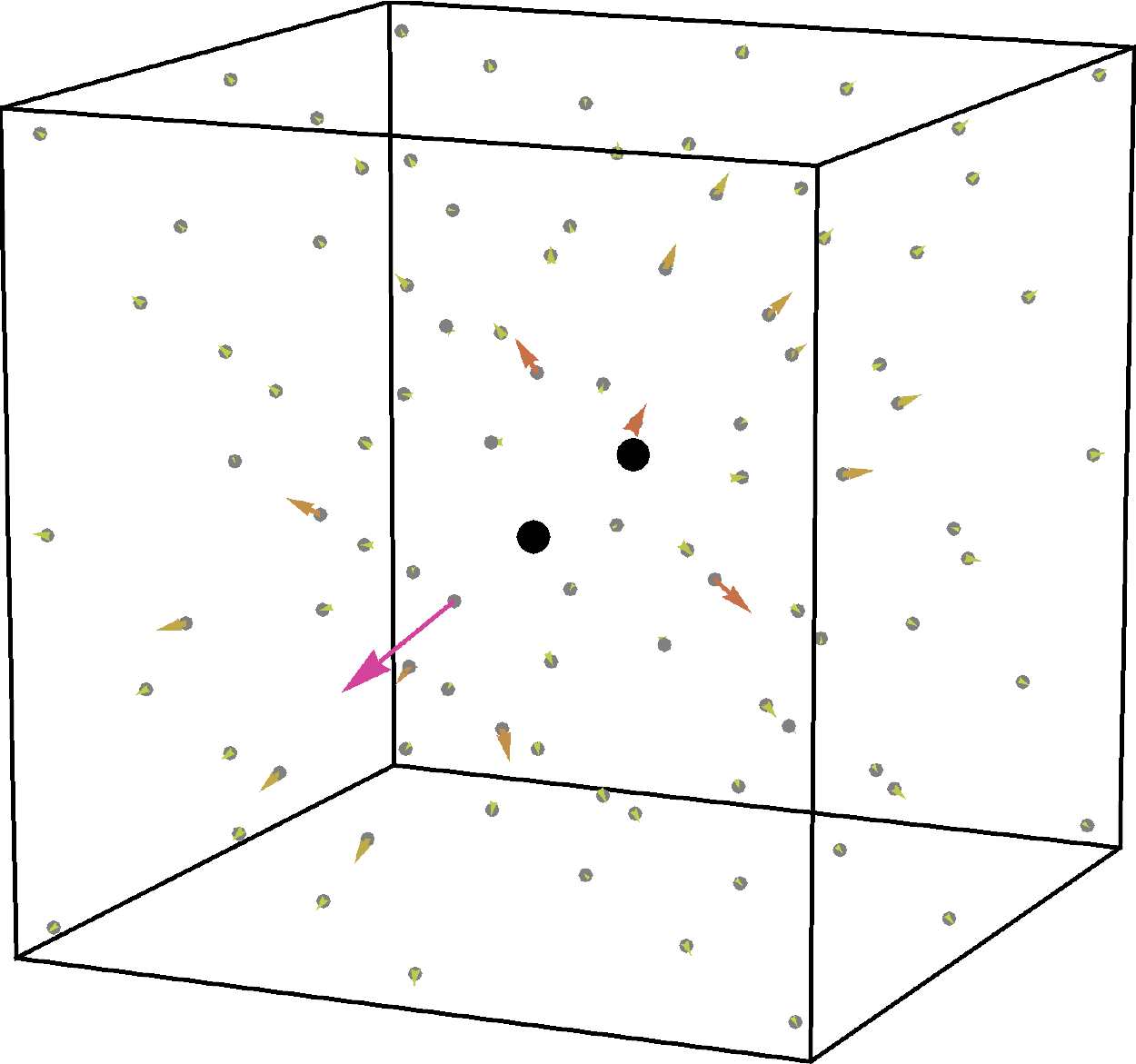} 
    & \includegraphics[width=\figwidthB, trim= 0cm -2.1cm 0.6cm 0.0cm,clip]{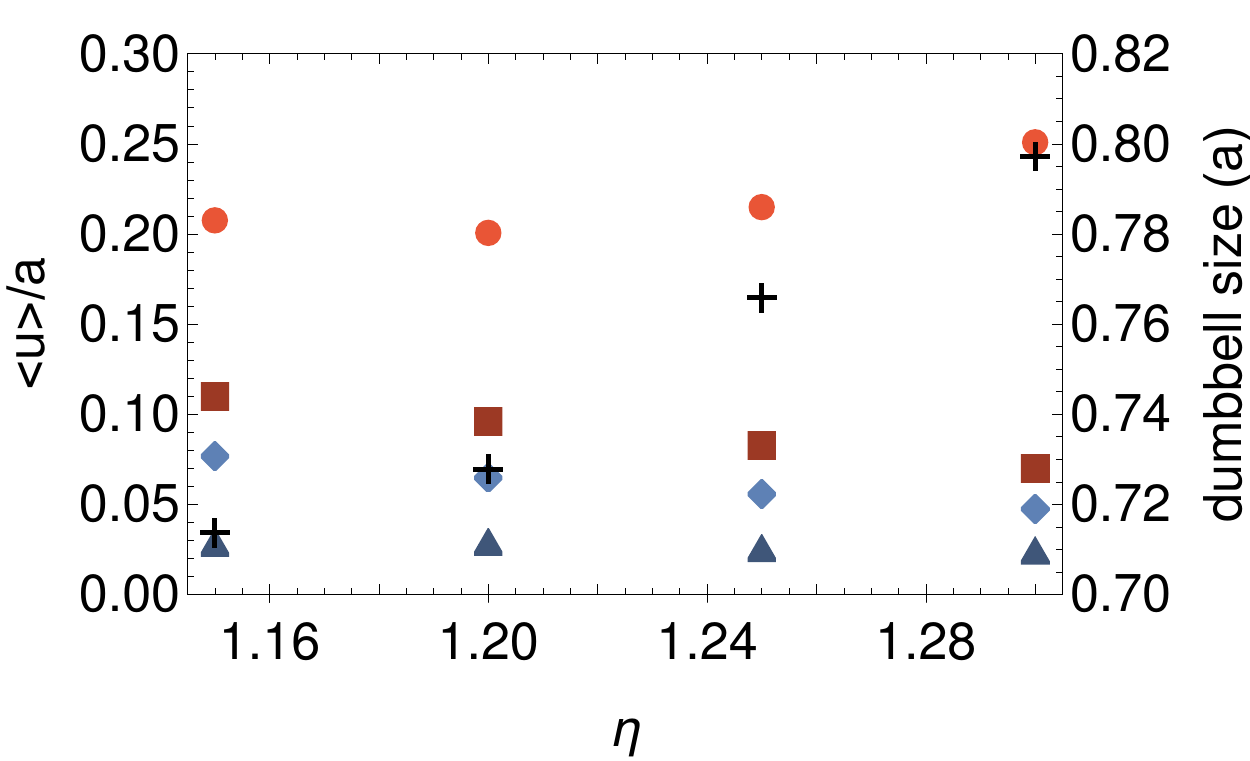}
    & \includegraphics[width=\figwidthB, trim= 0.6cm -2.1cm 0.0cm 0.0cm,clip]{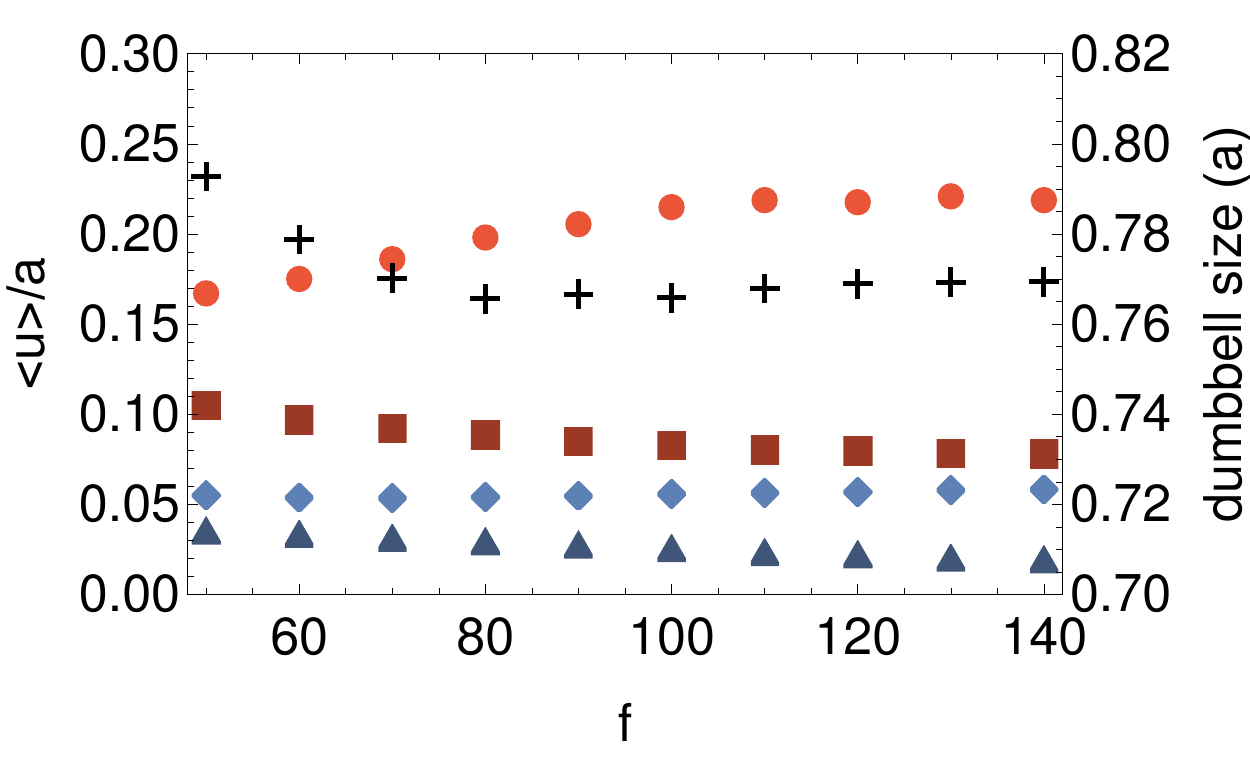}
    & \includegraphics[width=\legendsize,trim= 0.0cm -7.8cm 0.0cm 0.0cm]{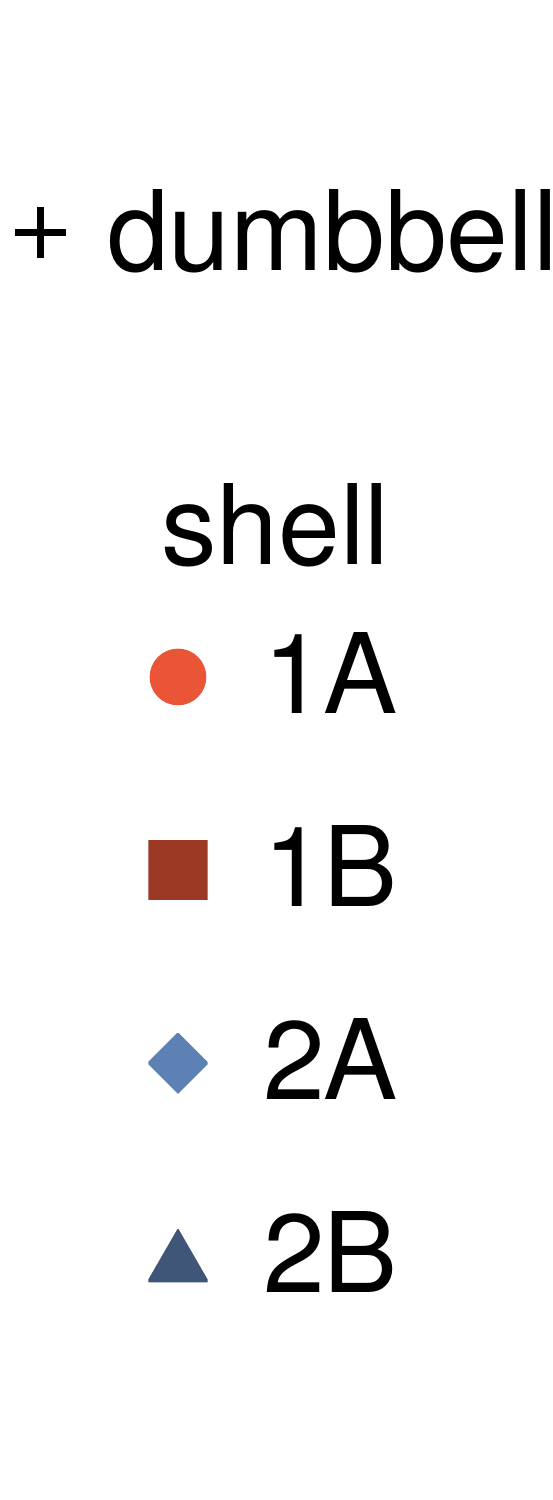}
	\end{tabular}
	\caption[width=1\linewidth]{
	Average deformation of the diamond crystal of star polymers associated with a-c) a vacancy and d-f) an interstitial at packing fraction $\eta=1.45$ and arm number $f=100$. 
	Figures and legends as explained in the caption of Fig. \ref{fig:starpolfcc}.
	}
	\label{fig:starpoldiamond}
\end{figure*}

\begin{figure*}
\begin{tabular}{lll}
	a) & \hspace{0.5cm} & b) \\[-0.3cm]
	\includegraphics[width=\figwidthvc]{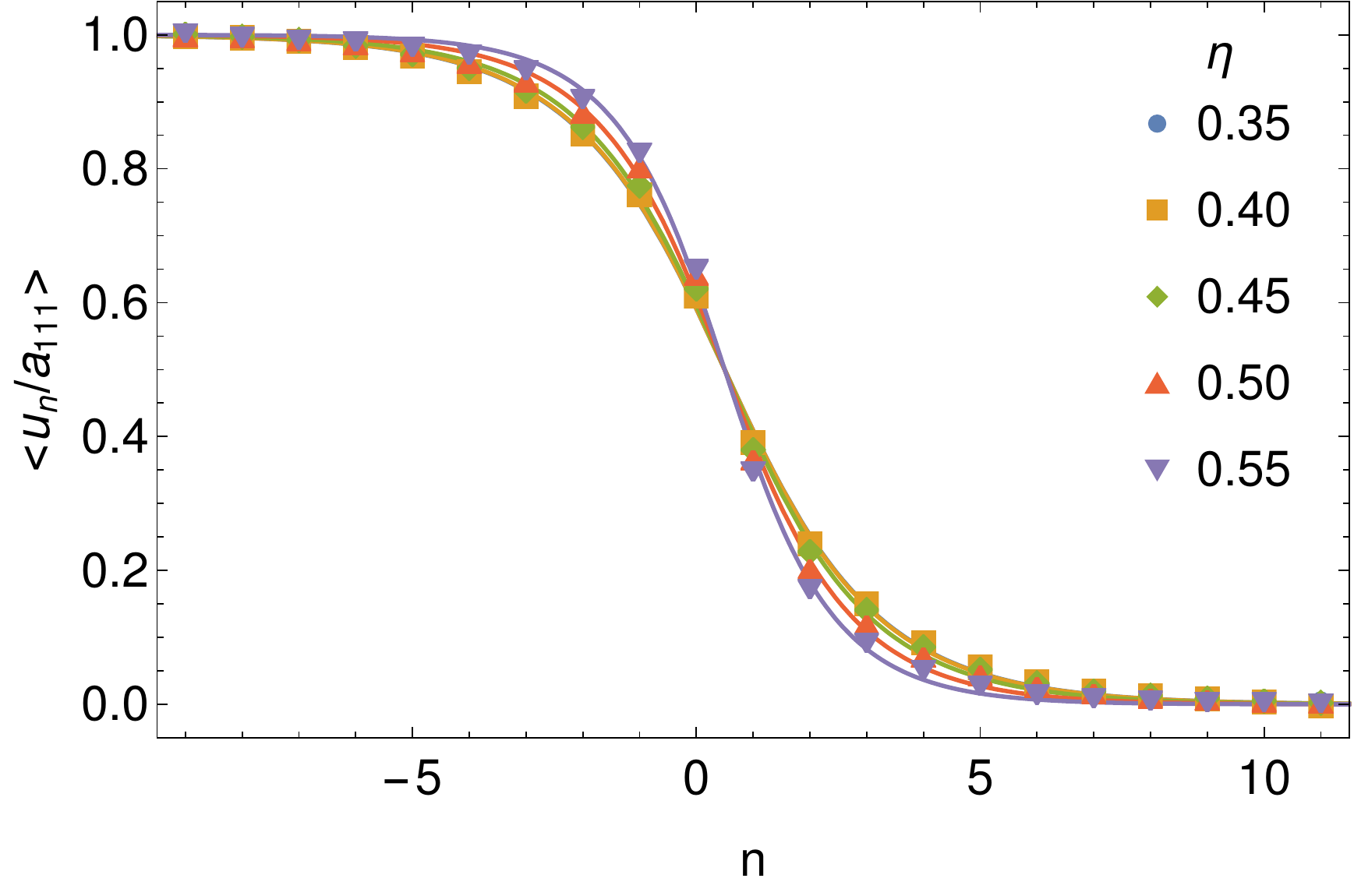}
	& & \includegraphics[width=\figwidthvc]{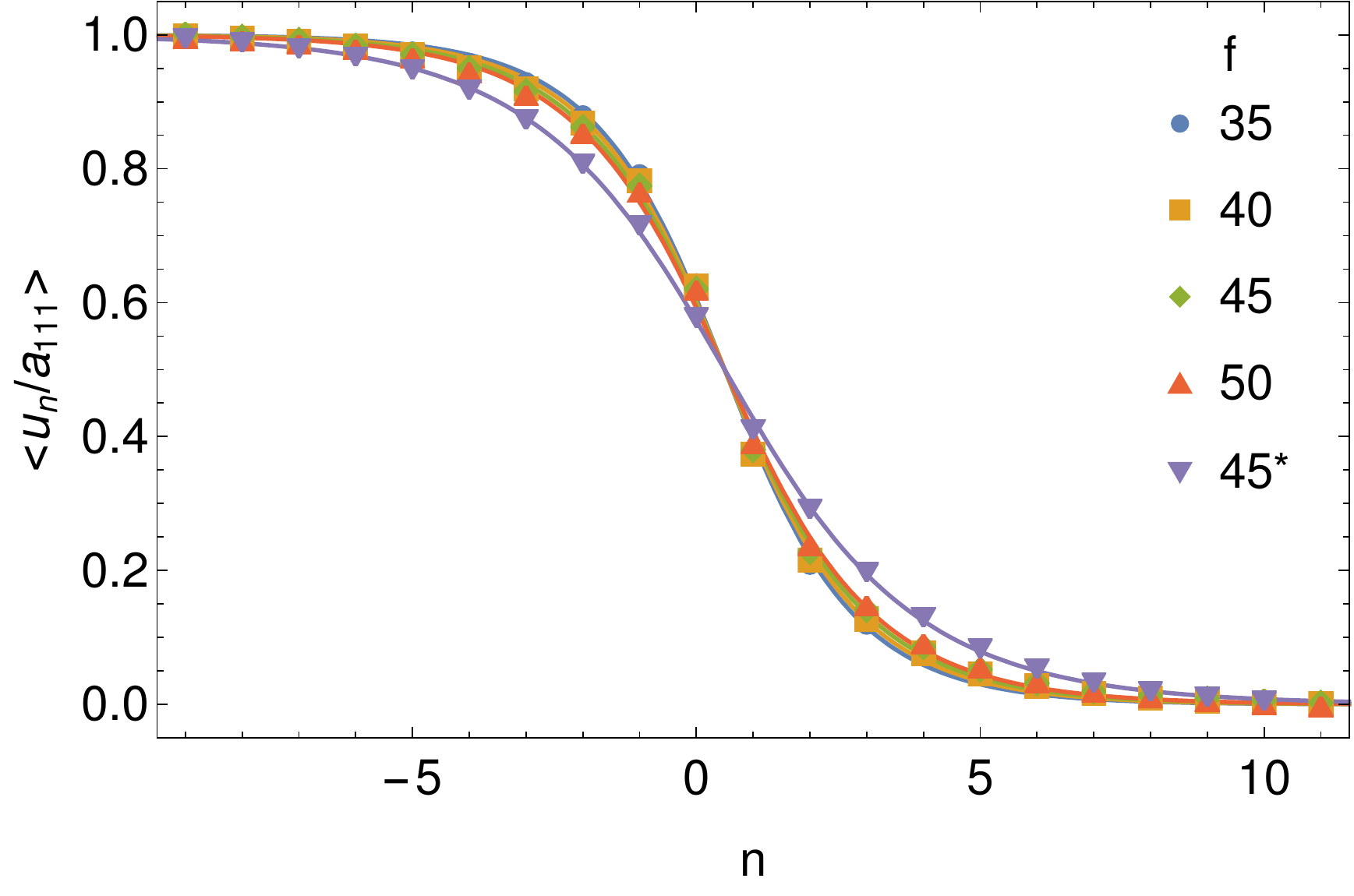}
	\end{tabular}
	\caption[width=1\linewidth]{
	Average displacement along the $\langle 111\rangle$ direction of a crowdion in the BCC crystal of star polymers for a) different packing fractions (at $f=45$) and b) different arm numbers (at $\eta=0.45$). The lines represent the corresponding fitted soliton solutions. $ ^*$Quenched result.
	}
	\label{fig:starpolbcccrowdion}
\end{figure*}

\begin{figure*}
\begin{tabular}{lll}
     a) & b) & c) \\[-0.25cm]
     \includegraphics[width=\widthpdb]{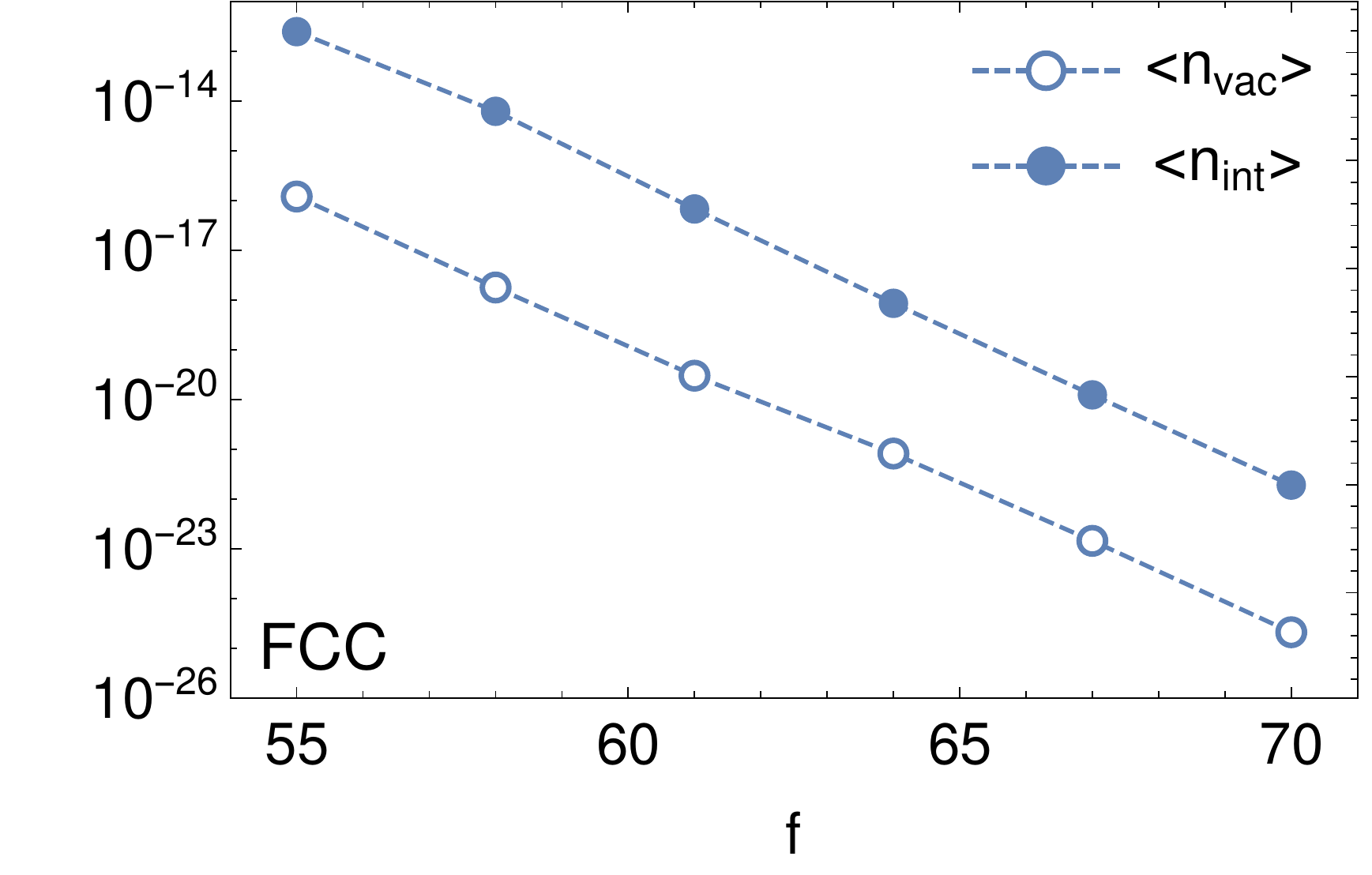}
     & \includegraphics[width=\widthpdb]{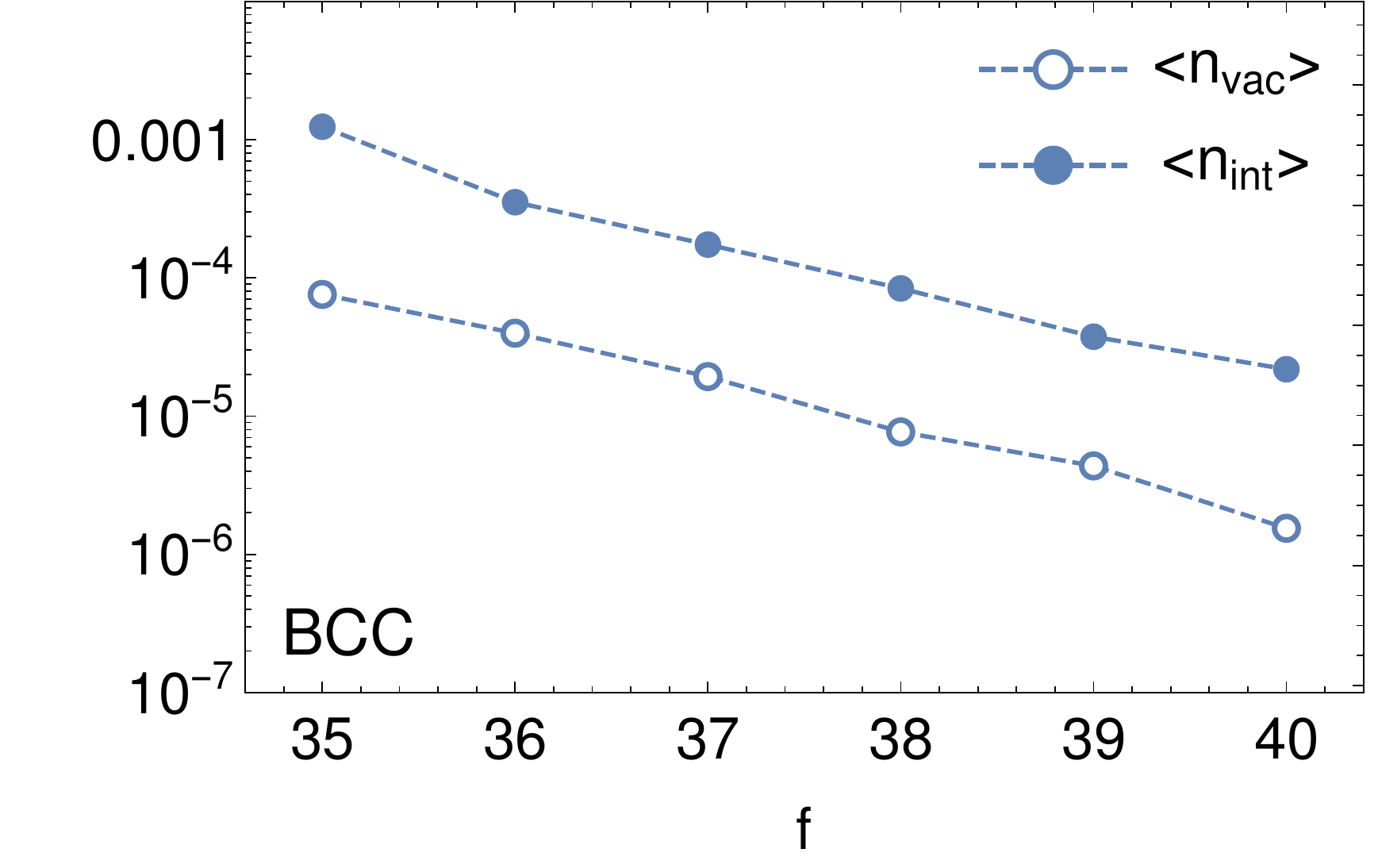}
     & \includegraphics[width=\widthpdb]{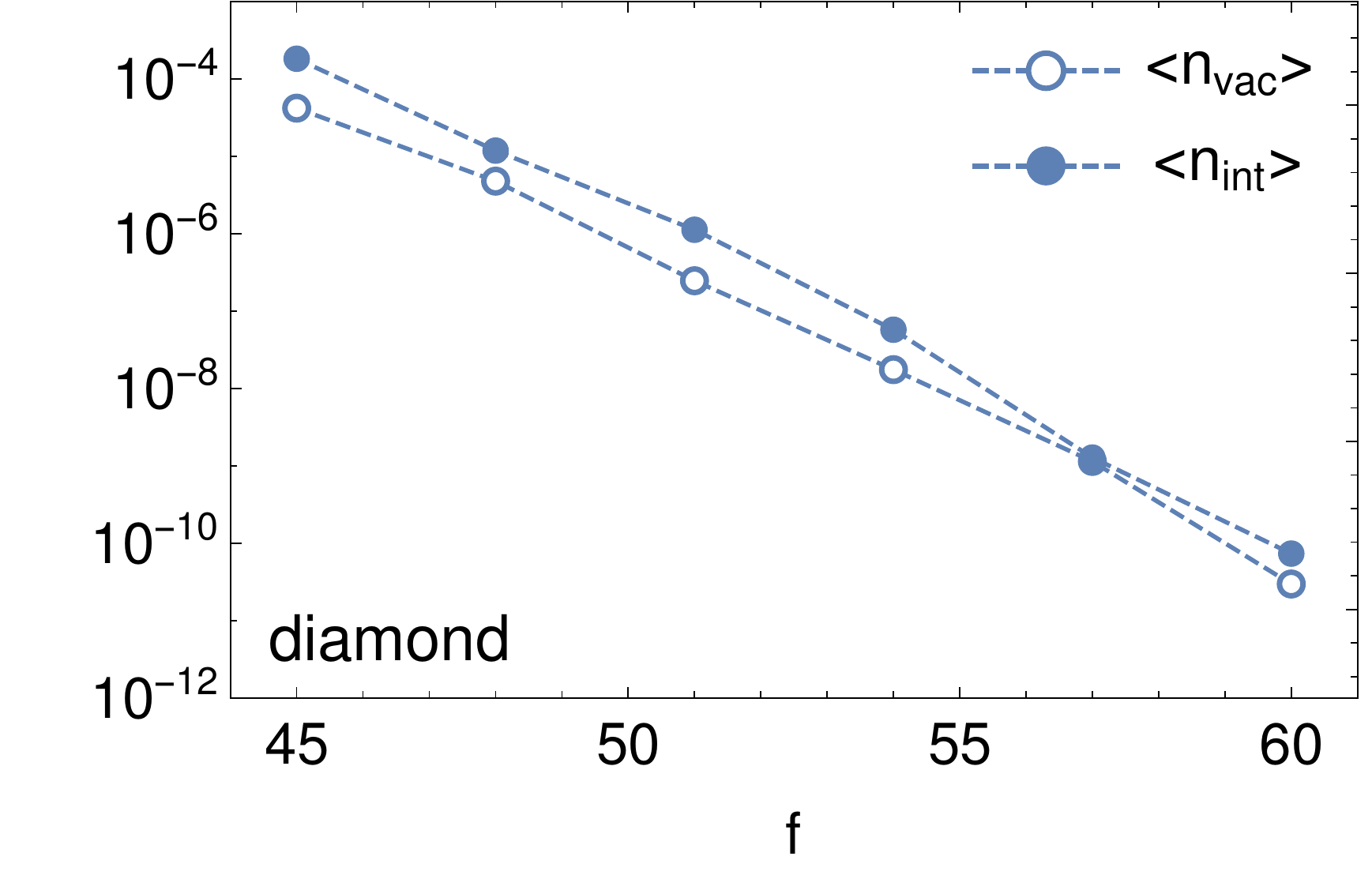}
\end{tabular}
    \caption[width=1\linewidth]{
	Vacancy and interstitial concentrations as a function of the arm number $f$ for the different crystal phases of star polymers.
	The packing fraction of each crystal is given in Tab. \ref{tab:starpolstatepoints}.
	Note that the arm numbers on the left side of a) and b-c) are close to the FCC-BCC phase transition and melting, respectively.
	}
	\label{fig:starpolconcentrations}
\end{figure*}

Figures \ref{fig:starpolfcc}-\ref{fig:starpoldiamond} show the average deformations of the FCC, BCC, and diamond crystals of star polymers. 
In these figures, a) and d) show the deformation for a typical vacancy and interstitial, respectively, while b-c,e-f) give the displacement for the first three neighbor shells as a function of the packing fraction and arm number. Additionally, e-f) give the distance between the two dumbbell particles, i.e. the interstitial and its companion.

Comparing Figs. \ref{fig:starpolfcc} and \ref{fig:starpolbcc} with Figs. \ref{fig:hertzfcc} and \ref{fig:hertzbcc}, we see that the structure of the deformation of the FCC and BCC crystals of star polymers is essentially the same as that for Hertzian spheres. Similar deformations were also seen for the FCC crystal of hard spheres \cite{van2017diffusion} and the FCC and BCC crystals of repulsive point Yukawa particles \cite{alkemade2021point}; hence, we conclude that the nature of the lattice deformation is mainly determined by the crystal structure and not by the specifics of the interaction potential.

Note that the deformation of the diamond crystal associated with a vacancy is similar to that of the BCC crystal. This is not entirely surprising, as the diamond crystal can be constructed from the BCC crystal by removing a particular set of particles. Nonetheless, the deformation of the diamond crystal caused by the interstitial is very different from that of the BCC crystal. The dumbbell does align itself along the $\langle111\rangle$ direction, but, since there are particles ``missing'' along this direction, the deformation is not extended. 

Furthermore, as for Hertzian spheres, we generally find that increasing the packing fraction increases the deformation caused by a vacancy and decreases that caused by an interstitial. Notice, however, that for the interstitials in the FCC and BCC crystals, the deformation first slightly increases before significantly decreasing.
Additionally, just as increasing the temperature makes it easier for Hertzian spheres to overlap, decreasing the arm number makes it easier for star polymers to overlap; hence, decreasing the arm number generally increases the deformation caused by a vacancy and decreases that caused by an interstitial.

Next, to confirm that the interstitial also forms a crowdion in the BCC crystal of star polymers, we compute the average displacement along the defect direction and compare the result to the soliton solution of the sine-Gordon equation. 
Figure \ref{fig:starpolbcccrowdion} shows the resulting displacement for different packing fractions and arm numbers. As for Hertzian spheres, we observe excellent agreement and find that the structure and range of the crowdion is practically independent of the packing fraction and arm number. 
Notice that the quenched crowdion (Fig. \ref{fig:starpolbcccrowdion}a) has a slightly longer range than the averaged crowdions. The range of the averaged crowdions is most likely slightly reduced by the (noisy) transitions from one orientation to another.

Lastly, we predicted the equilibrium concentration of vacancies and interstitials in the different crystals of star polymers.
Similar to Hertzian spheres, we choose one packing fraction (see Tab. \ref{tab:starpolstatepoints}) and varied the arm number to get some insight into the concentrations as a function of the proximity of a phase transition. For the BCC and diamond crystals this was the melting transition, while for the FCC crystal this was the FCC-BCC phase boundary. The resulting concentrations are shown in Fig. \ref{fig:starpolconcentrations}. 
We observe that all concentrations increase with decreasing arm number, meaning that they are highest closest to the phase transition. 
Notice, however, that the concentrations in the FCC crystal are extremely low, i.e. more than 8 orders of magnitude lower than in the BCC and diamond crystals. The most probable reason for this is the different nature of the phase transitions: solid-fluid for BCC and diamond crystals versus solid-solid for FCC crystal.

As for most of the Hertzian sphere crystals, we find that the concentration of interstitials is higher than that of vacancies in all three star polymer crystals. This is even the case for the FCC crystal, for which hard spheres and Hertzian spheres have higher vacancy concentrations. 
Yet, most noteworthy, is the very high concentration of interstitials in the BCC crystal: on the order of 0.1\% near melting.


\section{Conclusions}
In conclusion, we have characterized the point defects that appear in the wide range of crystal structures that occur in Hertzian spheres and star polymers. We found that the defects caused similar deformations in the FCC and BCC crystals of different models with isotropic repulsion, and thus concluded that the nature of the lattice distortion is mainly determined by the crystal structure and not by the specifics of the repulsive (isotropic) interaction potential.  Hence, even though not all of the models we study here  are  directly realizable in the lab, this work provides new insights into how defects manifest in crystals of soft particles.  

For most crystal structures, we found three-dimensional and local deformations, with three interesting exceptions: i) the vacancy in the SC crystal, which causes a lattice distortion that is extended in two dimensions, ii) the vacancy in the H crystal, which forms a voidion, and iii) the interstitial in the BCC crystals, which forms a crowdion. 
We showed that the structure of these voidion and crowdions is essentially independent of the system parameters.

Interestingly, in all cases outside of the FCC crystal of Hertzian spheres, the interstitial concentrations were found to be higher than the vacancy concentrations.  The largest defect concentrations arose as interstitials in the BCC crystals, which reached as high as 1\% in the Hertzian sphere model.


\section{Acknowledgements}
We would like to thank Frank Smallenburg, Berend van der Meer and Rinske Alkemade for many useful discussions. L.F. and M.d.J. acknowledge funding from the Vidi research program with project number VI.VIDI.192.102 which is financed by the Dutch Research Council (NWO).

\bibliography{paper}

\end{document}